\documentclass[twocolumn]{aastex62}
\pdfoutput=1 
\usepackage{amsmath,amstext}
\usepackage[T1]{fontenc}
\usepackage{apjfonts} 
\usepackage[figure,figure*]{hypcap}

\usepackage{graphicx}
\usepackage{graphics}
\usepackage{color}
\usepackage{lipsum}
\usepackage[caption=false]{subfig}
\definecolor{dartmouthgreen}{rgb}{0.05, 0.5, 0.06}
\usepackage{afterpage}
\usepackage{natbib}

\def\arcsec{\hbox{$^{\prime\prime}$}}
\def\deg{\hbox{$^\circ$}}
\def\hr{\textsuperscript{h}}
\def\min{\textsuperscript{m}}
\def\sec{\textsuperscript{s}\hspace{-0.7mm}}

\def\nh{$N_{\rm H}$}

\def\nh{N_\mathrm{H}}

\def\chandra{\textit{Chandra}}
\def\xmm{\textit{XMM-Newton}}
\def\WISE{\textit{WISE}}

\graphicspath{ {Images/} }

\shorttitle{Buried Black Hole Growth in Mergers}
\shortauthors{R.W. Pfeifle et al.}

\begin{document}

\title{Buried Black Hole Growth in IR-Selected Mergers: New Results from Chandra}

\author{Ryan W. Pfeifle}
\affiliation{Department of Physics and Astronomy, George Mason University, MS3F3, 4400 University Drive, Fairfax, VA 22030, USA}
\author{Shobita Satyapal}
\affiliation{Department of Physics and Astronomy, George Mason University, MS3F3, 4400 University Drive, Fairfax, VA 22030, USA}
\author{Nathan J. Secrest}
\affiliation{U.S. Naval Observatory, 3450 Massachusetts Avenue NW, Washington, DC 20392, USA}
\author{Mario Gliozzi}
\affiliation{Department of Physics and Astronomy, George Mason University, MS3F3, 4400 University Drive, Fairfax, VA 22030, USA}
\author{Claudio Ricci}
\affiliation{N\'ucleo de Astronom\'ia de la Facultad de Ingenier\'ia, Universidad Diego Portales, Av. Ej\'ercito Libertador 441, Santiago, Chile}
\affiliation{Kavli Institute for Astronomy and Astrophysics, Peking University, Beijing 100871, China}
\affiliation{Chinese Academy of Sciences South America Center for Astronomy and China-Chile Joint Center for Astronomy, Camino El Observatorio 1515, Las Condes, Santiago, Chile}
\author{Sara L. Ellison}
\affiliation{Department of Physics and Astronomy, University of Victoria, Victoria, BC V8P 1A1, Canada}
\author{Barry Rothberg}
\affiliation{Department of Physics and Astronomy, George Mason University, MS3F3, 4400 University Drive, Fairfax, VA 22030, USA}
\affiliation{LBT Observatory, University of Arizona, 933 N. Cherry Ave., Tuscan, AZ 85721, USA}
\author{Jenna Cann}
\affiliation{Department of Physics and Astronomy, George Mason University, MS3F3, 4400 University Drive, Fairfax, VA 22030, USA}
\author{Laura Blecha}
\affiliation{Department of Physics, University of Florida, P.O. Box 118440, Gainesville, FL 32611-8440, USA}
\author{James K. Williams}
\affiliation{Department of Physics and Astronomy, George Mason University, MS3F3, 4400 University Drive, Fairfax, VA 22030, USA}
\author{Anca Constantin}
\affiliation{Department of Physics and Astronomy, James Madison University, PHCH, Harrisonburg, VA 22807, USA}

\begin{abstract}
Observations and theoretical simulations suggest that a significant fraction of merger-triggered accretion onto supermassive black holes (SMBHs) is highly obscured, particularly in late-stage galaxy mergers, when the black hole is expected to grow most rapidly. Starting with the \textit{Wide-Field Infrared Survey Explorer} all-sky survey, we identified a population of galaxies whose morphologies suggest an ongoing interaction and which exhibit red mid-infrared colors often associated with powerful active galactic nuclei (AGNs).  In a follow-up to our pilot study, we now present \chandra{}/ACIS and \xmm{} X-ray observations for the full sample of the brightest 15 IR-preselected mergers. All mergers reveal at least one nuclear X-ray source, with 8 out of 15 systems exhibiting dual nuclear X-ray sources, highly suggestive of single and dual AGNs. Combining these X-ray results with optical line ratios and with near-IR coronal emission line diagnostics, obtained with the near-IR spectrographs on the Large Binocular Telescope, we confirm that 13 out of the 15 mergers host AGNs, two of which host dual AGNs. Several of these AGN are not detected in the optical. All X-ray sources appear X-ray weak relative to their mid-infrared continuum, and of the nine X-ray sources with sufficient counts for spectral analysis, eight reveal strong evidence of high absorption with column densities of $N_\mathrm{H} \gtrsim 10^{23}$~cm$^{-2}$. These observations demonstrate a significant population of single and dual AGNs are missed by optical studies due to high absorption, adding to the growing body of evidence that the epoch of peak black hole growth in mergers occurs in a highly obscured phase.
\end{abstract}


\section{Introduction}

Based on both observations and theoretical simulations, it is clear that galaxy interactions are ubiquitous and play a crucial role in the formation and evolution of galaxies \citep{toomre1972,mihos1996,schweizer1982,schweizer1996,hibbard1996,rothberg2004}.  Numerical simulations predict that gravitational instabilities during galaxy interactions cause large radial gas inflows that can fuel the central black holes \citep{barnes1996,hopkins2008a,dimatteo2005}. After three decades of extensive research, however, the observational connection between black hole growth, traced by active galactic nuclei (AGNs), and mergers is still a topic of vigorous debate. Several morphological studies of AGN hosts suggest that, by number, most AGNs are not associated with mergers \citep[e.g.,][]{kocevski2012,cisternas2011,schawinski2012,simmons2012,villforth2014,rosario2015,bruce2016,mechtley2016, villforth2017}. On the other hand, recent studies of kinematic pairs have shown that mergers exhibit a clear enhancement of AGN activity relative to a control sample of isolated galaxies \citep[e.g.,][]{ellison2011,silverman2011,satyapal2014}, demonstrating that mergers do trigger some AGNs, though recent simulations carried out in \citet{steinborn2018} suggest that mergers are not the statistically dominant AGN triggering mechanism. However, at the highest luminosities, many studies suggest that \textit{most} AGNs are triggered by mergers \citep[e.g.,][]{guyon2006,urrutia2008,koss2012,treister2012,glikman2015,fan2016,goulding2018,donley2018,sanders1988,rothberg2013,canalizo2001}, a result that is consistent with previous simulations \citep{barnes1991,hopkins2006,hopkins2008a,hopkins2014}.
\par

A major impediment that limits our ability to {\it quantify} the role of mergers in SMBH growth is that heavily obscured AGNs are not well sampled, because the vast majority of studies are conducted at optical wavelengths.  Obscuration from gas and dust is expected during the merger, since the inflowing material that can potentially feed the black hole can also obscure the activity. The greatest obscuration is expected precisely when the black hole accretion rates are the highest and dual AGNs with kiloparsec scale pair separations are expected to be found, as predicted by recent simulations \citep{blecha2018}. Indeed, recent observations demonstrate a rising fraction of highly buried AGNs with merger stage, and a prevalence of advanced mergers in samples of heavily obscured AGNs \citep[e.g.,][]{koss2010,urrutia2012,satyapal2014,kocevski2015,fan2016,weston2017,lansbury2017,ricci2017,donley2018}. The few contradictory studies are based on soft X-ray selection, which are biased against the most obscured AGNs \citep[e.g.,][]{villforth2014,villforth2017}. This suggests that highly obscured AGNs represent a key stage in the coevolution of galaxies and BHs, and may represent the hotly debated missing link between mergers and BH growth. It also suggests that there may be significant large-scale obscuration that is not directly associated with the tori of each individual AGN.

In addition to predicting that the heaviest obscuration occurs during the period of peak black hole growth in late-stage mergers, potentially limiting the detection and characterization of such AGNs when the accretion rates are highest, simulations also predict that accretion onto \textit{both} SMBHs occurs at this stage \citep{vanwassenhove2012,blecha2013,blecha2018}. Therefore  {\em dual} AGNs with separations $<$ 10~kpc likely coincide with the period of most rapid black hole growth and therefore represent a key stage in the evolution of galaxies which contributes significantly to the SMBH accretion history of the universe. Furthermore, dual AGNs represent the likely forerunner of SMBH binaries and mergers $\--$ the origin of the most titanic gravitational wave events in the universe \citep{merritt2005} $\--$ the frequency of which is of great importance to future gravitational wave searches in this mass regime. Observationally confirmed cases of dual AGNs are extremely rare, despite strong theoretical reasons for their existence and extensive observational campaigns, and until recently most have been discovered serendipitously. In recent years a small but growing number of dual AGN candidates have been discovered through systematic searches using the Sloan Digital Sky Survey (SDSS) and double-peaked emission lines as a pre-selection strategy \citep{comerford2011,comerford2015,barrows2017,mullersan2015}, although follow-up observations confirm duals in only a small fraction \citep{comerford2015,mullersan2015,fu2012}.
\par
Motivated by the possibility that black hole activity may be obscured in the most advanced merger stages where dual AGNs are expected to be found, we have been conducting a multiwavelength campaign of a sample of morphologically identified advanced mergers that display red mid-infrared colors often associated with powerful AGNs \citep{stern2012,satyapal2014,assef2013}. Based upon their optical spectroscopic classifications, the vast majority of these advanced mergers are expected to be dominated by star-formation rather than AGN activity, suggesting that they may represent an obscured population of AGNs that cannot be found through optical studies. In \citet{satyapal2017} (hereafter, Paper I), we presented {\it Chandra/ACIS} observations and near-infrared spectra with the {\it Large Binocular Telescope} (LBT) of six advanced mergers with projected pair separations less than $\approx$ 10 kpc. The combined X-ray, near-infrared, and mid-infrared properties of these mergers provided confirmation that four out of the six mergers host at least one AGN, and four of the six mergers possibly host dual AGNs, despite showing no firm evidence for AGNs based on optical spectroscopic studies. In \citet{ellison2017}, an additional mid-infrared selected merger was also confirmed as a dual using multiwavelength observations. These observations strongly suggested that optical studies miss a significant fraction of single and dual AGN candidates in advanced mergers, and that infrared selection is potentially an extremely effective way to identify them. All of the AGN candidates identified in Paper I appeared X-ray weak relative to their mid-infrared luminosities, suggesting that the buried AGNs in these mergers are highly absorbed, with intrinsic column densities of $N_\mathrm{H} \gtrsim 10^{24}$~cm$^{-2}$, consistent with the aforementioned numerical studies.

In this paper, we extend our study to nine new mid-infrared selected advanced mergers for which we were awarded {\it Chandra/ACIS} observations. Together with Paper I, we present a comprehensive X-ray investigation of 15 infrared-selected advanced mergers. In Section 2, we describe our sample selection, followed by a discussion of our observations and data analysis in Sections 3 and 4. In Section 5, we describe our results. We discuss the nature of the nuclear sources in Section 6. We present our final conclusions in Section 7. We provide a detailed description of our results for each interacting system in the Appendix. The full near-IR investigation will be presented in the next paper, Constantin et al., in prep.

Throughout this paper we adopt the following cosmological values: $H_0$ = 70 km s$^{-1}$ Mpc$^{-1}$, $\Omega_M=0.3$, and $\Omega_\Lambda = 0.7$. Angular distances and luminosities were calculated following \citet{wright2006}.

\section{Sample Selection}

As described in Paper I, we assembled a large sample of interacting galaxies using the Galaxy Zoo project \citep{lintott2008},\footnote{\url{http://www.galaxyzoo.org}} from the Sloan Digital Sky Survey (SDSS) DR7 \citep{abazajian2009}. We refer to Paper I for the details of the sample selection, although we provide a brief overview. We used the weighted-merger-vote-fraction, $f_{m}$, to quantify the interaction status of the sample. This parameter varies from 0 to 1, where 0 represents clearly isolated galaxies and a value of 1 represents a definite merger \citep{darg2010MNRAS}, with 0.4 representing a high likelihood of being a strongly disturbed merger \citep{darg2010MNRAS}. Here and in Paper I, we searched the AllWISE release of the {\it WISE} catalog,\footnote{\url{http://wise2.ipac.caltech.edu/docs/release/allwise/}} for galaxies with $f_{m} > 0.4$, {\it WISE} detections in the first 2 bands with a signal to noise ratio greater than 5$\sigma$ and W1-W2 colors in excess of 0.5. We adopted this color cut since simulations \citep{blecha2018} demonstrate that a color cut of W1-W2 > 0.5 yields a more complete selection of dual AGNs in mergers (see section 3.1-3.4 of \citealt{blecha2018}) than the more widely adopted W1-W2 > 0.8 color cut from \citet{stern2012}, which misses the majority of the lifetime of an AGN within an advanced merger. We then visually inspected the sample and selected all mergers with at least two distinct nuclei with nuclear separations of $<$ 10~kpc that are spatially resolvable by \chandra{} (angular resolution of 1\arcsec{}, or $\sim$ 1.3 kpc at z $=0.07$, the median redshift of our sample). This ensured that our selected mergers were likely to be strongly interacting and contain obscured AGNs \citep{stern2012,satyapal2014,satyapal2017}. These selection criteria resulted in a total of 178 candidates. In Paper I, we presented follow-up X-ray observations of the six brightest mergers in the W2 band that met our criterion. In this work, we present X-ray observations of the next nine brightest mergers, resulting in a total sample size of 15 mergers. Note that our adopted pair separation cutoff was chosen since confirmed dual AGNs at these pair separations are rare; this pairing phase allows us to probe not only the stage of most active black hole growth but also the only spatially observationally accessible precursors to the true binary AGN phase \citep{vanwassenhove2012,blecha2013}. 
\par Our working definition of a dual AGN in this paper  and Paper I corresponds to a merger with two confirmed nuclear AGNs with pair separations of less than 10~kpc. Note, however, due to the spatial resolution limit of \chandra{}, we cannot resolve pair separations of $<$ 1.3 kpc for the median redshift of our sample. In Figure~\ref{fig:sample_images} we show three-color SDSS images of our targets. The SDSS images show the targets are strongly disturbed systems, suggesting they are advanced mergers.  In Table \ref{table:sample}, we list the basic properties of the sources. Redshifts, stellar masses, and emission line fluxes for the galaxies in our sample were taken from the SDSS data release 7 (DR7) \citep{abazajian2009}, a result of the Max Planck Institut f\"{u}r Astrophysik/Johns Hopkins University (MPA/JHU) collaboration.\footnote{\url{http://www.mpa-garching.mpg.de/SDSS/}} SDSS spectra are available for both nuclei in only 6 out of 15 systems (SDSS fiber locations are displayed in  Figure~\ref{fig:sample_images}). The targets have highly disturbed morphologies, making it difficult to obtain meaningful estimates of their stellar masses and mass ratios because of blended photometry. The optical spectral class of each target was determined using the BPT line ratio diagnostics \citep{baldwin1981} following the classification scheme of \citet{kewley2001,kewley2006} for AGNs and \citet{kauffmann2003} for composites.  Only 1 of the 15 mergers in our full sample are identified in the optical regime as dual AGNs, while 7 out of 15 mergers contain at least one optical AGN (see the top panels in Figure~\ref{fig:sample_images}). One of our targets, J1356+1822, also known as Mrk\,463, is a well-known ULIRG  \citep{surace1999,surace2000} which met our selection criterion and was included in our sample. Mrk 463 hosts a previously discovered dual AGN system \citep{bianchi2008}.

\subsection{SED Decomposition}
In order to determine the 8-1000~\micron\ IR luminosities of our objects, we fit their spectral energy distributions (SEDs) using custom Python code employed in \citet{powell2018} for the \textit{Swift} BAT AGNs \citep{koss2017}. In brief, this code convolves the user's choice of SED templates with the system responses corresponding to their data, and the data is fit via weighted non-negative least-squares, with the weights being the inverse variances of the data. For our data, we combined an AGN template from \citet{fritz2006}, shown in Figure~1 of \citet{Hatziminaoglou2008}, with two templates from \citet{chary2001} corresponding to the lowest and highest IR luminosity star-forming galaxies, which differ primarily in the equivalent widths of their polycyclic aromatic hydrocarbon (PAH) features and the strength of the IR emission compared with the stellar emission. The AGN template has $W1-W2$/$W2-W3$ synthetic colors of 0.86/2.40, while the low-luminosity and high-luminosity star-forming galaxies have corresponding colors of 0.19/1.72 and 0.82/5.67, and so our templates have WISE colors typical of AGNs, spiral galaxies, and LIRGs/ULIRGs \citep[e.g.,][Figure~12]{wright2010}.
\afterpage{\clearpage}
\begin{figure*}[!ht]
    \centering
    \hspace{-2mm}
    \subfloat{\includegraphics[width=0.35\linewidth]{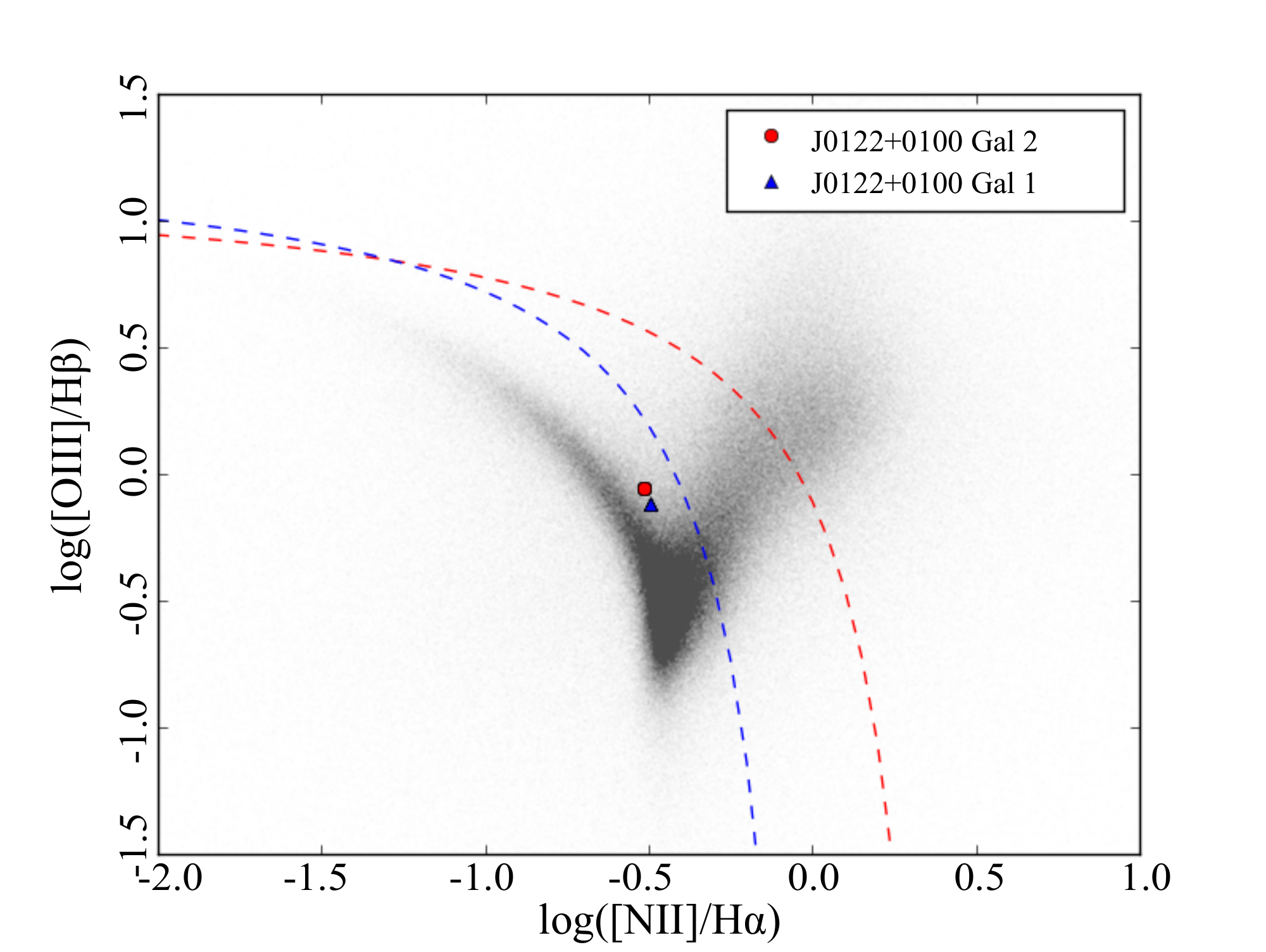}}
    \hspace{-5mm}
    \subfloat{\includegraphics[width=0.35\linewidth]{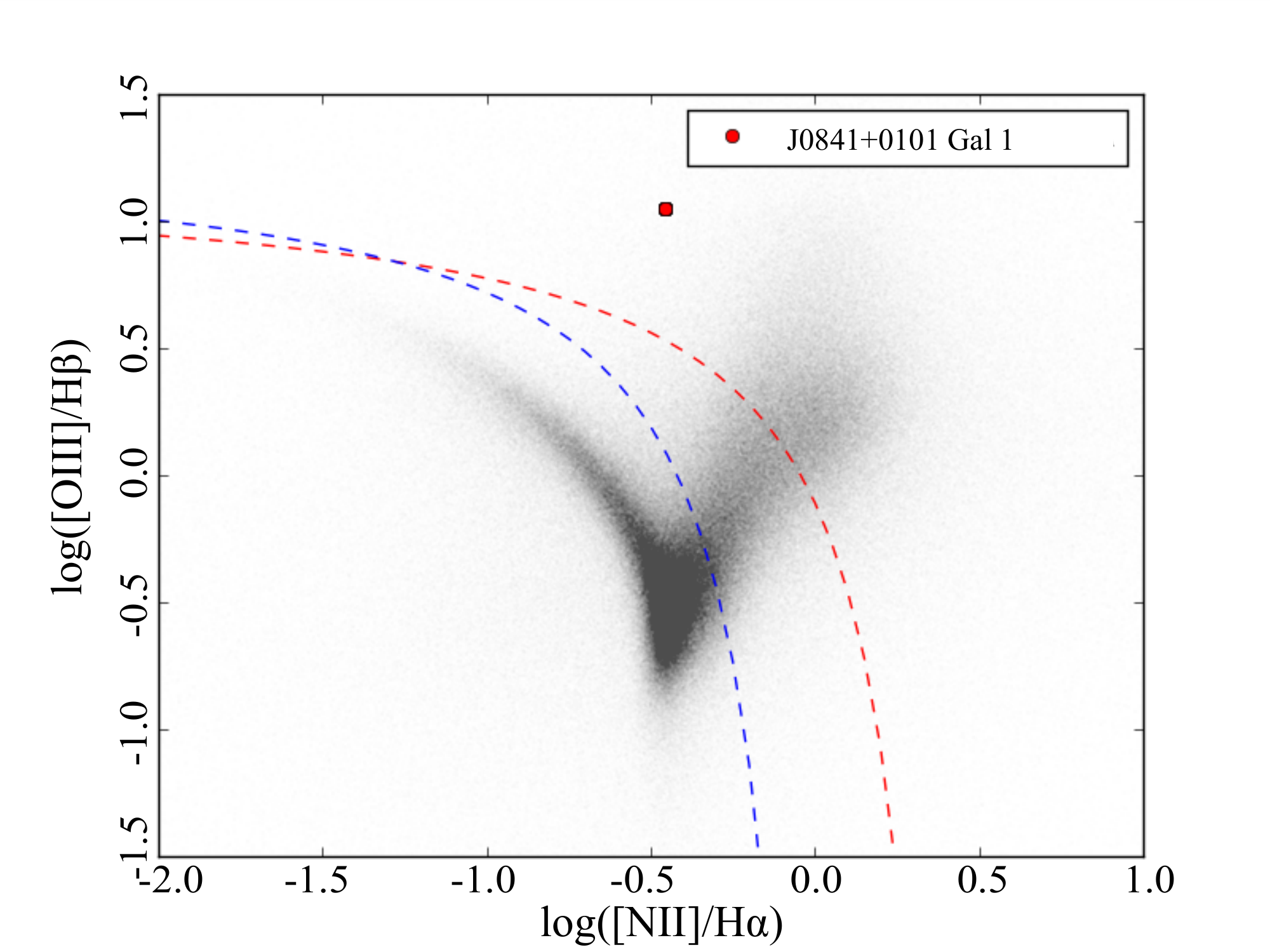}}
    \hspace{-5mm}
    \subfloat{\includegraphics[width=0.35\linewidth]{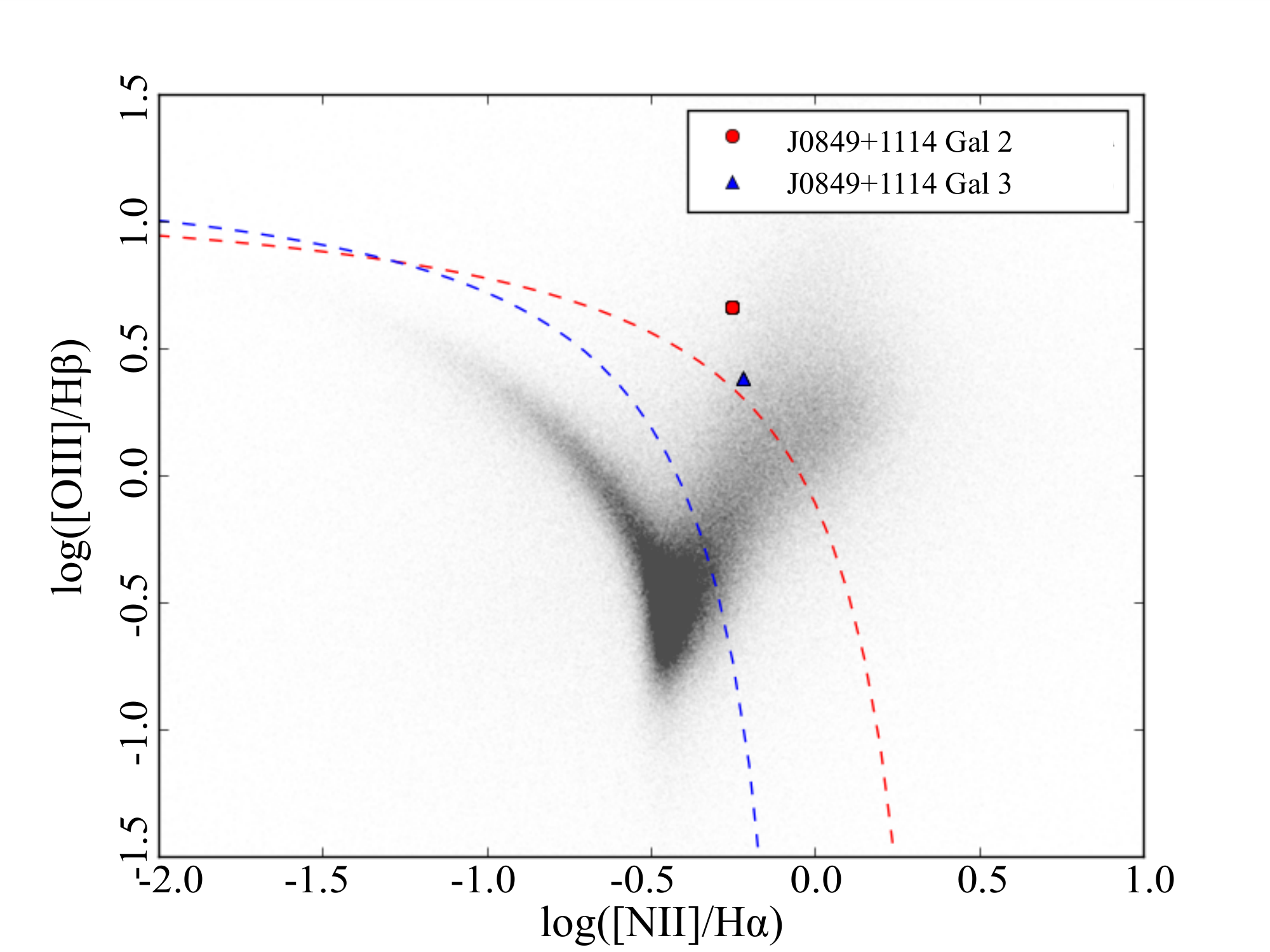}}\\
    \subfloat{\includegraphics[width=0.3\linewidth]{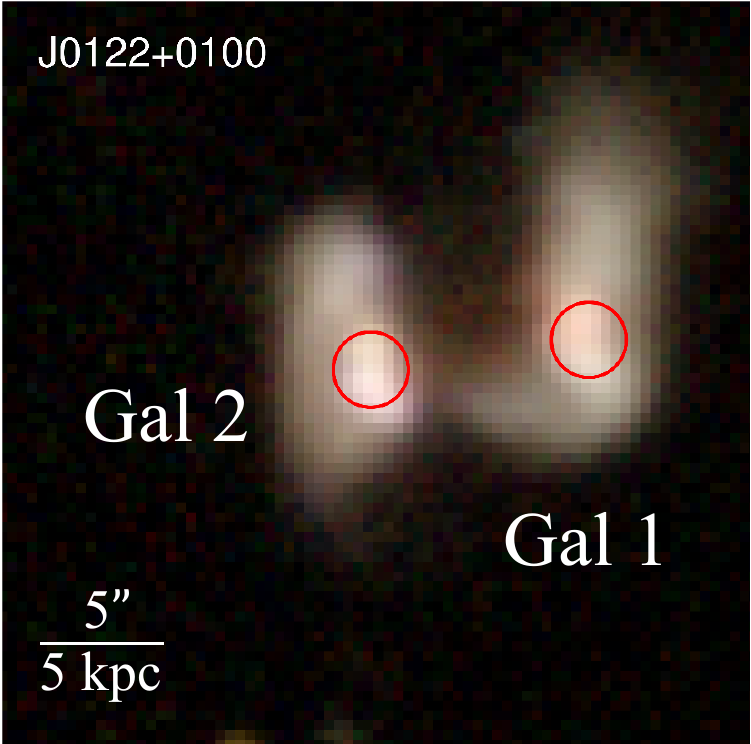} }
    \hspace{+2mm}
    \subfloat{\includegraphics[width=0.3\linewidth]{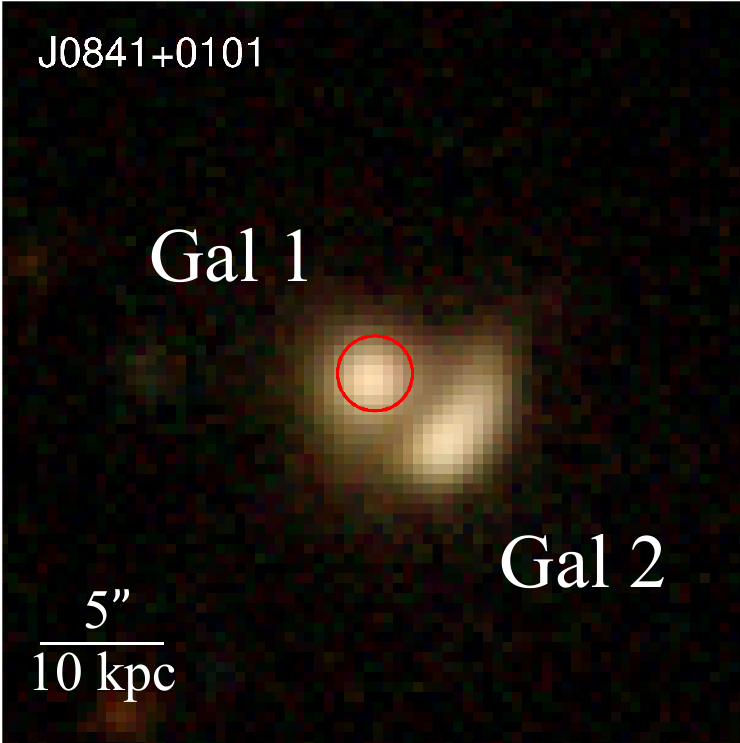}}
    \hspace{+2mm}
    \subfloat{\includegraphics[width=0.3\linewidth]{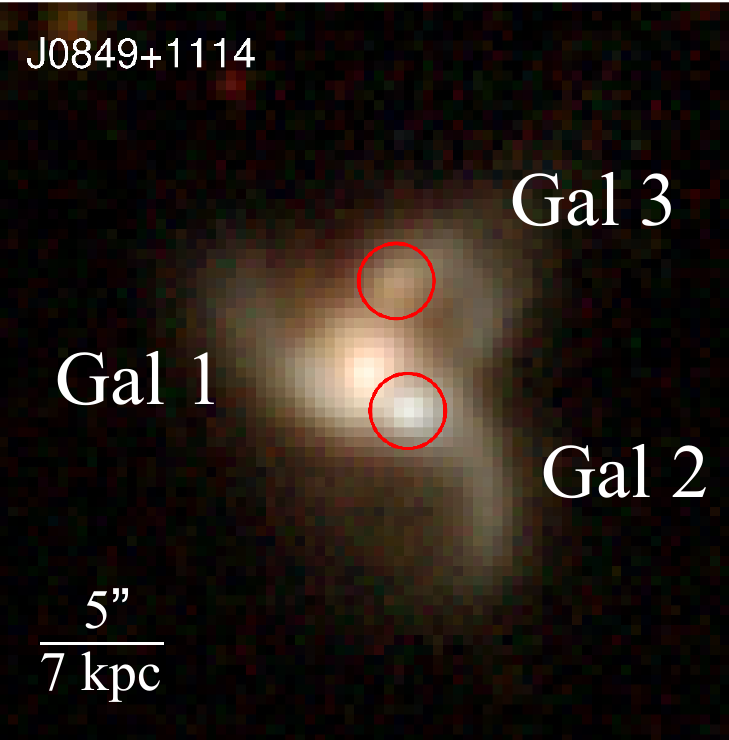} }\\
    \hspace{-2mm}
    \subfloat{\includegraphics[width=0.35\linewidth]{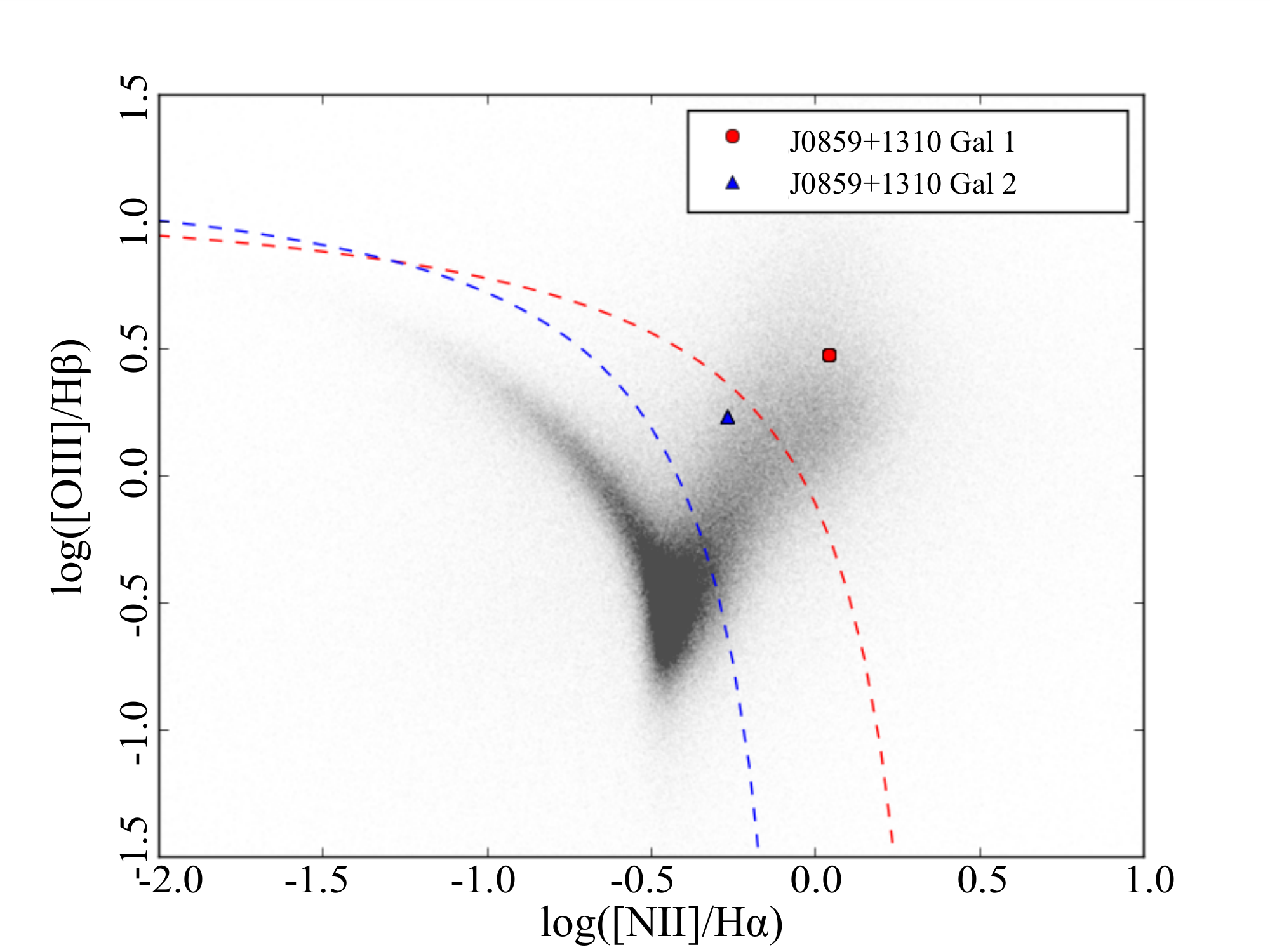}}
    \hspace{-5mm}
    \subfloat{\includegraphics[width=0.35\linewidth]{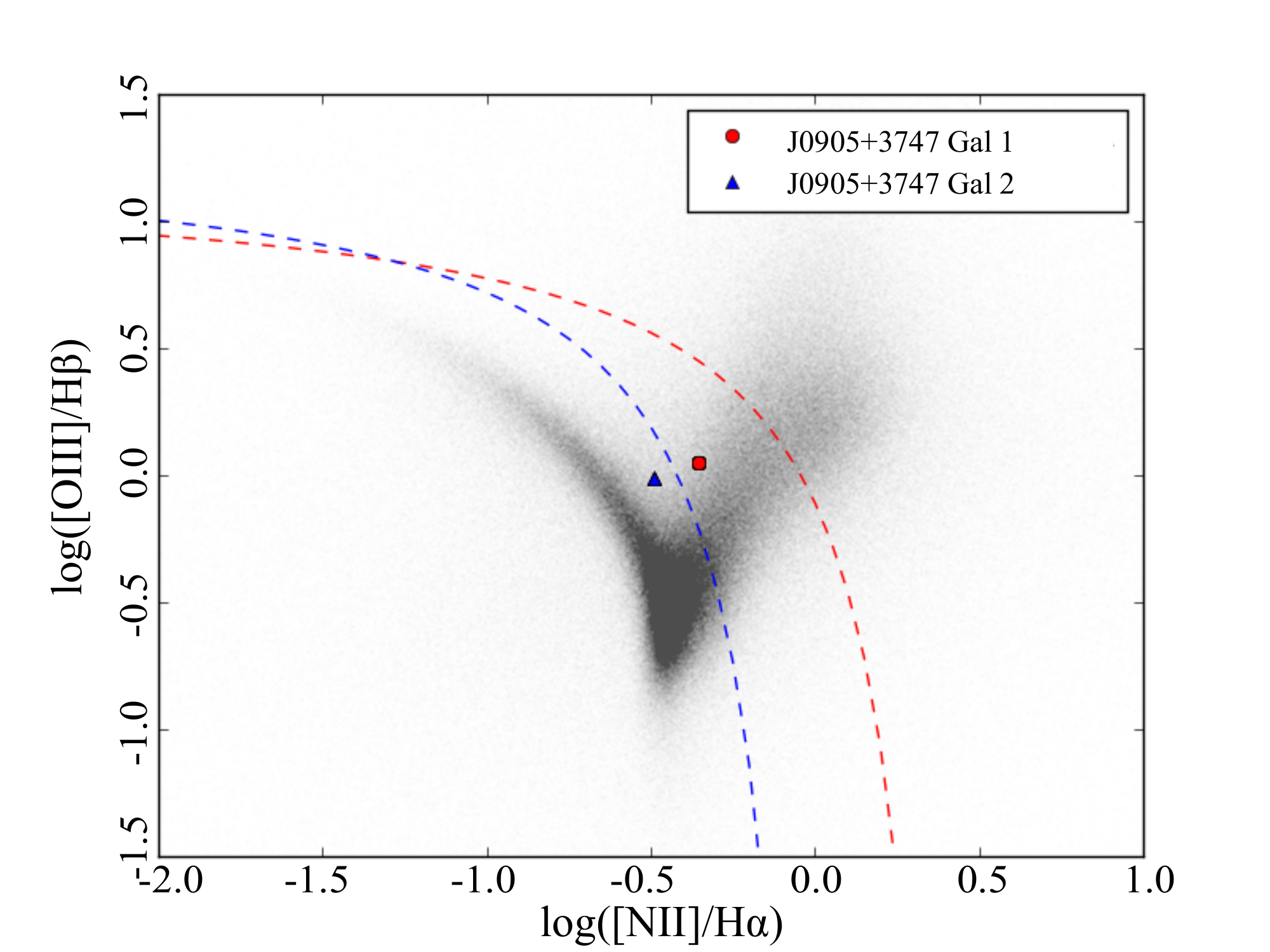} }
    \hspace{-5mm}
    \subfloat{\includegraphics[width=0.35\linewidth]{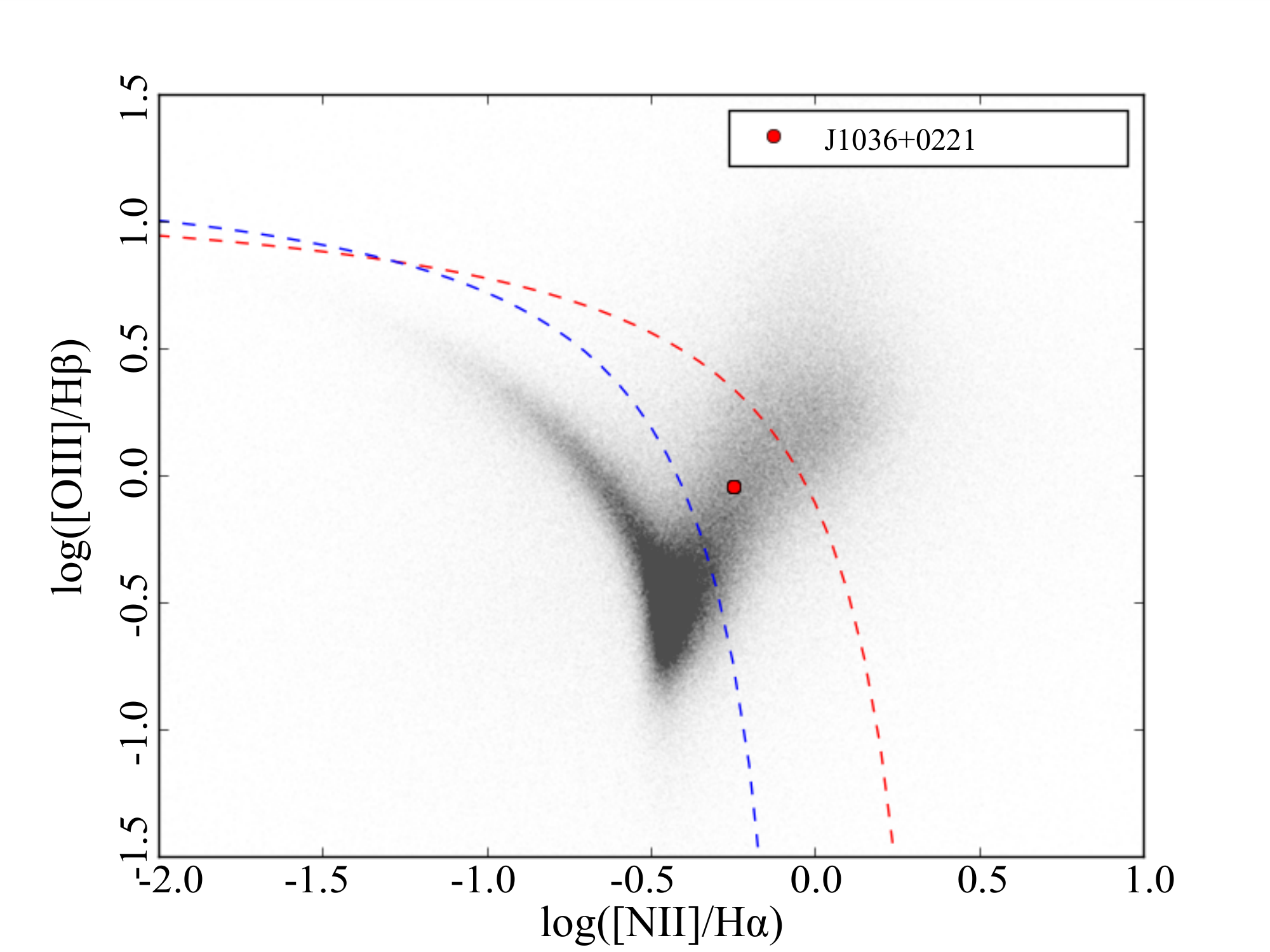}}\\
    \subfloat{ \includegraphics[width=0.3\linewidth]{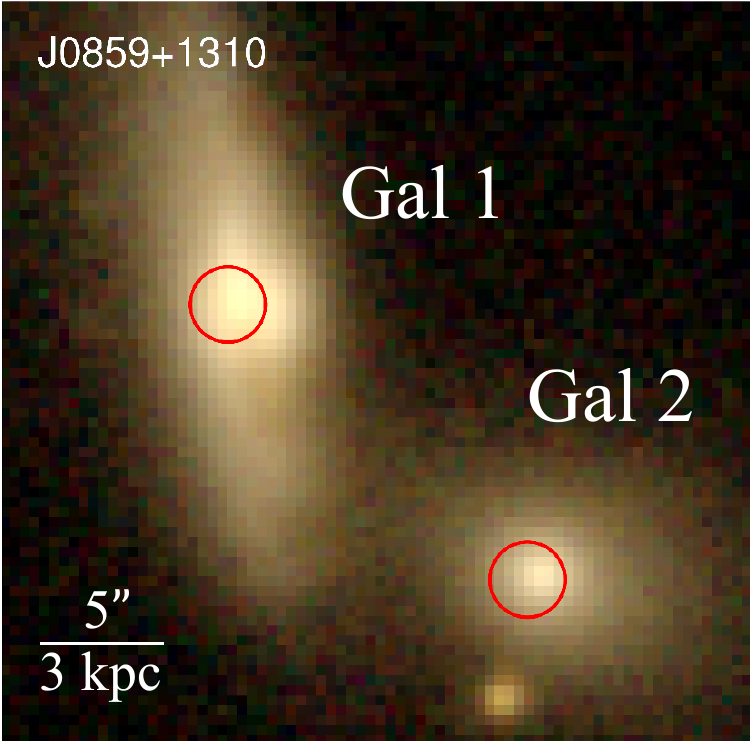}}
    \hspace{2mm}
    \subfloat{ \includegraphics[width=0.3\linewidth]{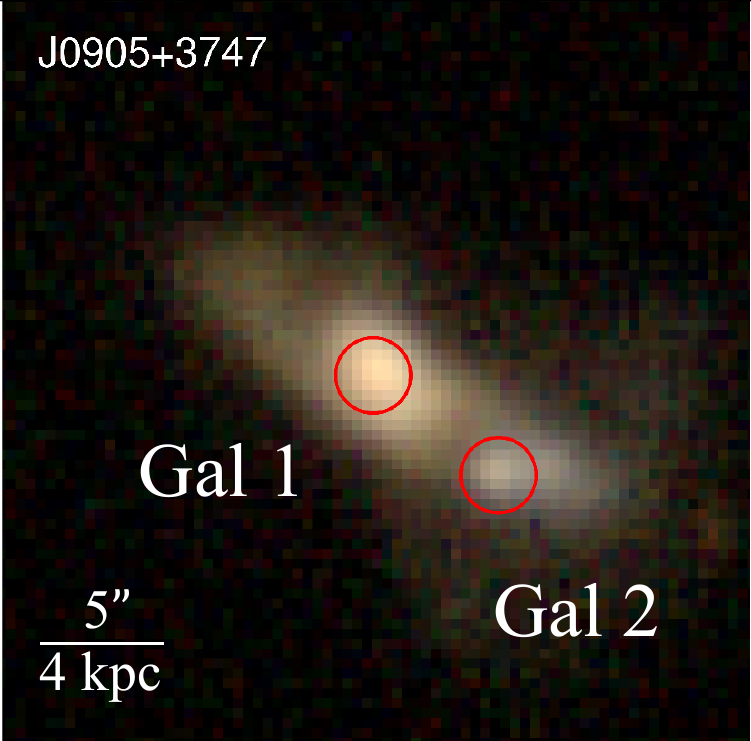}}
    \hspace{2mm}
    \subfloat{\includegraphics[width=0.3\linewidth]{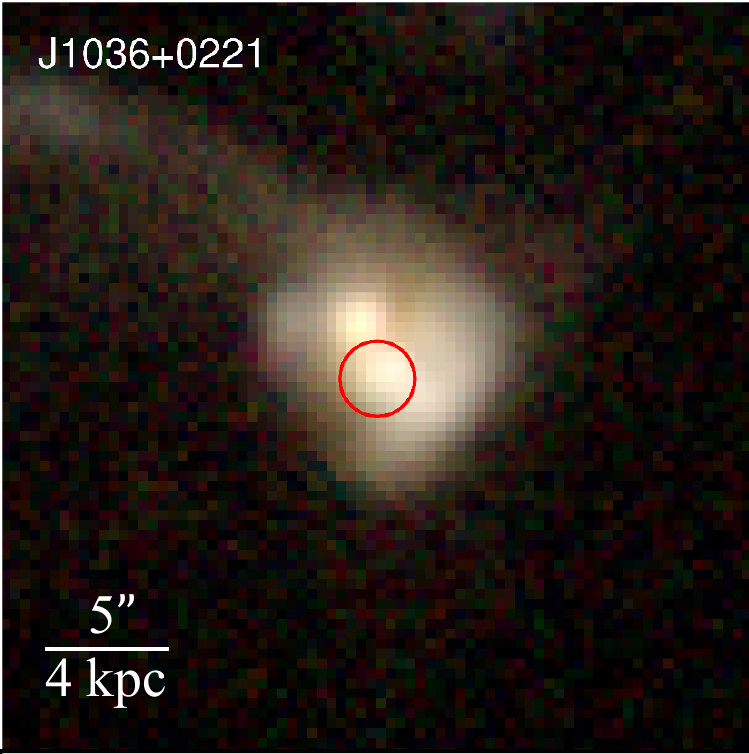}}\\
\caption{BPT and SDSS images of the full sample of 15 galaxy mergers observed during \chandra{} Cycles 15, 17, and 18. Red apertures with 1\farcs5 radii represent the SDSS fiber positions. These panels are not all to the same scale. The grey data points come from the MPA/JHU DR7 catalog \citep{abazajian2009}. The dashed red and blue line demarcations, from \citet{kewley2001} and \citet{kauffmann2003}, respectively, separate star-forming galaxies from AGNs. From the figure it is clear that not all X-ray sources and nuclei would be optically classified as AGNs. Note: some of these targets were published previously in Paper I (see Figure 1 of \citealp{satyapal2017}) but are included here for completeness of the sample; J0122+0100, J1045+3519, J1221+1137, and J1306+0735 were observed across both cycles 15 and 18.}
\label{fig:sample_images}
\end{figure*} 
\begin{figure*}[!ht]
    \ContinuedFloat
    \centering
    \hspace{-2mm}
    \subfloat{\includegraphics[width=0.35\linewidth]{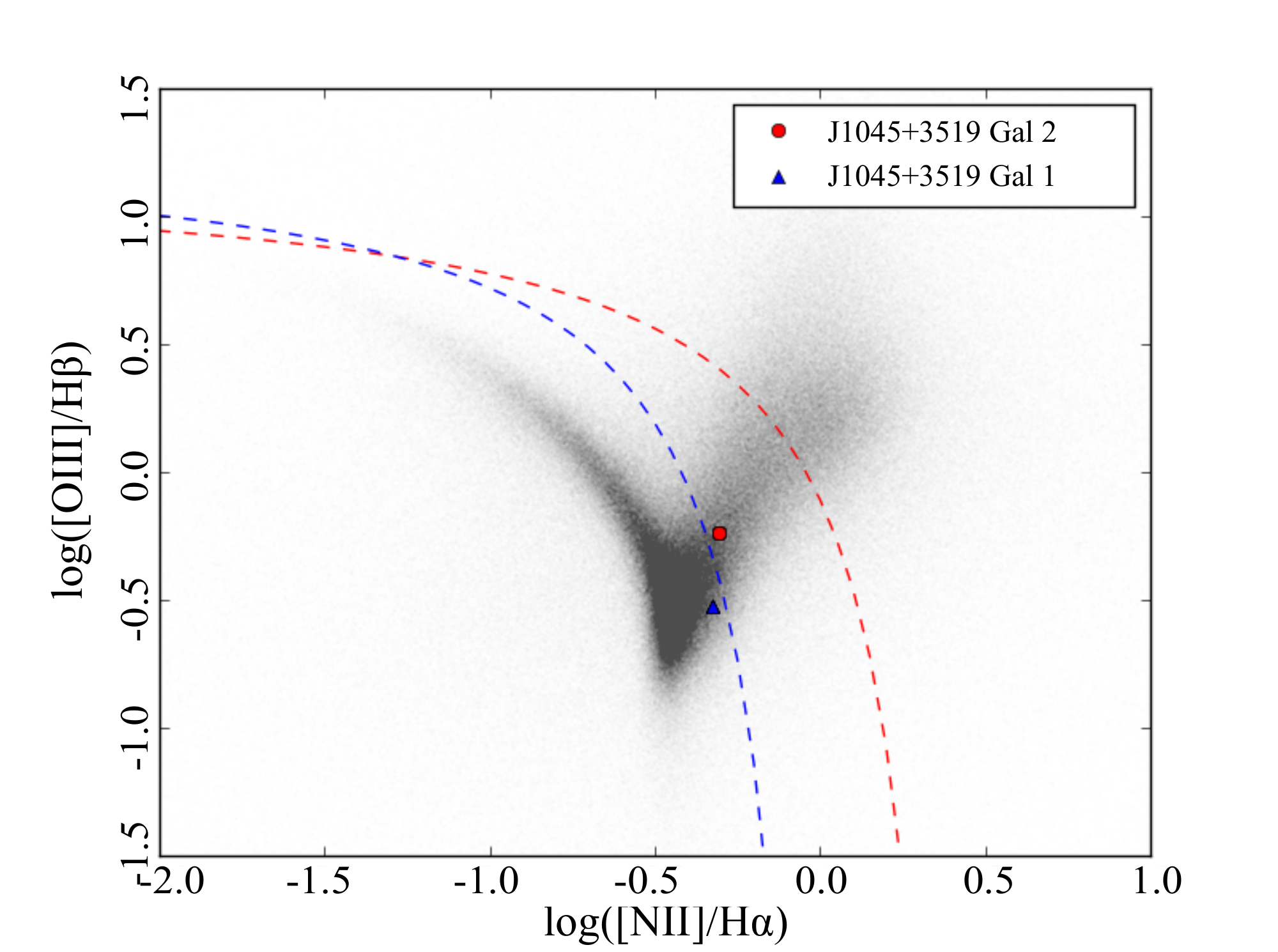}}
    \hspace{-5mm}
    \subfloat{\includegraphics[width=0.35\linewidth]{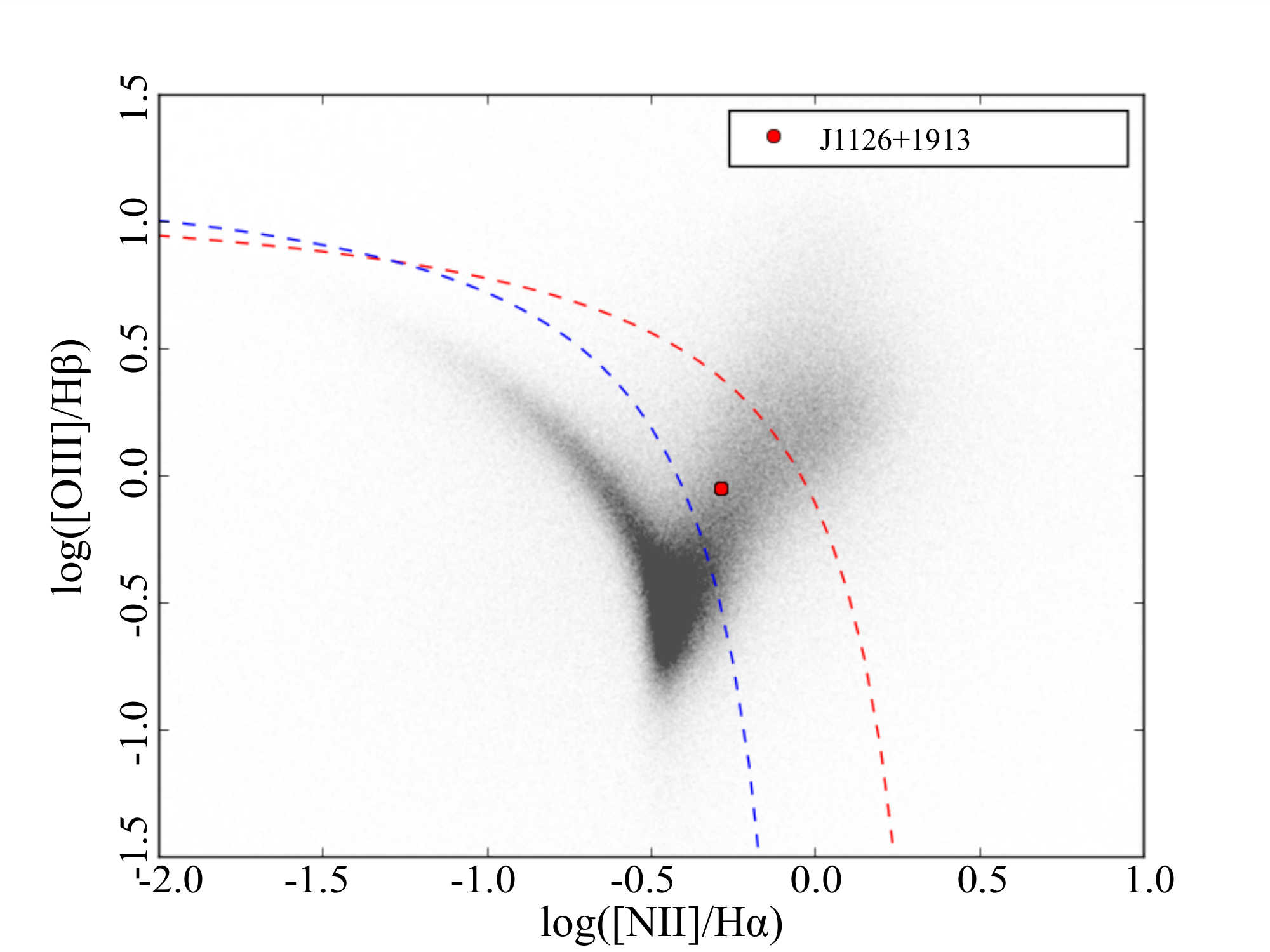} }
    \hspace{-5mm}
    \subfloat{\includegraphics[width=0.35\linewidth]{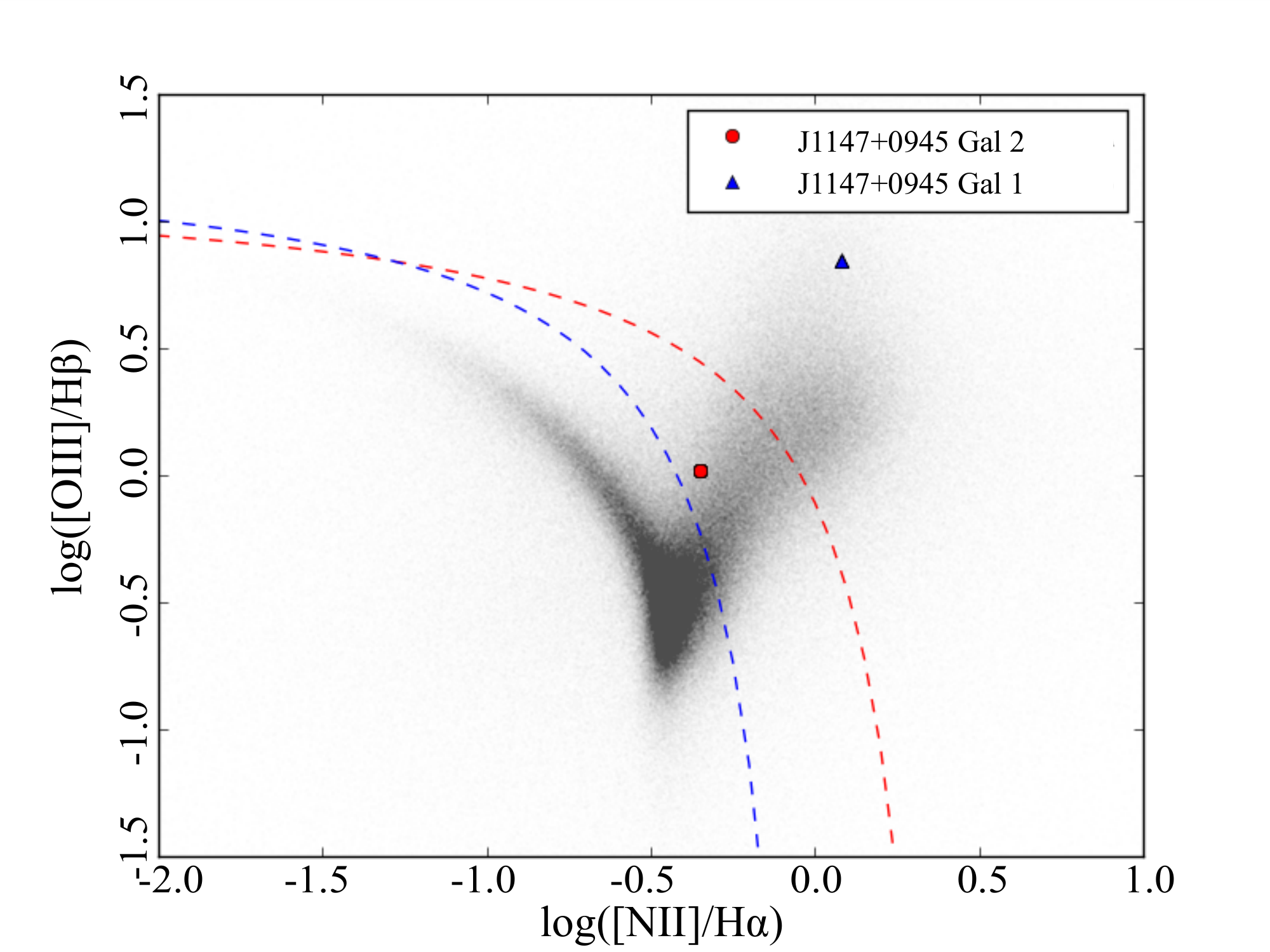}}\\
    \subfloat{\includegraphics[width=0.305\linewidth]{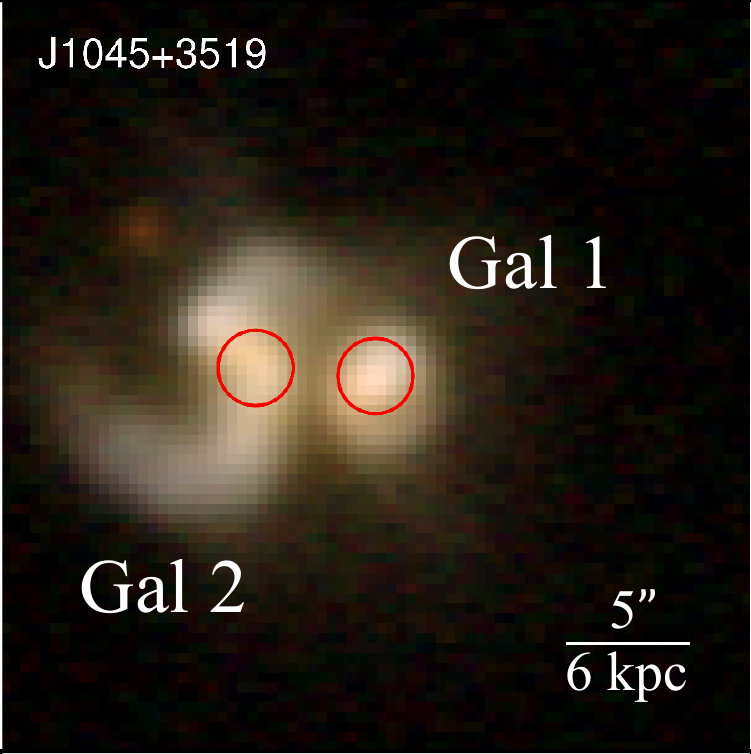}}
    \hspace{+2mm}
    \subfloat{\includegraphics[width=0.296\linewidth]{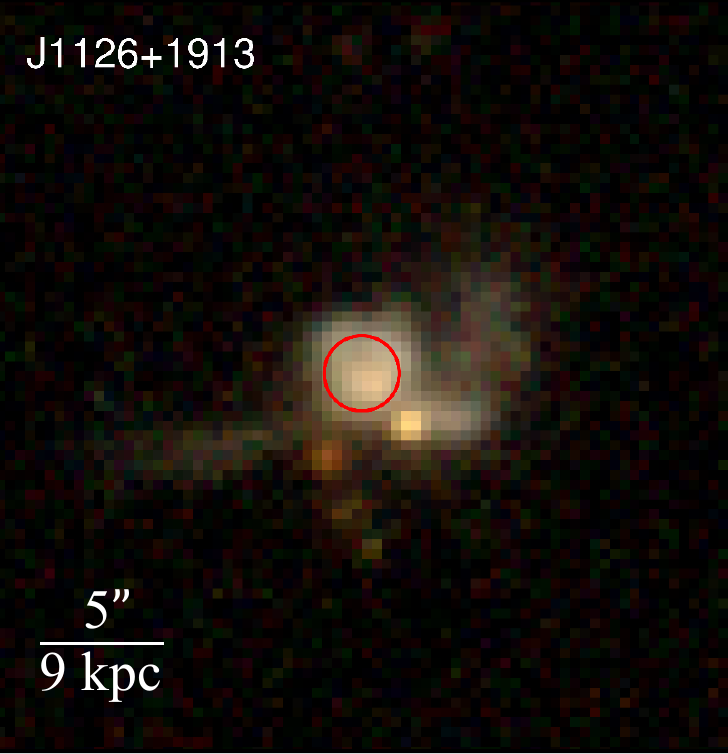} }
    \hspace{+2mm}
    \subfloat{\includegraphics[width=0.30\linewidth]{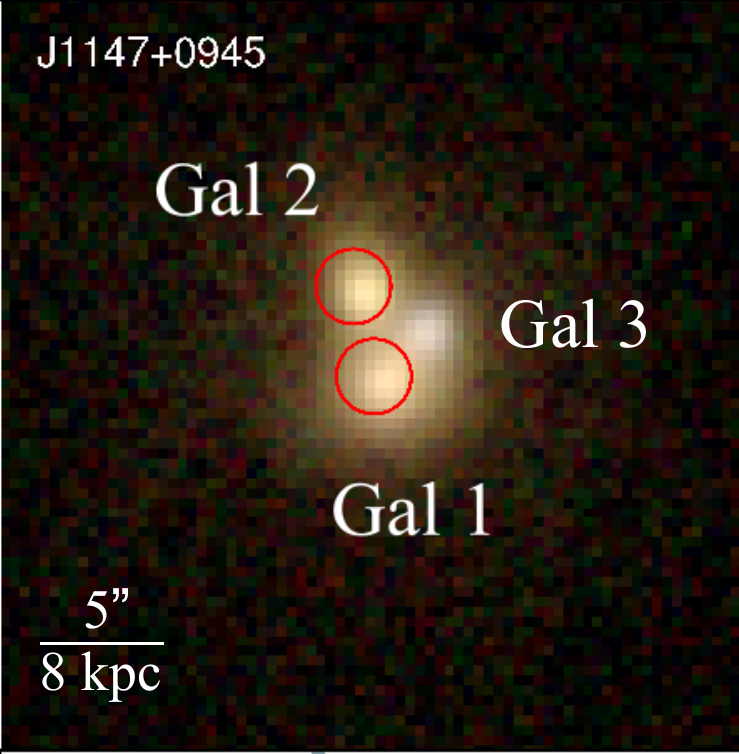} }\\
    \hspace{-2mm}
    \subfloat{\includegraphics[width=0.34\linewidth]{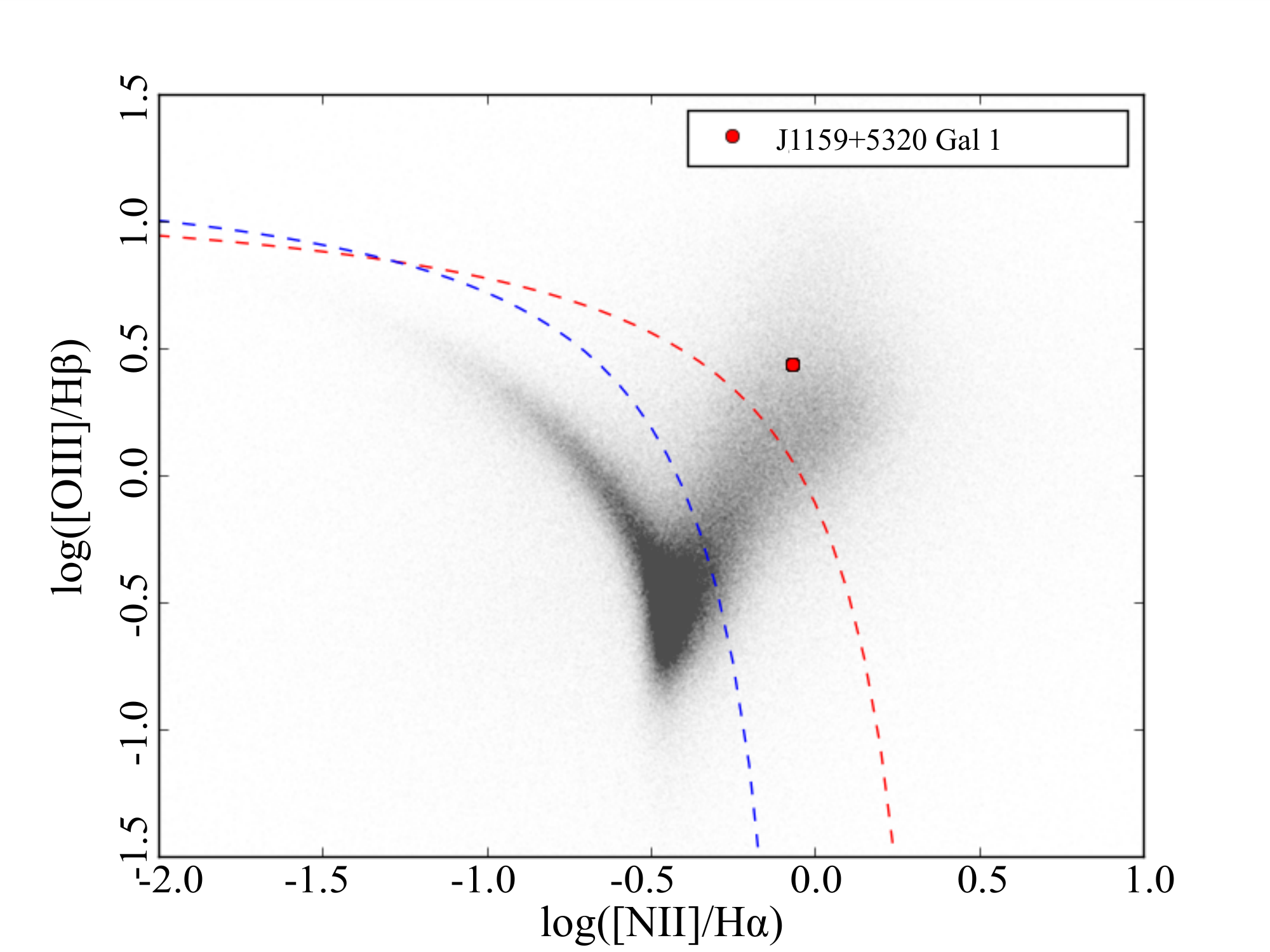}}
    \hspace{-5mm}
    \subfloat{\includegraphics[width=0.34\linewidth]{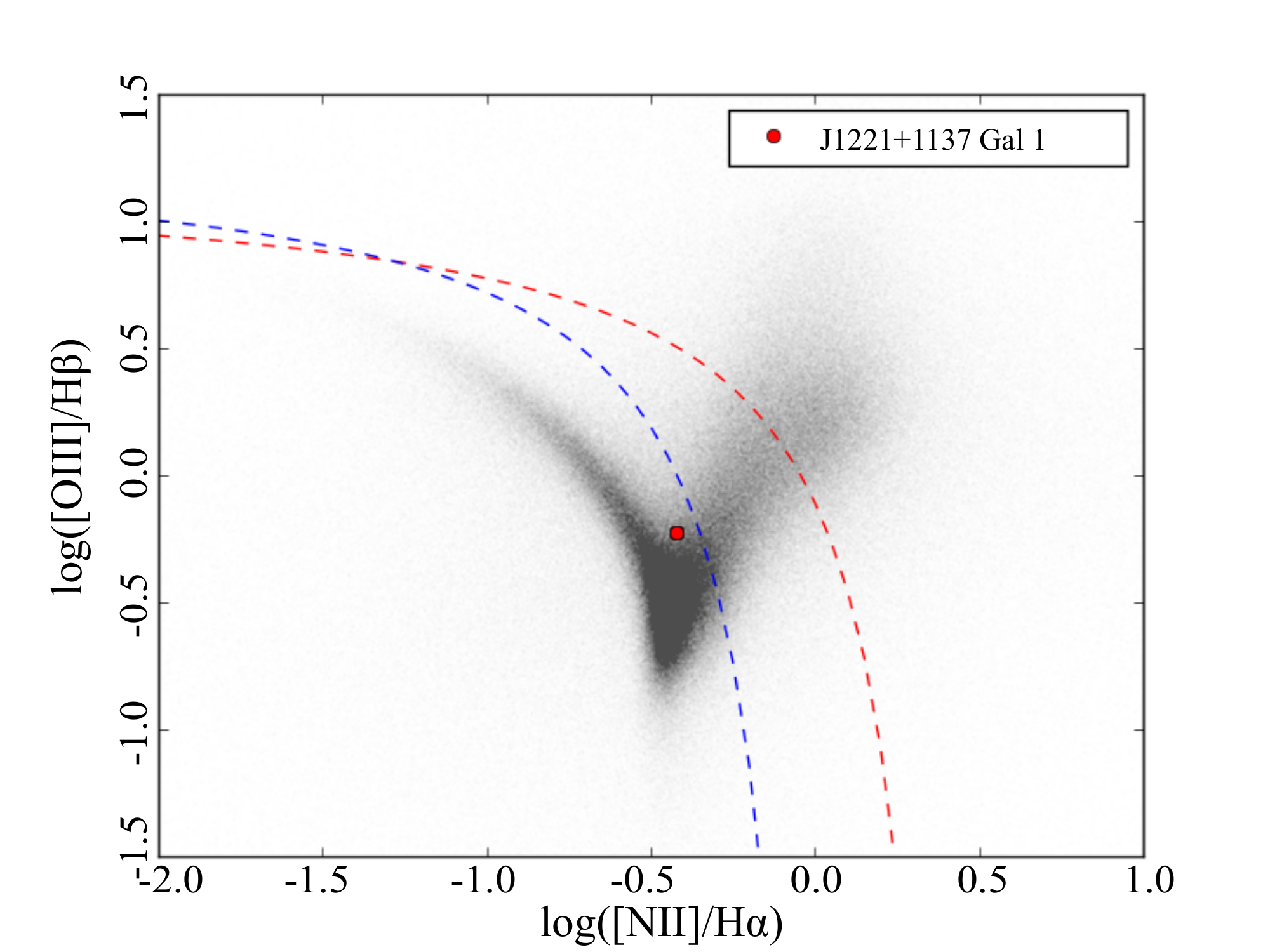} }
    \hspace{-5mm}
    \subfloat{\includegraphics[width=0.34\linewidth]{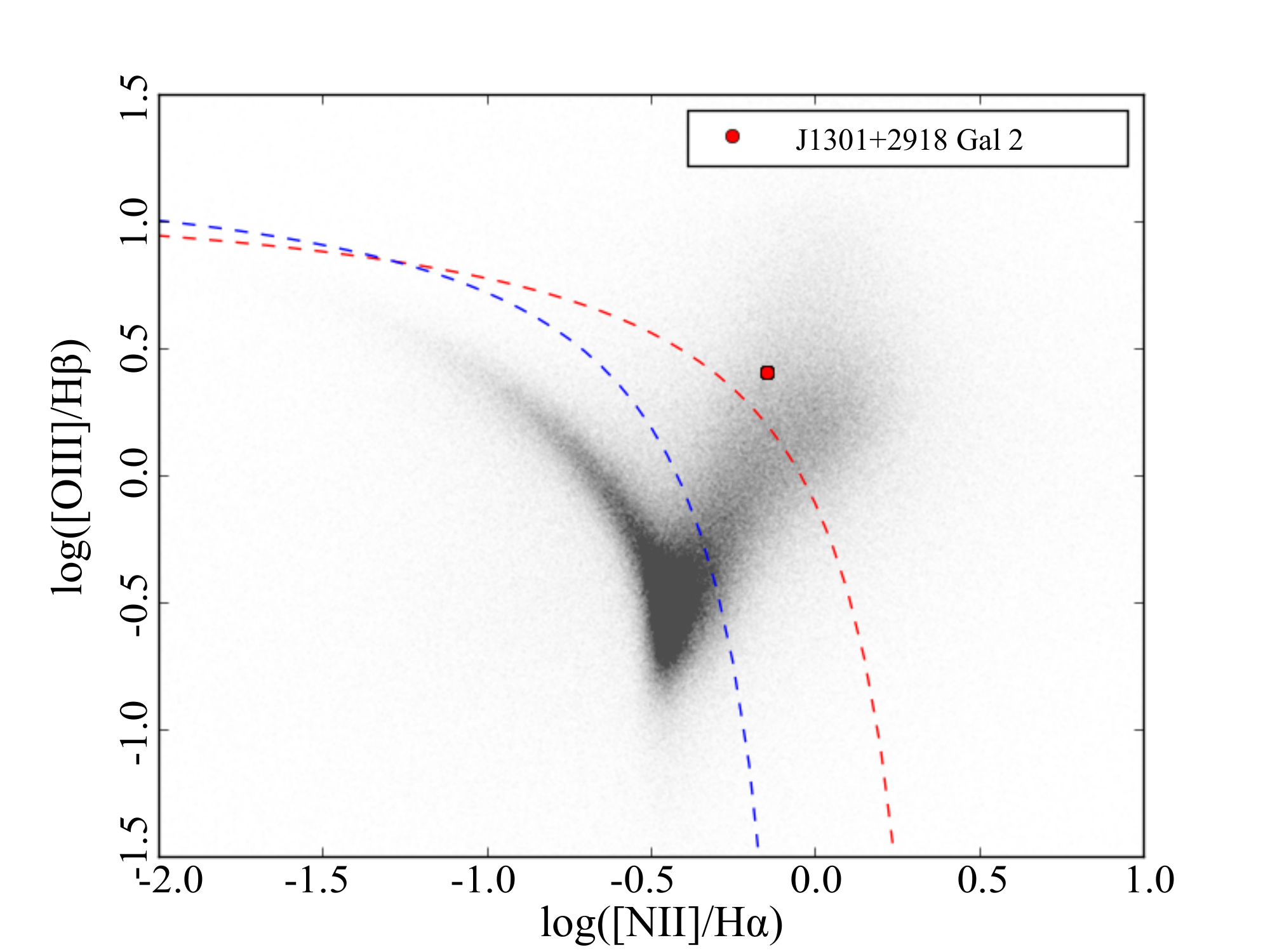} }\\
    \subfloat{\includegraphics[width=0.3\linewidth]{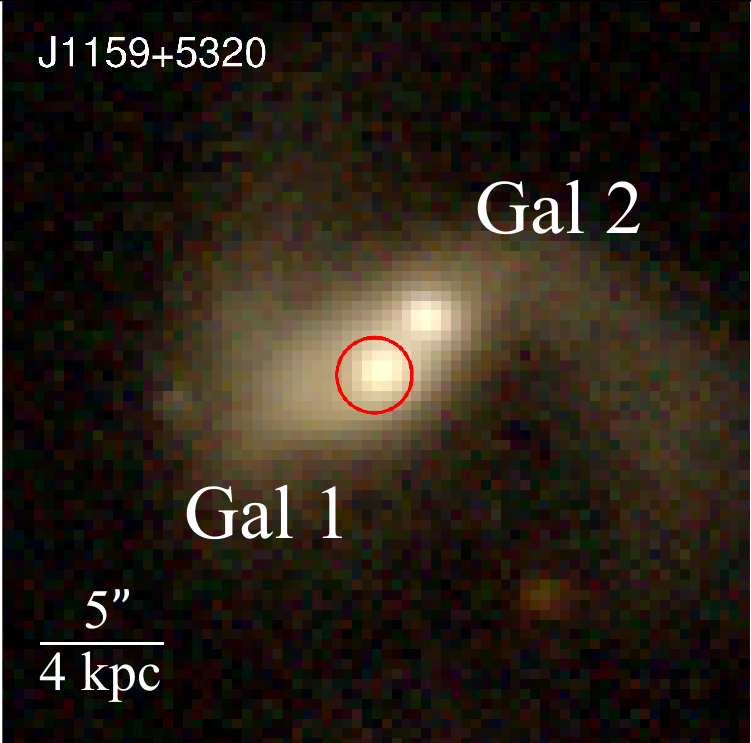}}
    \hspace{+2mm}
    \subfloat{\includegraphics[width=0.29\linewidth]{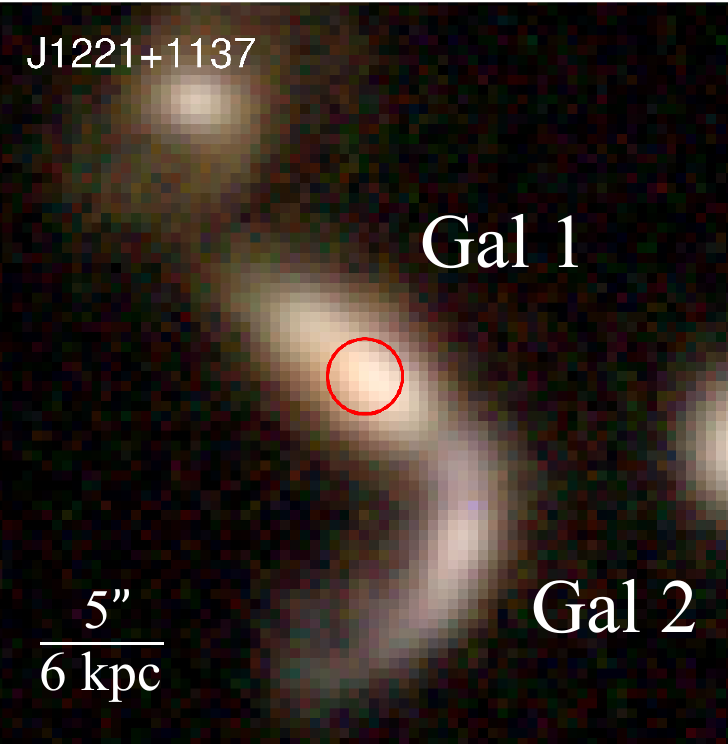} }
    \hspace{+2mm}
    \subfloat{\includegraphics[width=0.29\linewidth]{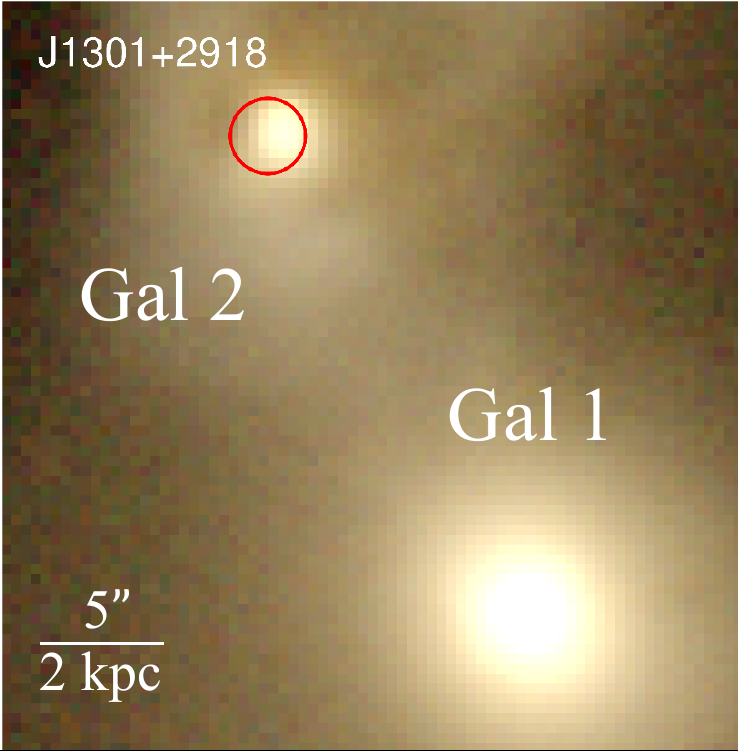}}\\
\caption{Figure 1 - Continued}
\end{figure*}
\begin{figure*}[!ht]
    \ContinuedFloat
    \centering
    \hspace{-2mm}
    \subfloat{\includegraphics[width=0.35\linewidth]{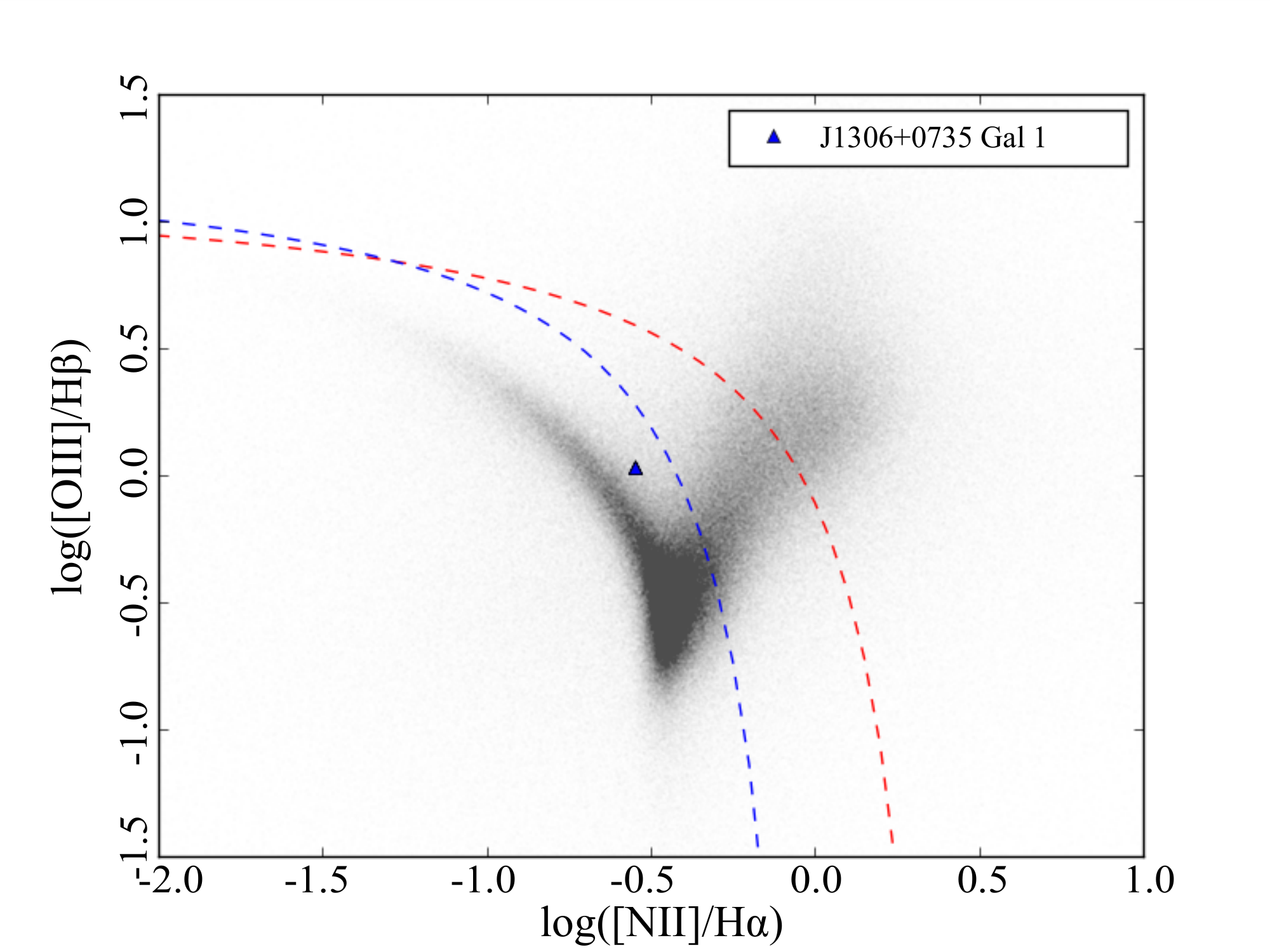}}
    \hspace{-5mm}
    \subfloat{\includegraphics[width=0.35\linewidth]{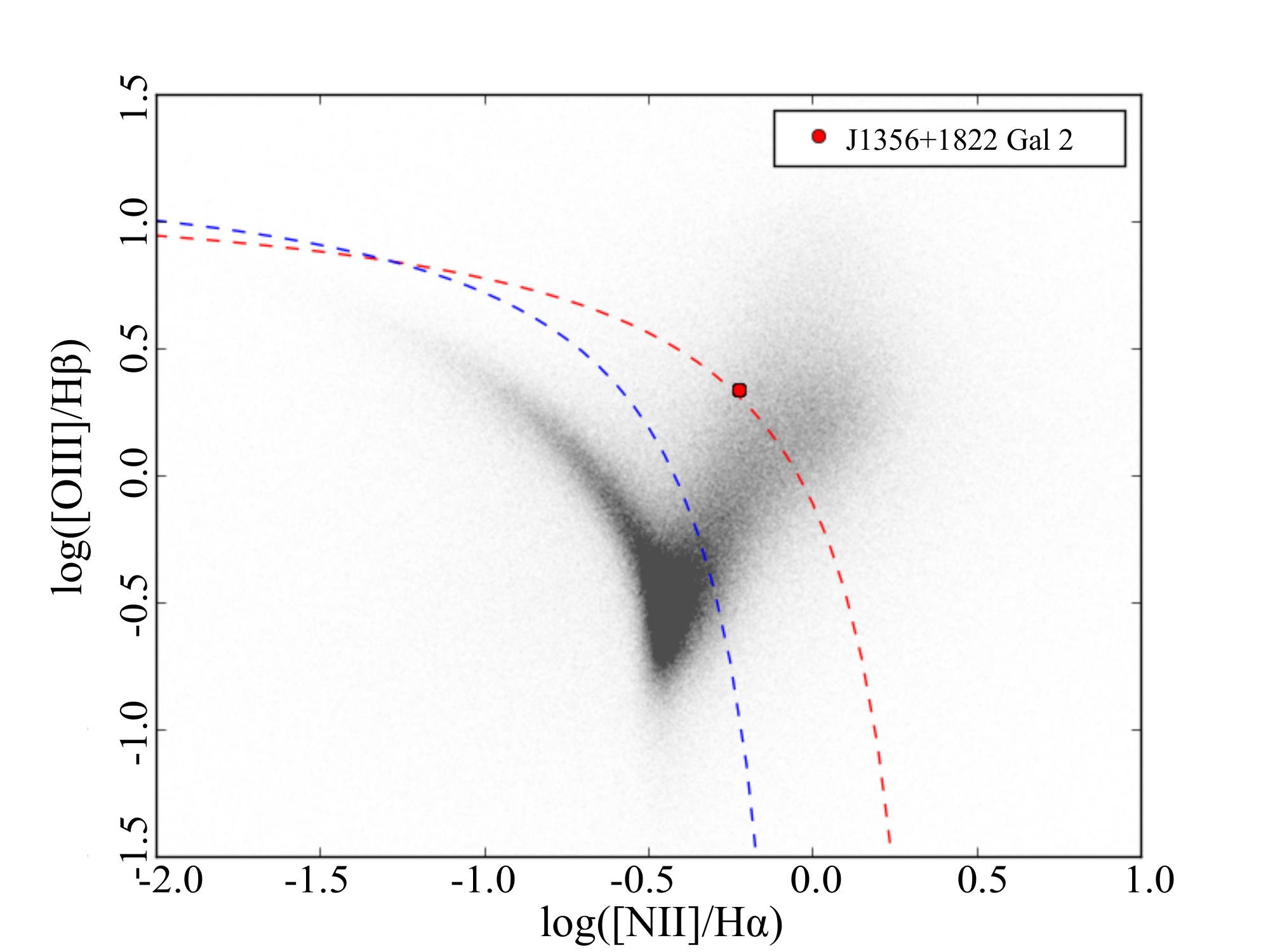}}
    \hspace{-5mm}
    \subfloat{\includegraphics[width=0.35\linewidth]{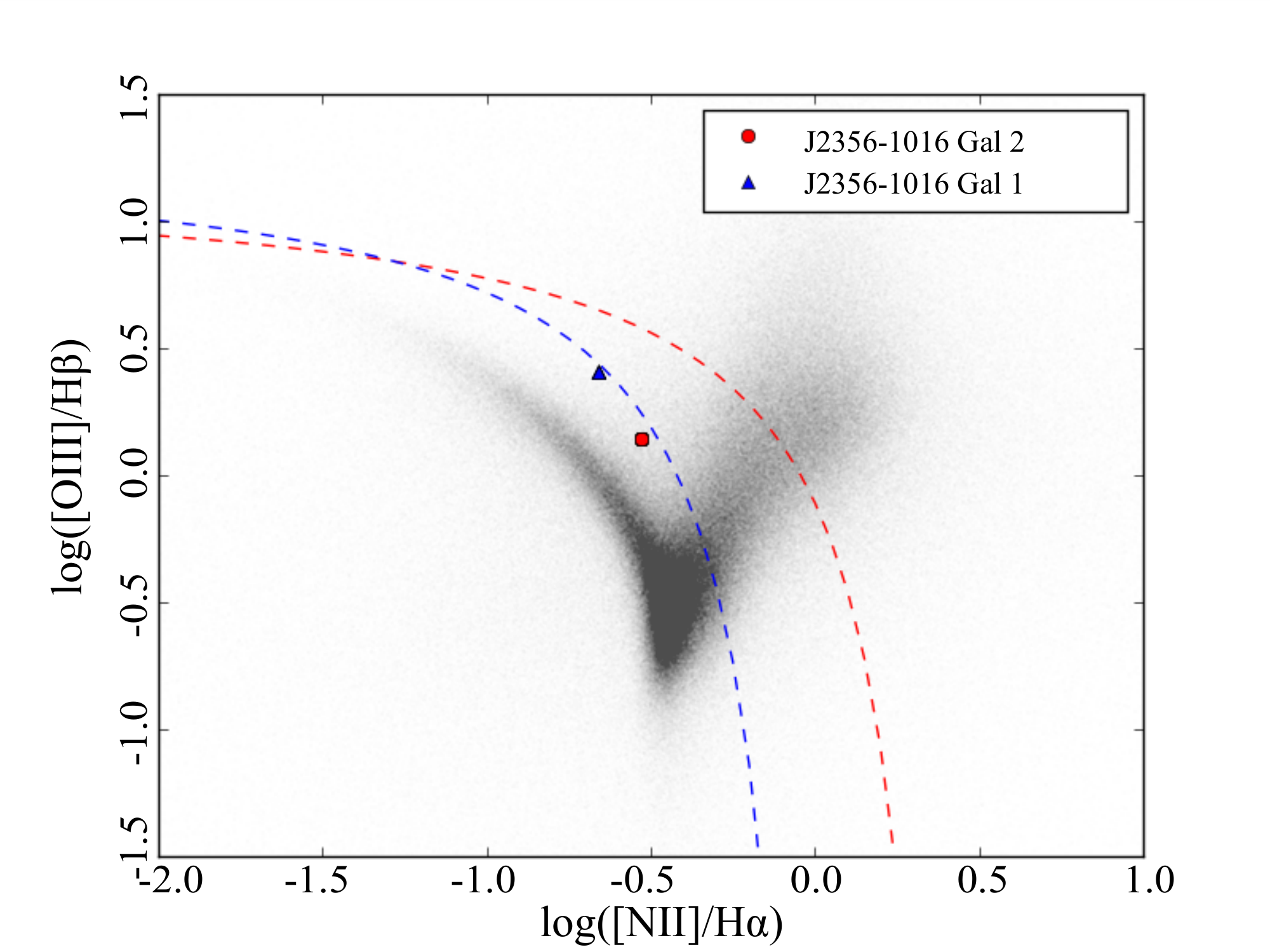}}\\
    \subfloat{\includegraphics[width=0.3\linewidth]{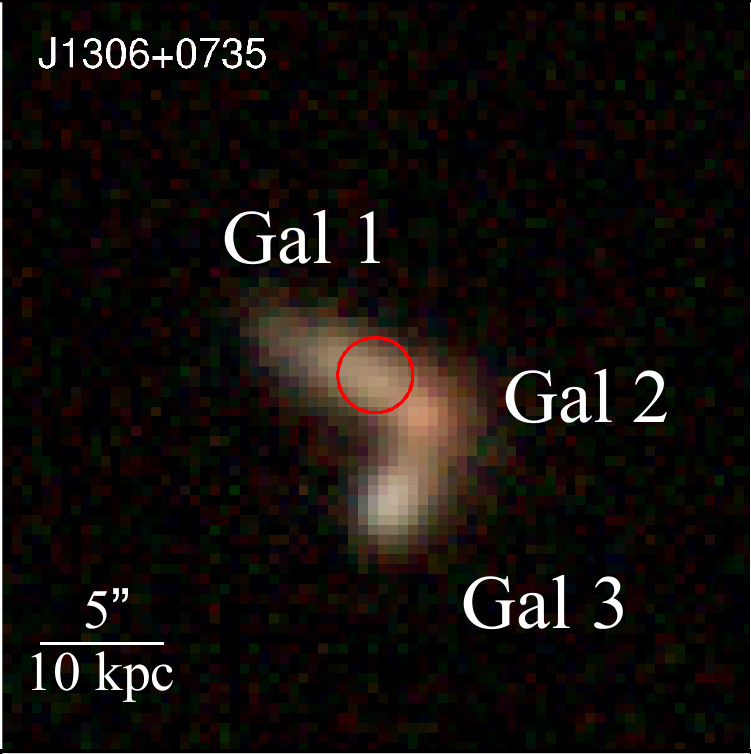} }
    \hspace{+4mm}
    \subfloat{\includegraphics[width=0.3\linewidth]{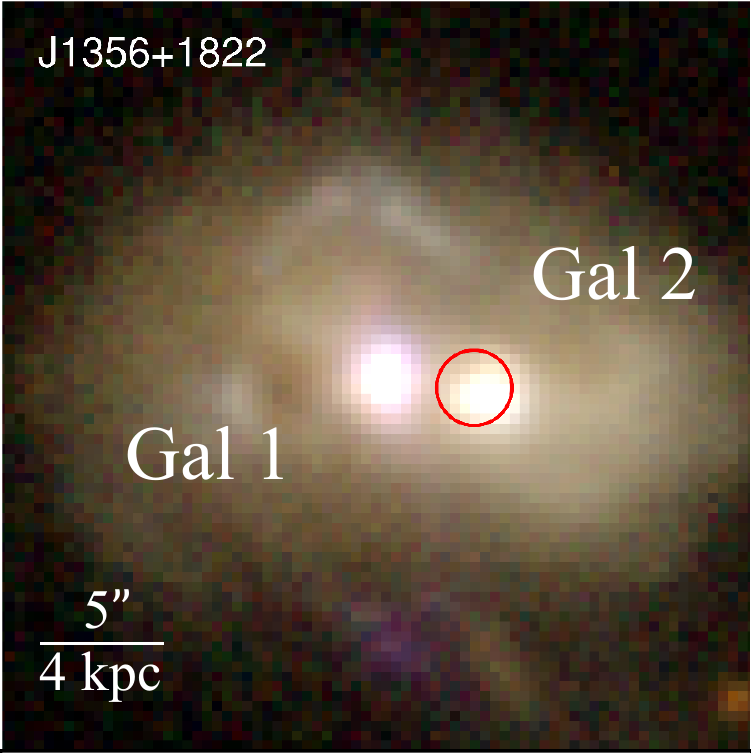}}
    \hspace{+4mm}
    \subfloat{\includegraphics[width=0.3\linewidth]{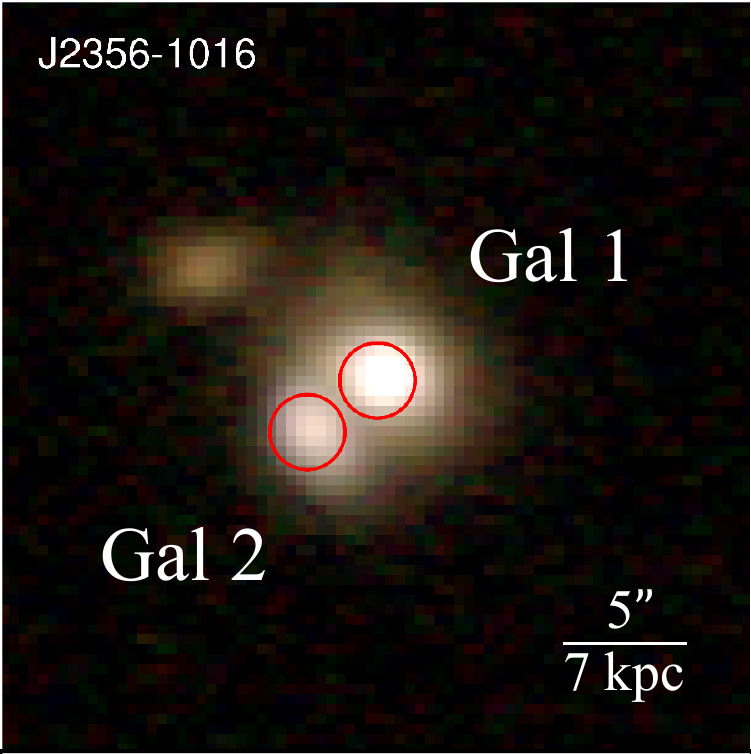} }\\
\caption{Figure 1 - Continued}
\end{figure*}
\\
\clearpage

We used photometry from the SDSS~DR12, \textit{The Two Micron All Sky Survey} \citep[2MASS;][]{skrutskie2006}, WISE, and the \textit{Infrared Astronomical Satellite} \citep[IRAS;][]{neugebauer1984} appropriate for extended systems. For SDSS, we used the \texttt{modelMag} values, with the exception of the $u$ band, which we exclude due to uncertainties arising from sky level estimates and the known ''red leak''/scattered light issues with the band.\footnote{\url{https://www.sdss.org/dr12/imaging/caveats/}} For 2MASS, we use the Extended Source Catalog (XSC) magnitudes where available, and the Point Source Catalog magnitudes otherwise. We do not use the 2MASS data for SDSS~J130125.26+291849.5, as the 2MASS catalog photometry for this object is severely at odds with both the SDSS and the WISE data. For WISE, we use the elliptical \texttt{gmag} magnitudes where available, the point-spread-function (PSF)-fit \texttt{mpro} magnitudes otherwise, with the exception of SDSS~J130125.26+291849.5 where we use the large aperture \texttt{mag\_8} magnitudes. For sources with IR flux densities either in the IRAS Point Source Catalog (PSC) or the IRAS Faint Source Catalog (FSC), we additionally use the 60 and 100 \micron\ flux densities, preferentially from the PSC. To convert to the AB system, we added 0.02~mag to the SDSS $z$ band,\footnote{\url{https://www.sdss.org/dr12/algorithms/fluxcal/\#SDSStoAB}} we used the 2MASS Vega/AB offsets available in \textsc{topcat},\footnote{version 4.6-1; \url{http://www.star.bris.ac.uk/~mbt/topcat}} and we used the standard Vega/AB offsets listed in the WISE documentation.\footnote{\url{http://wise2.ipac.caltech.edu/docs/release/allsky/expsup/sec4_4h.html\#conv2ab}} Finally, we corrected the $g$ through $W2$ magnitudes for Galactic dust extinction using $E(B-V)$ values following \citet{schlafly2011}. We added in quadrature 0.05 mag to all formal magnitude errors to account for realistic flux calibration offsets between the facilities, and we fit the AGN component along a grid of $E(B-V)$ values ranging from 0.0 to 30. We used the \citet{gordon1998} extinction curve for UV wavelengths and the \citet{cardelli1989} extinction curve otherwise. To estimate formal errors, we fit each system 100 times, each time permuting the magnitudes by their uncertainties. 

\begin{figure}[h]
\centering
\includegraphics[width=\linewidth]{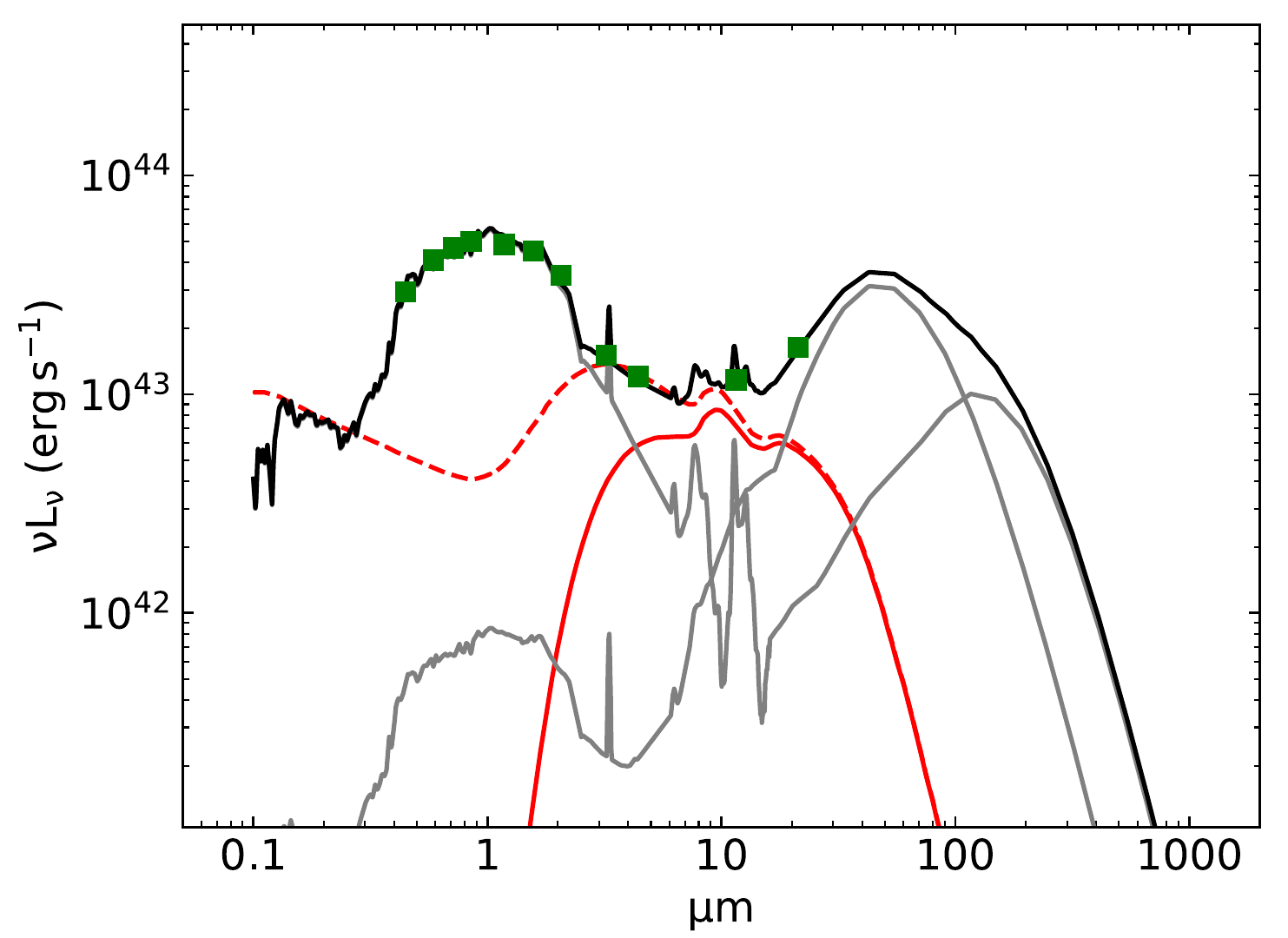}
\caption{Example of one of our SED fits, SDSS J115930.29+532055.7. The two gray subcomponents are the star-forming templates, while the red solid/dashed subcomponent is the reddened/intrinsic AGN. The line-of-sight best-fit extinction of the AGN component is $E(B-V)=7.1$. The wavelength scale is rest-frame.}
\label{fig:J1159+5320sed}
\end{figure}

We find that our systems have $8-1000~\micron$ luminosities from star formation between $1.4\times10^{10}~L_\sun$ and $6.1\times10^{11}~L_\sun$, with a mean value of $2.7\times10^{11}~L_\sun$, and $80\%$ of our systems are above $10^{11}~L_\sun$, placing them predominantly in the class of luminous infrared galaxies (LIRGs). We show an example of one of our SED fits in Figure~\ref{fig:J1159+5320sed}.

\begin{table*}
\caption{WISE Full Merger Sample Properties}
\begin{center}
\scalebox{0.95}{
\hspace{-2.2cm}
\begin{tabular}{ccccccccccc}
\hline
\hline
\noalign{\smallskip}
Name  & Redshift & D$_{\rm{L}}$  & $\Delta \theta$  & $r_{\rm{p}}$ & log$(\rm{M/M}_\sun)_1$ & log$(\rm{M/M}_\sun)_2$ & log$(\rm{M/M}_\sun)_3$ & W$1\--$W$2$ & W$2\--$W$3$ & $\rm{log}(\rm{L}_{\rm{IR}}/\rm{L}_{\sun})$ \vspace{-4mm}\\
\noalign{\smallskip}
 &  &  &  &  &  &  &  &  &  &\vspace{-4mm}\\
\noalign{\smallskip}
(SDSS) & &  (Mpc) &  (\arcsec) & (kpc) &  & & & (mag) & (mag) & \\
\noalign{\smallskip}
\hline
\noalign{\smallskip}
J012218.11+010025.7 & 0.05546 & 247.5 & 8.7 & 8.7 & 10.35 & 9.97 & \dots$^{(\dagger)}$ & 1.54 & 3.87 & $11.26\pm0.02$\\
J084135.08+010156.2 & 0.11060 & 512.6 & 3.9 & 7.9 & 10.55 & \dots & \dots$^{(\dagger)}$ & 1.77 & 3.96 & $11.79\pm0.07$\\
J084905.51+111447.2 & 0.07727 & 350.2 & 2.2$^{(a)}$ & 3.3$^{(a)}$ & \dots & 10.19 & 9.63 & 1.69 & 3.55 & $11.43\pm0.03$ \\ 
 &  &  & 4.0$^{(b)}$ & 5.8$^{(b)}$ &  &  &  &  &  &\\
J085953.33+131055.3 & 0.03083 & 135.1 & 16.1 & 9.9 & 10.63 & 10.14 & \dots$^{(\dagger)}$ & 0.90 & 2.73 & $10.19\pm0.02$ \\
J090547.34+374738.2 & 0.04751 & 210.8 & 6.2 & 5.8 & 10.54 & 7.92 & \dots$^{(\dagger)}$ & 1.16 & 3.70 & $11.10\pm0.02$ \\
J103631.88+022144.1 & 0.05040 & 224.0 & 2.8 & 2.8 & 10.47  & \dots & \dots$^{(\dagger)}$ & 1.32 & 4.06 & $11.66\pm0.02$\\
J104518.03+351913.1 & 0.06758 & 304.1 & 7.0 & 9.0 & 10.64 & 10.56 & & 0.60 & 4.49 & $11.56\pm0.02$\\
J112619.42+191329.3 & 0.10299 & 474.9 & 2.3 & 4.5 & 10.24 & \dots & \dots$^{(\dagger)}$ & 0.81 & 4.24 & $11.40\pm0.05$\\
J114753.62+094552.0 & 0.09514 & 436.4 & 3.8$^{(c)}$ & 6.8$^{(c)}$ & 10.26 & 10.98 & \dots & 0.83 & 2.54 & $10.52\pm0.05$\\
&  &  & $2.4^{(d)}$ & $4.3^{(d)}$ &  &  &  &  & &\\ 
J115930.29+532055.7 & 0.04498 & 199.2 & 2.7 & 2.4 & 10.44 & \dots & \dots$^{(\dagger)}$ & 0.84 & 3.33 & $10.15\pm0.05$ \\
J122104.98+113752.3 & 0.06820 & 307.1 & 7.1 & 9.3 & 10.87 & \dots & \dots$^{(\dagger)}$ & 0.55 & 4.60 & $11.70\pm0.02$\\
J130125.26+291849.5 & 0.02340 & 102.0 & 21.8 & 10.3 & \dots & 10.70 & \dots$^{(\dagger)}$ & 1.26 & 4.06 & $11.25\pm0.02$\\
J130653.60+073518.1 & 0.11111 & 515.1 & 2.0$^{(e)}$ & 4.0$^{(e)}$ & 10.25 & \dots & \dots & 0.67 & 4.36 & $11.61\pm0.02$\\
 &  &  & 3.7$^{(f)}$ & 7.4$^{(f)}$ &  &  &  &  &  &\\
J135602.89+182218.2 & 0.05060 & 224.9 & 4.0 & 4.0 & \dots & 10.73 & \dots$^{(\dagger)}$ & 1.16 & 2.80 & $11.44\pm0.05$ \\
J235654.30-101605.3 & 0.07390 & 334.1 & 3.6 & 5.0 & 10.75 & 9.35 & \dots$^{(\dagger)}$ & 1.02 & 3.05 & $11.57\pm0.03$\\ 
\noalign{\smallskip}
\hline
\end{tabular}
}
\end{center}
\tablecomments{Column 1: SDSS target designation. Column 2: Redshifts. Column 3: Distance to merger in megaparsecs. Columns 4$\--$5: Angular separation of the galaxy nuclei in arcseconds and kiloparsecs, respectively. Columns 6$\--$8: Logarithmic masses of galaxy nuclei, where the subscripts 1, 2, and 3 denote the galaxy number; nuclei without a mass listing are denoted with an ellipsis. A dagger ($\dagger$) is appended in cases where a third nucleus is not present. Columns 9$\--$10: WISE color cuts for these systems. Column 11: Logarithm of $(\rm{L}_{\rm{IR}}/\rm{L}_{\sun})$, where $\rm{L}_{\rm{IR}}$ is the integrated 8-1000 \micron\ luminosity of the star-forming templates. The (formal) error margins are $1~\sigma$. (a) Angular separation between the SW and SE X-ray sources. (b) Angular separation between the SE and N X-ray sources. (c) Angular separation between the S and NE nuclei. (d) Angular separation between the S and NW nuclei. (e) Angular separation between the NE and SW X-ray sources. (f) Angular separation between the SW and SE X-ray sources.}
\label{table:sample}
\end{table*}

\section{Observations and Data Reduction}
\subsection{\chandra{}/ACIS Imaging Observations}
\chandra{} observations of the galaxy mergers from Cycle 17 and Cycle 18 were carried out with the ACIS-S instrument between 2015 October and 2018 January, all of which were performed with the sources at the aimpoint of the S3 chip. Details regarding the observations of the \chandra{} Cycle 15 targets are discussed  in Paper I. Exposure times for the targets ranged from 5.7 ks to 36 ks. Table 2 lists the information for the \chandra{} observations, while Table 3 lists the \xmm{} observations (see Section 3.2). 

All \chandra{} data were reduced and analyzed using version 4.9 of the \chandra{} Interactive Analysis of Observations (\textsc{ciao}) software package \citep{fruscione2006} along with the Chandra Calibration Database (\textsc{caldb}) version 4.7.6. Source aperture positions were determined through two methods: (1) the \textsc{ciao} module \textsc{wavdetect} was employed initially to pick out sources from the background; (2) $0.3\--8$ keV images were then smoothed using a $2$ and $3$ pixel Gaussian kernel to aid in the placement of apertures, particularly for cases of low-count sources. In most cases, these methods of aperture placement were used as a check against one another to ensure the most accurate placement. For counts extraction we utilized the \textsc{dmextract} package in \textsc{ciao}. Circular region apertures of 1.5\arcsec{} in radius were used for source count extraction while background counts were extracted from either circular or annular regions in areas free of any X-ray signatures near or around the sources. 

Due to the low-count nature of many of the X-ray sources, we employed the binomial no-source probability ($P_{\rm{B}}$) to verify the statistical significance of each X-ray source detection. $P_{\rm{B}}$ is proportional to the probability that the measured counts (Table 4) are the result of spurious background activity. The no-source probability $P_{\rm{B}}$, adopted from \citet{lansbury2014}, is calculated through the expression:
\begin{equation*}
P_{\rm{B}} (x \geqslant S) = \sum_{x=S}^{T} \frac{T!}{x!(T-x)!}p^x(1-p)^{T-x}
\end{equation*}

\noindent where we take T to be the sum of the total source (S) and total background  (B) counts in the full $0.3\--8$ keV energy band, and $p=1/(1+B/B_{\rm{src}})$, where $B_{\rm{src}}$ is the total background counts scaled by the ratio of the source and background region $(A_{\rm{S}}/B_{\rm{S}})$. Adopting the significance metric used in Paper I, we require that X-ray sources must possess $P_{\rm{B}} < 0.002$ to be considered a real X-ray detection rather than spurious background activity. 

We employ a combination of Gaussian and Gehrels statistics when computing the uncertainties in the source photon counts. For sources with fewer than 20 counts (see Table 4), and for all normalized background counts, we use Gehrels statistics to compute the error \citep{gehrels1986}. For sources with 20 counts or more (see Table 4), we use Gaussian statistics. The upper Gehrels error bound is computed as $1+\sqrt{x+0.75}$ while the lower bound is computed as $\sqrt{x-0.25}$, where $x$ is the counts. In Table~\ref{table:chandracounts} we quote the appropriate error for the counts of each source.\footnote{We quote symmetric error bounds as we took into account only the upper Gehrels error bound to be conservative.} In computing the error for background subtracted values such as fluxes and luminosities, we added the Gehrels error for the normalized background and the appropriate error of the source in quadrature.

The column densities shown in Table~\ref{table:chandracounts} are (foreground) weighted Galactic total hydrogen column densities and were generated via the \textit{Swift} Galactic $\nh$ tool, based upon the work of \citet{willingale2013}. 
To calculate hardness ratios for targets with sufficient counts, we use:
\begin{equation*}
HR = \frac{H-S}{H+S}
\end{equation*}
where H and S represent the counts from the $2\--8$ keV and $0.3\--2$ keV bands, respectively.

\begin{table*}
\caption{Target and \chandra~Observation Information}
\begin{center}
\begin{tabular}{ccccccc}
\hline
\hline
\noalign{\smallskip}
Name & $\alpha$ & $\delta$ & Cycle & Obs. Date & ObsID & Exp (ks) \\
\noalign{\smallskip}
\hline
\noalign{\smallskip}
J0122+0100  & 01\hr22\min18\sec.11  & +01\deg00\arcmin25\arcsec.76  & 18  & 2016 Sept 17  & 19505  & 65.2\\
J0841+0101  & 08\hr41\min35\sec.08  & +01\deg01\arcmin56\arcsec.20  & 17  & 2016 Jan 10  & 18199  & 21.9\\
J0849+1114  & 08\hr49\min05\sec.51  & +11\deg14\arcmin47\arcsec.26  & 17  & 2016 Mar 3  & 18196  & 21.0\\
J0859+1310  & 08\hr59\min53\sec.33  & +13\deg10\arcmin55\arcsec.39  & 17  & 2016 Jan 7  & 18200  & 16.2\\
J0905+3747  & 09\hr05\min47\sec.34  & +37\deg47\arcmin38\arcsec.24  & 17  & 2016 Jan 8  & 18197  & 17.2\\
J1045+3519 & 10\hr45\min18\sec.00& +35\deg19\arcmin13\arcsec.2& 18 & 2018 Jan 1 & 19506 & 23.8\\
 & & & 18 & 2018 Jan 7 & 20911 & 14.9\\
J1147+0945  & 11\hr47\min53\sec.68  & +09\deg45\arcmin55\arcsec.48  & 17  & 2016 Nov 7  & 18198  & 22.9\\
J1159+5320  & 11\hr59\min30\sec.29  & +53\deg20\arcmin55\arcsec.76  & 17  & 2016 Jul 19  & 18193  & 14.3\\
J1221+1137  & 12\hr21\min04\sec.98  & +11\deg37\arcmin52\arcsec.34  & 18  & 2017 May 1  & 19504  & 23.2\\
J1301+2918  & 13\hr01\min25\sec.26  & +29\deg18\arcmin49\arcsec.53  & 17  & 2016 Mar 6  & 18201  & 5.8\\
J1306+0735  & 13\hr06\min53\sec.60  & +07\deg35\arcmin18\arcsec.18  & 18  & 2017 Apr 25  & 19507  & 29.2\\
  &   &   & 18 & 2017 Apr 27  & 20064  & 24.7\\
  &   &   & 18 & 2017 Apr 30  & 20065  & 36.1\\
J1356+1822  & 13\hr56\min02\sec.89  & +18\deg22\arcmin18\arcsec.29  & 17  & 2016 Mar 10  & 18194  & 9.6\\
J2356-1016  & 23\hr56\min54\sec.49  & -10\deg16\arcmin07\arcsec.40  & 17  & 2015 Oct 30  & 18195  & 8.6\\
\noalign{\smallskip}
\hline
\end{tabular}
\end{center}
\tablecomments{Column $1$: Truncated merger designation. Columns $2\--3$: Coordinates of \chandra\ observations. Column $4\-- 5$: \chandra{} observation cycle and UT date of \chandra{}/ACIS observations. Column $6\--7$: \chandra\ observation ID and exposure time. }
\label{table:chandraobservations}
\end{table*}

\begin{table*}
\caption{Target and \xmm \ Observation Information}
\begin{center}
\begin{tabular}{ccccccc}
\hline
\hline
\noalign{\smallskip}
Name & $\alpha$ & $\delta$ & Cycle & Obs. Date & ObsID & Exp (ks) \\
\noalign{\smallskip}
\hline
\noalign{\smallskip}
J0122+0100 & 01\hr22\min18\sec.11 & +01\deg00\arcmin25\arcsec.76 & AO-15 & 2016 Jun 20 & 782010101 & 71 \\
J1221+1137 & 12\hr21\min04\sec.98 & +11\deg37\arcmin52\arcsec.34 & AO-16 & 2016 Jun 10 & 782010201 &  46\\
\noalign{\smallskip}
\hline
\end{tabular}
\end{center}
\tablecomments{Column $1$: Truncated merger designation. Column $2\--3$: Coordinates of \xmm{} observations. Columns $4\--5$: \xmm\ observation cycle and UT date of \xmm{} pn, MOS1, and MOS2 observations. Column $6\--7$: \xmm{} observation ID and exposure time.}
\label{table:xmmobservations}
\end{table*}

\subsection{XMM-Newton}
Table 3 shows details of the observations for the two merger systems observed by \xmm{} during the AO-15 and AO-16 observation cycles. Data calibration was performed using SAS, version 16.1.0, and the most up-to-date CCF calibration files. The \textsc{epic} events were screened to remove known hot pixels and data affected by background flaring.
To extract counts from our event files, we created $0.3\--10$ keV binned (bin factor = 32) images of the event files. Circular source apertures of $R = 30\arcsec$ were employed for source count extraction while background counts were extracted from apertures of $R = 60\arcsec$ in a region near the source and free of spurious sources. We constructed spectra for all three \textsc{epic} detectors for each data set using the \textsc{evselect} command and the same source and background apertures listed above. We also created the redistribution matrix and ancillary response files necessary for spectral modeling using the \textsc{rmfgen} and \textsc{arfgen} commands.

\subsection{Large Binocular Telescope Near-Infrared Spectroscopy}

As in Paper I, we obtained near-IR spectroscopic data to constrain whether the X-ray emission is consistent with AGN signatures or if it could instead be produced by high-mass X-ray binaries. We obtained near-IR ground-based spectra of their nuclei with the \emph{Large Binocular Telescope} Near Infrared Spectroscopic Utility with Camera Instruments (LBT LUCI; \citealp{seifert2003,seifert2010}). We obtained near-IR spectra for all 30 nuclei and regions of interest in our sample. The \emph{LBT} observations were conducted between November 2014 and January 2018, and were centered on the coordinates of the X-ray detections (listed in Table~\ref{table:chandracounts}). The configurations used were  the 1\farcs0 longslit or 1\farcs5 longslits, the G200 grating, and the HKspec filter. This gives an observed-frame wavelength coverage of $~1.37-2.37$ $\mu$m with a central wavelength of 1.93 $\mu$m, and a spectral resolution of $R \sim 858 - 1376$ (depending on the slitwidth) over this wavelength range.\footnote{\textsc{LBT} hosts two nearly identical LUCI spectrographs. LUCI-1 or LUCI-2 were used depending on availability. Unfortunately, no observations were obtained using both LUCIs in a binocular configuration.} The one-dimensional spectra were extracted using apertures which ranged in size from 0.5\arcsec{}x1\arcsec{} to 1.2\arcsec{}x1.5\arcsec{}; the extraction along the spatial direction of the spectra was determined by picking the smallest size based on the seeing conditions.

The observations, associated data reduction process, extracted 1-dimensional spectra, and measurements of six of these mergers have been presented in Paper I, and the rest will be discussed in detail in Constantin et al. (in prep.), along with a comprehensive analysis of the near-infrared kpc-scale properties of the whole sample of mid-infrared selected mergers. In this work we refer the reader to our discussion of the near-IR results in Section 6, Table~\ref{table:sfrlums},  Table~\ref{table:summary}, as well as the Appendix. Paper I discusses in detail the near-IR properties of the first six mergers followed-up with LBT. 

\section{Chandra and XMM Spectral Analysis}
\subsection{Chandra Spectra Extraction and Fit Significance}
Spectral extraction was performed for sources in each merger using the \textsc{ciao} \textsc{specextract} module, which provided spectra and their redistribution/response (RMF/ARF) files using source and background aperture inputs for use in the spectral fitting process. The 1\farcs5 source apertures as well as background apertures used for source and background counts extraction were reemployed for spectral extraction. Due to the low number of counts in the sources, spectra were not grouped and the corresponding ARF and RMF response files were not weighted. \\
\indent Spectral fitting was performed using the \textsc{xspec} \citep{arnaud1996} version 12.9.1 X-ray spectral fitting package. Due to the low-count nature of most of the X-ray sources, we employed C-stat statistics \citep{cash1979} during the fitting process. As discussed in \citet{tozzi2006} and \citet{brightman2014}, in the low-count regime ($\sim$ 100 counts or less) Cash statistics provide a more reliable metric for statistical significance than the traditional $\chi^2$ statistic. Consequently, we report the reduced C-stat value as the metric for goodness of fit, where the reduced C-stat is given by C-stat/dof and dof stands for the degrees of freedom of the fit. In order to obtain reliable fit results, we limited the spectral analyses to sources with $\geq$ 100 counts with one exception, SDSSJ1301+2918, which we fit simultaneously with archival data. We did not fit models to the remaining sources with less than 100 counts.\\

\subsection{Chandra Fitting Procedure}
Seven of the 15 observed mergers each possess a source with a sufficient number of counts ($\geq$100 counts) to enable direct spectral fitting. We took two approaches to the modeling: one phenomenological model and a physically-motivated model, BNTorus \citep{brightman2011}. 

The phenomenological model can be broken into four sub-models, each of which were fit independently to determine the best fit:
\begin{enumerate}
    \item The base model employed only an absorbed power-law, which took into account Galactic and extragalactic absorption, redshift, along with a \textsc{cabs} component to take into account Compton scattering. This model provided outputs for $\Gamma$ and $\nh{}$ and contained three free parameters. It is given in \textsc{xspec} as: $phabs \times [zphabs \times cabs \times pow$].
    \item The base model plus a scattered power-law component. This model contained four free parameters. It is given in \textsc{xspec} as: $phabs \times [pow + zphabs \times cabs \times pow$].
    \item The base model, a scattered power-law, and a Gaussian emission line component to account for potential Fe K$\alpha$ fluorescent line emission. This model contained five free parameters. It is given in \textsc{xspec} as: $phabs \times [zgauss + pow + zphabs \times cabs \times pow$].
    \item The base model with a Gaussian emission line component but without the scattered power-law component. This model contained four free parameters. It is given in \textsc{xspec} as: $phabs \times [zgauss + zphabs \times cabs \times pow$].
\end{enumerate}

The BNTorus model can also be broken into four sub-models which were independently fit:
\begin{enumerate}
    \item \textsc{BNTorus} with an opening angle of 60\deg{} and an edge-on inclination (87\deg{}). This model takes into account the extragalactic $\nh{}$,\ $\Gamma$, the redshift, and an additional component (phabs) was included to account for Galactic absorption. BNTorus self-consistency accounts for any fluorescent emission lines. This model contained three free parameters. We refer to this as the base BNTorus model, given in \textsc{xspec} components as: $phabs \times [atable\{torus1006.fits\}]$.
    \item The base model and an additional scattered power-law component, const $\times$ phabs, to account for soft X-ray emission, where const stands for the scattering fraction $f_S$. The normalization of the scattered power-law was tied to that of BNTorus. This model contained four free parameters and is given in \textsc{xspec} as: $phabs \times [const\times powerlaw + atable\{torus1006.fits\}]$.
    \item The base model and an additional \textsc{apec} component, which models spectral emission due to collisionally ionized diffuse gas. \textsc{apec} accounts for plasma temperature, redshift, elemental abundance, and possesses its own normalization. This model contained five free parameters and is given in \textsc{xspec} as: $phabs \times [apec + atable\{torus1006.fits\}]$.
    \item The base model with two additional components, a scattered power-law and \textsc{apec}. The normalization of the scattered power-law was tied to that of BNTorus. This model contained six free parameters, and is given in \textsc{xspec} as: $phabs \times [const\times powerlaw + apec + atable\{torus1006.fits\}]$
\end{enumerate}

To determine the best fitting model for each approach, we obtained the C-stat values for each given fit of a spectrum and examined the $\Delta$C-stat value between fits of different model permutations. We specify in the process of this analysis that all additional components added to either model approach possessed only one free parameter each (with the exception of \textsc{apec} which possessed two free parameters), and thus adding one component to each model approach represented adding one free parameter to the model. With this in mind, a statistically significant improvement with 90\% confidence for the addition of one free parameter for a fit is given by $\Delta$C-stat = C-stat$_{old}$ $\--$ C-stat$_{new}$ $>$ 2.71 \citep{brightman2014,tozzi2006,marchesi2016}. If the addition of a component with one free parameter to the model resulted in a $\Delta$C-stat $>$ 2.71, we identified it as a statistically relevant component and included it in the final model for the spectrum in question. For components with more than one free parameter, we required a $\Delta$C-stat twice as high or $>$ 5.42. The exception to these rules, of course, is if nonphysical values were pegged for other model components after the addition of a new component, at which point we deemed the new component insignificant to the fit.

All modeling approaches above shared the following commonalities during the fitting procedure:
\begin{enumerate}
    \itemsep0em
    \item For all fits, Galactic absorption was fixed to the value determined along the line of sight obtained via the Swift Galactic $\nh$  calculator \citep{willingale2013}. Values of Galactic $\nh$ are listed in Tables 4, 5, and 6. 
    \item Redshifts were fixed to the spectroscopic redshift value for the host galaxy in each merger (see Table~\ref{table:sample}).
    \item The Gaussian line component, when statistically significant and included in the phenomenological model, was frozen at the peak of the excess emission above the power-law in the range of $6\--7$ keV (with line peaks at either 6.4 or 6.7 keV). The line widths were frozen at a $\sigma$ of 0.1 keV. The normalization was free to vary.
    \item For models with Gaussian emission line components, we computed the equivalent width using the \textsc{eqw} and \textsc{err} (90\% uncertainty) commands in \textsc{xspec}. The equivalent width of these spectral lines provides crucial insight into the level of obscuration along the line of sight for each source (e.g. \citealp{brightman2011}).
    \item Unless otherwise stated, normalizations of the model components were allowed to vary freely.
    \item For fits incorporating multiple datasets, we appended an additional constant to the front of each model in order to account for inter-detector sensitivity.
\end{enumerate}
The results of these models are shown in Tables~\ref{table:phenspec} and ~\ref{table:bntspec}. Components which are absent from the best fitting model are denoted with an ellipsis. We discuss the general results of direct spectral fitting in Section 5.2 as well as the implementation of this model on a case-by-case basis in the Appendix.

\begin{figure*}[h!]
    \centering
    \includegraphics[width=0.9\linewidth]{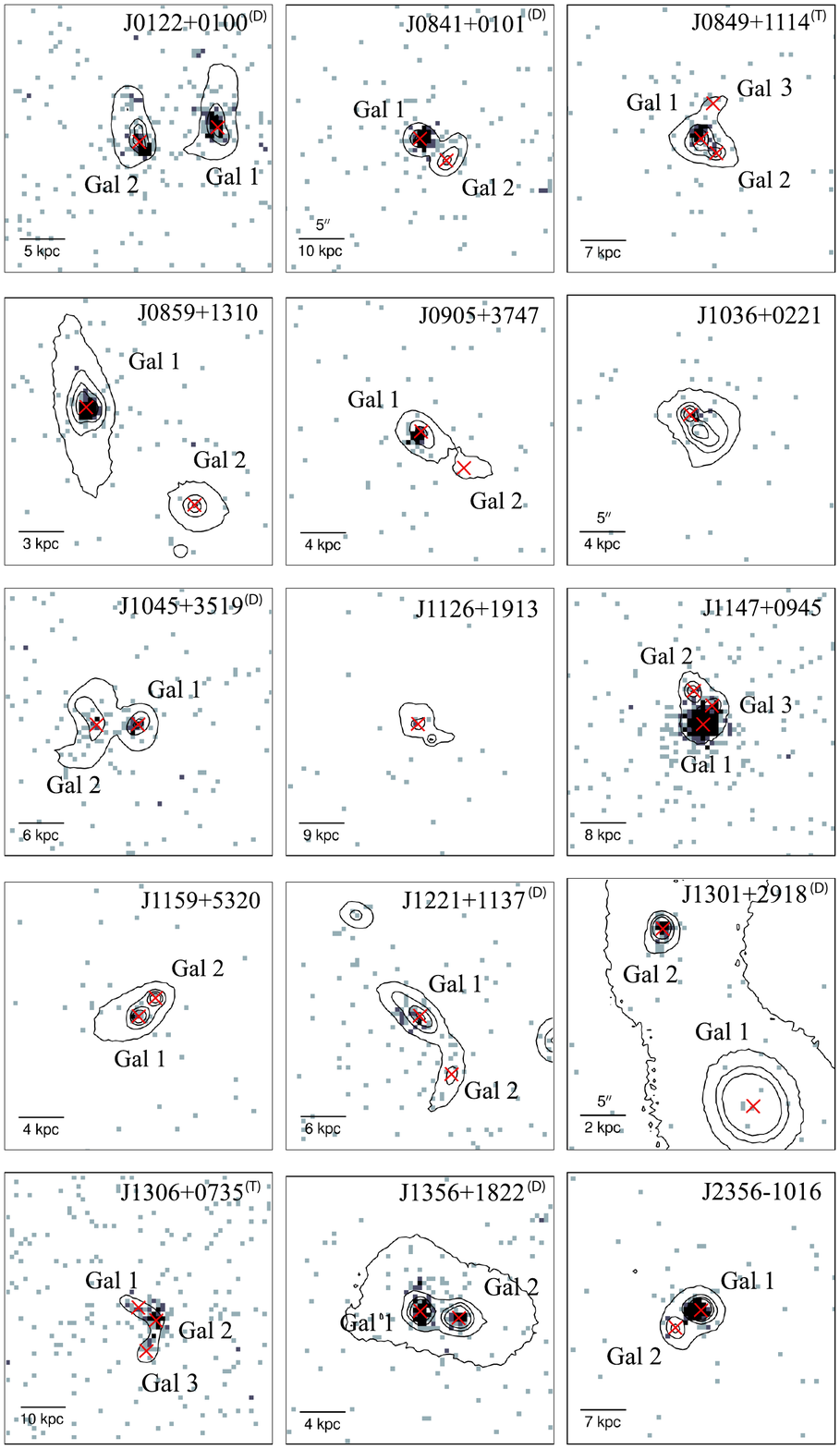}
    \caption{$0.3\--8$ keV X-ray images from \chandra{} Cycle 15, 17, and 18 observations with SDSS r-band contours overlaid for each merger. North is up and East is to the left. Note, the X-ray data for J1126+1319 and J1036+0221 were originally reported in \citet{satyapal2017} - we show the data here for completeness of the sample. Each red X indicates the approximate position of a galaxy nucleus or other region of interest. Mergers with dual or triple X-ray sources are denoted with a (D) or (T)  in the upper right hand corner of the panels.}
    \label{fig:xrays}
\end{figure*}

\subsection{XMM-Newton Fitting Procedure}
 In fitting the $0.3\--10$ keV X-ray spectra of J0122+0100 and J1221+1137 obtained from the \xmm{} pn camera, we followed identical procedures to those given in Section 4.1 and 4.2 with a few exceptions:
\begin{itemize}
    \item $\chi^2$ statistics were employed, rather than C-Stat statistics. Here, again, a statistically significant improvement to a fit with 90\% confidence must result in a change in the $\chi^2$ statistic greater than 2.71.
    \item For the case of the phenomenological model, we added a constant term in front of the scattered power-law component to represent the scattering fraction and tied the normalizations of the absorbed and scattered power-laws together. 
    \item To simplify the phenomenological and BNTorus models, we froze the scattered power-law constants to 0.1, or 10\%. 
\end{itemize}
This form of the full phenomenological model is given in \textsc{xspec} components as:
\begin{equation*}
    phabs \times [const \times pow + zgauss + zphabs \times cabs \times pow]
\end{equation*}
We discuss the spectral fitting results briefly in Section 5.3 and the detailed implementation of the model for both mergers in the Appendix. {Owing to the faintness of these sources, we were unable to use the generated spectra from the MOS1 and MOS2 cameras, as these spectra were background dominated and yielded nonphysical photon count rates.}

\begin{table*}
\caption{\chandra{} X-ray Sources}
\begin{center}
\scalebox{0.95}{
\hspace{-2cm}
\begin{tabular}{ccccccccccc}
\hline
\hline
\noalign{\smallskip}
Name (SDSS) & Source & Galaxy  & $\nh$ & $\alpha_\chi$ & $\delta_\chi$ & Counts  & Counts  & Counts  & HR & log$(P_{\rm{B}})$ \vspace{-4mm}\\
\noalign{\smallskip}
& & &  &  &  &  &  &  &  &  \vspace{-4mm}\\
\noalign{\smallskip}
 & &  & ($10^{20}$ cm$^{-2}$) &  &  &  $0.3\--8$ keV &  $0.3\--2$ keV &  $2\--8$ keV & &\\
\noalign{\smallskip}
\hline
\noalign{\smallskip}
J0122+0100 & NW$^{\dagger}$ & 1 & 3.50 & 1\hr22\min17\sec.555 & +1\deg00\arcmin27\arcsec.341 & 69$\pm9$ & 46$\pm7$ & 23$\pm5$ & -0.32 & -97.2\\
 & SE$^{\dagger\dagger}$ & 2 & 3.50 & 1\hr22\min18\sec.083 & +1\deg00\arcmin24\arcsec.723 & 60$\pm8$ & 37$\pm6$ & 23$\pm5$ & -0.22 & -81.4 \\
J0841+0101 & E$^{\dagger}$ & 1 & 4.68 & 8\hr41\min35\sec.054 & +1\deg01\arcmin56\arcsec.05 & 181$\pm14$ & 87$\pm10$ & 94$\pm10$ & 0.04 & -351.2\\
 & W$^{\dagger\dagger}$ & 2 & 4.68 & 8\hr41\min34\sec.775 & +1\deg01\arcmin54\arcsec.690 & 5$\pm4$ & 5$\pm4$ & 0$\pm3$ & \dots & -3.8\\
J0849+1114 & SE$^{\dagger}$ & 1 & 3.80 & 8\hr49\min05\sec.529 & +11\deg14\arcmin47\arcsec.876 & 108$\pm11$ & 57$\pm8$ & 51$\pm7$ & -0.06 & -206.0 \\
 & SW$^{\dagger\dagger}$ & 2 & 3.80 & 8\hr49\min05\sec.381 & +11\deg14\arcmin45\arcsec.747 & 11$\pm5$ & 10 $\pm5$ & 1$\pm3$ & -0.85 & -12.3\\
 & N$^{\dagger\dagger}$ & 3 & 3.80 & 8\hr49\min05\sec.448 & +11\deg14\arcmin51\arcsec.646 & 6$\pm4$ & 5$\pm4$ & 1$\pm3$ & \dots & -5.5\\
J0859+1310 & NE$^{\dagger,\dagger\dagger}$ & 1 & 3.72 & 8\hr59\min53\sec.299 & +13\deg10\arcmin55\arcsec.03 & 434$\pm21$ & 6$\pm4$ & 428$\pm21$ & 0.97 & -984.7\\
& \dots & 2 & 3.72 & \dots & \dots & \dots & \dots & \dots & \dots & \dots\\
J0905+3747 & NE$^{\dagger,\dagger\dagger}$ & 1 & 1.91 & 9\hr05\min47\sec.374 & +37\deg47\arcmin37\arcsec.88 & 69$\pm9$ & 19$\pm6$ & 50$\pm7$ & 0.45 & -124.3\\
& \dots & 2 & 1.91 & \dots & \dots & \dots & \dots & \dots & \dots & \dots\\
J1045+3519 & W$^{\dagger\dagger}$ & 1& 1.96& 10\hr45\min18\sec.051 & +35\deg19\arcmin12\arcsec.987  & 23$\pm5$& 18$\pm6$& 6$\pm4$& -0.50 & -27.20\\
 & E$^{\dagger\dagger}$ & 2& 1.96& 10\hr45\min18\sec.42 & +35\deg19\arcmin12\arcsec.93 & 13$\pm5$& 8$\pm4$& 6$\pm4$& -0.14 & -13.1 \\
J1147+0945 & S$^{\dagger\dagger}$ & 1 & 2.91 & 11\hr47\min53\sec.611 & +9\deg45\arcmin51\arcsec.66 & 3145$\pm56$ & 663$\pm26$ & 2483$\pm50$ & 0.58 & -8070.2\\
& \dots & 2 & 2.91 & \dots & \dots & \dots & \dots & \dots & \dots & \dots\\
& \dots & 3 & 2.91 & \dots & \dots & \dots & \dots & \dots &\dots & \dots\\
J1159+5320 & SE$^{\dagger}$ & 1 & 1.78 & 11\hr59\min30\sec.327 & +53\deg20\arcmin56\arcsec.030 & 19$\pm6$ & 2$\pm3$ & 17$\pm6$ & 0.80 & -27.7\\
& \dots & 2 & 1.78 & \dots & \dots & \dots & \dots & \dots & \dots & \dots\\
J1221+1137 & NE* & 1 & 2.83 & 12\hr21\min05\sec.042 & +11\deg37\arcmin52\arcsec.01 & 25$\pm5$ & 18$\pm6$ & 7$\pm4$ & -0.45 & -33.7\\
 & SW* & 2 & 2.83 & 12\hr21\min04\sec.776 & +11\deg37\arcmin47\arcsec.43 & 5$\pm4$ & 3$\pm4$ & 2$\pm3$ & \dots & -5.0 \\
J1301+2918 & NE$^{\dagger}$ & 2 & 0.97 & 13\hr01\min25\sec.255 & +29\deg18\arcmin49\arcsec.165 & 50$\pm7$ & 29$\pm6$ & 21$\pm5$ & -0.16 & -108.8\\
 & SW$^{\dagger\dagger}$ & 1 & 0.966 & 13\hr01\min24\sec.552 & +29\deg18\arcmin30\arcsec.036 & 3$\pm4$ & 3$\pm3$ & 0$\pm3$ & \dots & -3.7\\
J1306+0735 & NE* & 1 & 2.51 & 13\hr06\min53\sec.601 & +7\deg35\arcmin18\arcsec.85 & 18$\pm5$ & 12$\pm5$ & 6$\pm4$ & -0.35 & -15.1\\
 & SW* & 2 & 2.51 & 13\hr06\min53\sec.429 & +7\deg35\arcmin17\arcsec.17 & 61$\pm8$ & 34$\pm6$ & 27$\pm6$ & -0.12 & -74.7\\
 & SE$^{\dagger\dagger}$ & 3 & 2.51 & 13\hr06\min53\sec.550 & +7\deg35\arcmin14\arcsec.44 & 15$\pm6$ & 13$\pm5$ & 2$\pm4$ & -0.74 & -11.9\\
J1356+1822 & E$^{\dagger}$ & 1 & 2.20 & 13\hr56\min02\sec.887 & +18\deg22\arcmin18\arcsec.214 & 154$\pm13$ & 84$\pm9$ & 70$\pm9$ & -0.09 & -322.6\\
 & W$^{\dagger}$ & 2 & 2.20 & 13\hr56\min02\sec.619 & +18\deg22\arcmin17\arcsec.741 & 50$\pm7$ & 17$\pm6$ & 33$\pm6$ & 0.32 & -87.0\\
J2356-1016 & NW$^{\dagger\dagger}$ & 1 & 2.93 & 23\hr56\min54\sec.361 & -10\deg16\arcmin05\arcsec.666 & 522$\pm23$ & 54$\pm8$ & 468$\pm22$ & 0.79 & -1245.4\\
& \dots & 2 & 2.93 & \dots & \dots & \dots & \dots & \dots & \dots & \dots\\
\noalign{\smallskip}
\hline
\end{tabular}
}
\end{center}
\tablecomments{Column 1: Truncated merger designation. Column 2: X-ray source designation given in cardinal coordinates. Column 3: Galaxy nucleus hosting the X-ray source. Source positions were determined and verified with several methods: ($\dagger$) $\--$ Source position determined and/or verified using the \textsc{wavdetect} \textsc{ciao} package; ($\dagger\dagger$) $\--$ source position determined and/or verified with the aid of a smoothed $0.3\--8$ keV image using a $2\--3$ pixel Gaussian kernel; (*) $\--$ source positions adopted from \citet{satyapal2017}. Columns 4: Galactic $\nh$ is given in units of $10^{20}$ cm$^{-2}$. Columns $5\--6$: Right ascension and declination coordinates of source apertures. Columns $7\--9$: photons detected in each energy band. Columns $10\--11$: Hardness ratio and logarithm of the binomial no-source $P_{\rm{B}}$ statistic.}
\label{table:chandracounts}
\end{table*}

\section{X-ray Results}
\subsection{\chandra{}\textit{/ACIS-S} Imaging Results}
\chandra \ $0.3\--8$ keV X-ray images are shown with SDSS contours overlaid in Figure~\ref{fig:xrays} for all fifteen galaxy mergers in this sample. We report the count statistics, hardness ratios, and no-source probabilities $P_{\rm{B}}$ in Table~\ref{table:chandracounts} for each source identified.  In the 15 mergers, a total of 25 X-ray sources coincident with the galaxy nuclei are detected in the full $0.3\--8$ keV band using the $P_{\rm{B}} < 0.002$ metric discussed in Section 3.1, while 18 out of the 25 sources are also detected at this threshold in the hard $2\--8$ keV band. We find a single X-ray source in 7 out of 15 mergers while the remaining 8 show dual X-ray signatures coincident with the nuclei of the mergers. In 2 out of the 8 systems with dual X-ray sources, we also note the presence of a third X-ray source (see Tables 4, 7, and 8). 

Sufficient counts ($\geq$ 100 counts) were obtained to perform direct spectral fitting for sources in 7 out of 15 mergers observed with \chandra{}. We discuss the spectral analysis of these 7 systems below and list the results in Table~\ref{table:phenspec} and 6 for the two different modeling approaches outlined in Section 4. Since spectral analysis for all 15 mergers was not possible, we took a uniform approach for estimating the absorbed X-ray luminosities using the \chandra{} \textsc{PIMMS} for all sources. We list these luminosities in Table~\ref{table:sfrlums}, assuming  a simple power-law model with $\Gamma=1.8$ \citep{mushotzky1993,ricci2017apj} and corrected for Galactic absorption along the line of sight. We use the term absorbed luminosity to refer to luminosities which are corrected for Galactic absorption, but which not corrected for intrinsic absorption of the X-ray source. 

\subsection{\chandra{}\textit{/ACIS-S} Spectral Analysis Results}
The results of the phenomenological modeling approach are listed in Table~\ref{table:phenspec}. We find for $\Gamma$ a range of $1.5\--3.0$ across all seven sources and statistically significant scattered power-laws, which account for the soft X-rays in the spectra, in five sources. Direct fitting reveals high $\nh{}$, on the order $\gtrsim 10^{23}$ cm$^{-2}$, in four out of seven sources while a fifth source, J2356-1016NW, shows this level of $\nh$ within its determined uncertainties. We identify statistically significant Fe K$\alpha$ fluorescent emission lines in five of the seven sources. As discussed in \citet{ghisellini1994} and \citet{brightman2011}, equivalent widths in excess of 150 eV cannot be obtained for AGNs with a toroidal geometry along unobscured lines of sight and require column densities in excess of $\nh{} \sim 10^{23}$ cm$^{-2}$. Four of the five sources with Fe K$\alpha$ lines exhibit equivalent widths in excess of 150 eV, further suggesting a majority of the modeled X-ray sources are buried under high column densities (if we assume a toroidal geometry explains the nature of the obscuring material). We note one source, J0849+1114SE, exhibits both an iron line with a very high equivalent width and a low level of $\nh{}$. While contradictory, further modeling with BNTorus, discussed below, lends greater evidence to a scenario in which the system is indeed buried behind a high obscuring column as well. Thus, we identify with this modeling method a total of 6 sources which exhibit signs of high obscuration. All sources modeled with this approach exhibit unabsorbed luminosities in excess of $\rm{L}_{2\--10\ \rm{keV}} \sim 10^{42}$ erg s$^{-1}$, and thus we conclude all seven sources are bona fide AGN. 

The results of the BNTorus modeling approach are reported in Table~\ref{table:bntspec}. For this approach direct fitting reveals a range in $\Gamma$ of $1.4\--2.5$ and statistically significant scattered power-laws for five sources with a range of $0.6\--4.7\%$ for the scattering fractions. We also find five sources exhibit high $\nh{}$, on the order $\gtrsim10^{23}$ cm$^{-2}$, consistent with that predicted by the phenomenological model above with one exception: J2356-1016NW no longer reaches a level of $\sim 10^{23}$ cm$^{-2}$ within its uncertainties. As noted previously, BNTorus reveals high obscuration in J0849+1114SE. As before, all sources exhibit unabsorbed luminosities in excess of $\rm{L}_{2\--10\ \rm{keV}} \sim 10^{42}$ erg s$^{-1}$, from which we conclude again that all sources are AGNs.

Briefly, we note that we also compared these results to that obtained via the MYTorus model \citep{murphy2009} and found largely consistent results with regard to the levels of $\nh{}$ and the equivalent widths discussed above and listed in Tables~\ref{table:phenspec} and ~\ref{table:bntspec}. We discuss the spectral results of these systems on a case-by-case basis in the Appendix and include brief comparisons between the phenomenological, BNTorus, and MYTorus approaches. All spectral plots are shown in the Appendix.

\subsection{XMM-Newton Results}
Examined with the phenomenological approach, both J0122+0100 and J1221+1137 are best fit with absorbed power-laws with scattered power-law components. While we found \textsc{apec} components to be statistically significant for both models, the inclusion of \textsc{apec} resulted in nonphysical values for either $\Gamma$ (in the case of J0122+0100) or $\nh$ (for J1221+1137), and we therefore rejected the addition of an \textsc{apec} component to the best fitting models. The models for each system reveal high obscuration, with $\nh > 10^{23}$ cm$^{-2}$, and unabsorbed luminosities in excess of $\rm{L}_{2\--10\ \rm{keV}} \sim 10^{42}$ erg s$^{-1}$ after correcting for absorption, indicating both systems contain at least a single AGN - consistent with the results of Paper I. 

Spectral fitting with the BNTorus approach yielded similar results for these two systems: we find both are best fit with the base BNTorus model plus a scattered power-law component, and the use of the \textsc{apec} component provided a statistically significant improvement to the BNTorus model for both X-ray sources. \textsc{Apec} returned plasma temperatures of $\rm{kT} \sim 0.5$ keV and $\rm{kT} \sim 0.9$ keV for J0122+0100 and J1221+1137, respectively. For both systems, the BNTorus approach shows high obscuration, with $\nh > 10^{23}$ cm$^{-2}$. After correcting for absorption, we find for J1221+1137 an unabsorbed luminosity in excess of $\rm{L}_{2\--10\ \rm{keV}} \sim 10^{42}$ erg s$^{-1}$, while the best fit model for J0122+0100 returns a value of $\rm{L}_{2\--10\ \rm{keV}} = 9.7^{+9.3}_{-4.3}\times10^{41}$ erg s$^{-1}$. We find the BNTorus and phenomenological model results for both mergers agree within the uncertainties.

We discuss the fitting and results of each system in the Appendix. All spectral plots are shown in the Appendix.

\begin{table*}
\caption{Spectral Fitting Results for the Phenomenological Model}
\begin{center}
\begin{tabular}{ccccccccc}
\hline
\hline
\noalign{\smallskip}
 Target & Reduced & $N^{\rm{Gal.}}_{\rm{H}}$ & $\Gamma$ & $N_{\rm{H}}$  &  Fe K$\alpha$ line & Line peak & Equiv. width  & $\rm{L}^{\rm{Unabs.}}_{2\--10\ \rm{keV}}$ \vspace{-4mm}\\
\noalign{\smallskip} 
 & &  &  &    &  &   &\vspace{-4mm}\\
\noalign{\smallskip} 
 & C-Stat & ($10^{20}$ cm$^{-2}$) &  & ($10^{22}$ cm$^{-2}$) &  &  (keV) & (keV) & (erg s$^{-1}$)\\
 \noalign{\smallskip}
\hline
\noalign{\smallskip}
J0841+0101E & $412.45 / 512$ & 4.68 & $2.5^{+0.4}_{-0.4}$ & $29.8^{+11.0}_{-9.0}$ & Y & 6.4 & $0.75^{+2.59}_{-0.47}$ & $2.3^{+0.7}_{-1.5}\times10^{43}$ \vspace{+1mm}\\
J0849+1114SE & $302.87 / 521$ & 3.80 & $3.0^{+0.8}_{-0.9}$ & $4.0^{+1.8}_{-1.5}$ & Y & 6.4 & $4.37^{11.78}_{-2.82}$ & $1.4^{+0.5}_{-0.9}\times10^{42}$ \vspace{+1mm}\\
J0859+1310NE & $430.08 / 522$ & 3.72 & $2.4^{+0.9}_{-0.8}$ & $17.4^{+5.0}_{-4.3}$ & Y & 6.7 & $0.23^{+\dots}_{-\dots}$ & $7.9^{+1.1}_{-8.5}\times10^{42}$ \vspace{+1mm}\\
J1147+0945S & $540.60 / 521$ & 2.91 & $1.5^{+0.2}_{-0.1}$ & $2.6^{+0.3}_{-0.3}$ & Y & 6.4 & $0.11^{+0.09}_{-0.09}$ & $8.9^{+1.3}_{-1.5}\times10^{43}$ \vspace{+1mm}\\
J1301+2918NE & $583.42 / 1571$ & 0.966 & $1.7^{+0.2}_{-0.2}$ & $438.3^{+945.7}_{-294.6}$ & Y & 6.4 & $2.06^{+7.58}_{-0.67}$ & $6.9^{+0.7}_{-5.4}\times10^{45}$ \vspace{+1mm}\\
J1356+1822E & $376.14 / 522$ & 2.20 & $2.1^{+0.4}_{-0.3}$ & $75.2^{+42.7}_{-28.3}$ & N & \dots & \dots & $3.9^{+1.4}_{-3.6}\times10^{43}$ \vspace{+1mm}\\
J2356-1016NW & $510.86 / 522$ & 2.93 & $2.0^{+0.4}_{-0.4}$ & $8.6^{+1.7}_{-1.6}$ & N & \dots & \dots  & $4.7^{+0.8}_{-2.5}\times10^{43}$  \vspace{+1mm}\\
\noalign{\smallskip}
\hline
\end{tabular}
\end{center}
\tablecomments{Column 1: Target designation with cardinal coordinate designation. Column 2: Reduced C-Stat value, given by C-Stat/dof. Column 3: Galactic $\nh$. Column 4: Photon index. Column 5: $\nh$ within merger system along line of sight of X-ray source in question. Columns $6\--8$: Results for presence or absence of Iron K$\alpha$ emission lines. Column 9: Unabsorbed luminosity $\rm{L}_{2\--10\ \rm{keV}}$ corrected for the absorption reported in  columns 3 and 5.}
\label{table:phenspec}
\end{table*}

\begin{table*}
\caption{Spectral Fitting Results from the BNTorus Model}
\begin{center}
\begin{tabular}{ccccccccc}
\hline
\hline
\noalign{\smallskip}
 Target & Reduced & $N^{\rm{Gal.}}_{\rm{H}}$ & $\Gamma$ & $f_s$ & $N_{\rm{H}}$ & $\rm{L}^{\rm{Unabs.}}_{2\--10\ \rm{keV}}$ \vspace{-4mm}\\
\noalign{\smallskip} 
 &  &  &  &  &   &  &  \vspace{-4mm}\\
\noalign{\smallskip} 
 & C-Stat & ($10^{20}$ cm$^{-2}$) & & (\%) & ($10^{22}$ cm$^{-2}$) & (erg s$^{-1}$)\\
 \noalign{\smallskip}
\hline
\noalign{\smallskip}

J0841+0101E & $417.63 / 522$ & 4.68 & $2.5^{+\dots}_{-0.4}$ & $1.4^{+2.0}_{-0.8}$ & $35.2^{+15}_{-9.5}$ & $3.1^{+0.9}_{-2.6}\times10^{43}$ \vspace{+1mm}\\
J0849+1114SE & $315.25 / 522$ & 3.80 & $1.5^{+0.6}_{-\dots}$ & $4.7^{+51.4}_{-4.4}$ & $114^{+\dots}_{-93}$ & $1.4^{+0.7}_{-0.1}\times10^{43}$ \vspace{+1mm}\\
J0859+1310NE & $432.20 / 523$ & 3.72 & $2.2^{+0.6}_{-0.6}$ & \dots & $14.9^{+3.9}_{-2.7}$ & $6.7^{+0.9}_{-0.7}\times10^{42}$ \vspace{+1mm}\\
J1147+0945S & $544.72 / 523$ & 2.91 & $1.5^{+0.1}_{-0.1}$ & \dots & $2.3^{+0.2}_{-0.2}$ & $8.9^{+1.2}_{-1.7}\times10^{43}$ \vspace{+1mm}\\
J1301+2918NE & $593.50 / 1572$ & 0.966 & $1.8^{+0.3}_{-0.2}$ & $1.2^{+2.3}_{-0.7}$ & $148^{75}_{-53}$ & $5.2^{+2.0}_{-0.5}\times10^{42}$ \vspace{+1mm}\\
J1356+1822E & $378.79 / 522$ & 2.20 & $2.1^{+0.4}_{-0.4}$ & $1.7^{+3.0}_{-1.2}$ & $56^{+33}_{-20}$ & $1.9^{+0.6}_{-0.2}\times10^{43}$ \vspace{+1mm}\\
J2356-1016NW & $510.36 / 522$ & 2.93 & $2.2^{+0.5}_{-0.4}$ & $0.6^{+0.8}_{-0.4}$ & $8.2^{+1.4}_{-1.6}$  & $4.7^{+0.7}_{-2.4}\times10^{43}$ \vspace{+1mm}\\
\noalign{\smallskip}
\hline
\end{tabular}
\end{center}
\tablecomments{Column 1: Target designation with cardinal coordinate designation. Column 2: C-Stat value over the degrees of freedom. Column 3: Galactic $\nh$. Column 4: Photon index. Column 5: Fraction of scattered photons. Column 6: $\nh$ within merger system along line of sight of X-ray source in question. Column 9: Unabsorbed luminosity $\rm{L}_{2\--10\ \rm{keV}}$ corrected for the absorption reported in columns 3 and 5.}
\label{table:bntspec}
\end{table*}

\section{The Nature of the Nuclear Sources}
In our sample of 15 mergers, we detect at least one X-ray source at the $2\sigma$ level or higher in all mergers, with 13 out of the 15 showing  $>$  $3\sigma$ detections, suggestive of the presence of at least one AGN per interacting system in the entire sample.  Out of the 15 mergers, 8 display dual X-ray sources coincident with the optical nuclei, 6 of which are detected at the $2\sigma$ level or higher. Two of these systems display triple X-ray sources with SDSS counterparts. For 7 of the detected X-ray sources, there are sufficient counts for a spectral analysis. The unabsorbed hard X-ray luminosities where available, or the absorbed hard X-ray luminosities for sources with $<$ 100 counts, range from $\rm{L}_\mathrm{2-10~keV}\sim4\times10^{39}$~erg~s$^{-1}$ to $\sim4\times10^{43}$~erg~s$^{-1}$. These X-ray luminosities are within the range of the absorbed hard X-ray luminosities reported in the literature for confirmed dual AGNs (see Table 8 in Paper I and references therein). Apart from the well studied dual, Mrk 463 \citep{bianchi2008}, we find no evidence for statistically significant variability of the X-ray sources in this sample compared to archival data. We discuss briefly the variability of Mrk 463 in the Appendix.

While X-ray emission coincident with the galaxy nuclei is highly suggestive of AGN activity, we investigated, as in Paper I, the possibility that the X-ray emission could be produced by a population of high-mass X-ray binaries (HMXBs) using the LBT near-IR data. We obtained near-IR spectra for all 30 nuclei or other regions of interest within our sample. These spectra yielded Pa$\alpha$ line fluxes for 23 out of the 25 X-ray sources. The observations for the two nuclei in SDSSJ1301+2918 did not show confident detections of Pa$\alpha$ because the line was redshifted to 1.92$\mu$m, which lies within a telluric atmospheric absorption band barely accessible even in the very driest of conditions. The near-infrared spectral analysis will be described in a Constantin et al., in prep.

Though outlined thoroughly in Paper I, we briefly discuss the calculation of the predicted X-ray emission from XRBs here. We first assumed all of the Pa$\alpha$ flux arises due to gas ionized by star formation alone, although in reality it is possible some of this emission could arise from gas ionized by AGN activity. To compute the H$\alpha$ line fluxes, we took the near-IR Pa$\alpha$ fluxes and assumed an intrinsic H$\alpha$ to Pa$\alpha$ line flux ratio of 7.82 \citep{osterbrock2006}. With these H$\alpha$ fluxes, we used the relation between the star formation rate (SFR) and H$\alpha$ flux obtained in \citet{kennicutt1994} to compute the SFRs at the locations of the Chandra X-ray sources and remaining nuclei. Finally, we computed the expected X-ray contribution from XRBs by employing the relation given in \citet{lehmer2010} which relates the SFR, stellar mass, and X-ray emission for a galaxy. It is important to note that the infrared luminosities of the mergers in our sample are similar to the luminosities of the sample of local LIRGs used in \citet{lehmer2010} to derive this global, galaxy-wide relation. For cases where a mass was not available for one of the nuclei within a merger we used the mass from the companion nucleus; note, the mass dependent term within the \citet{lehmer2010} relation has little affect on the general result. the In Table~\ref{table:sfrlums}, we list the predicted X-ray luminosities from HMXBs. In all cases but one the absorbed X-ray luminosities exceed that predicted from star formation, highly suggestive that the X-ray emission requires the presence of an AGN. In Constantin et al., we also report the detection of coronal line emission, a robust indicator of an AGN (see Paper I) in 8 out of the 15 mergers.

If we adopt a strict definition of an AGN in this work as 1) requiring  $\rm{L}_\mathrm{2-10~keV} > 10^{42}$~erg~s$^{-1}$, either absorbed or unabsorbed when a spectral analysis was performed, 2) the detection of a coronal line, 3) the detection of a statistically significant Fe K$\alpha$ fluorescent emission line, or 4) optical spectroscopic classification as an AGN, we confirm the presence of at least one AGN in 13 out of the 15 mergers, with two of the systems hosting dual AGNs (J0849+1114 and the previously known dual system in Mrk 463). All X-ray sources that do not meet our strict definition of an AGN we classify as AGN candidates. We provide a summary classification for all targets in Table~\ref{table:summary}. We note that while a ULX origin for the X-ray detections is a possibility, the vast majority of ULXs have total unabsorbed 0.2$\--$10 keV luminosities between $10^{39}-10^{40}$~erg~s$^{-1}$ \citep{sutton2012}, significantly below most of the absorbed luminosities of our targets, which are themselves lower limits to the actual absorption-corrected luminosities. Furthermore, our targets were selected using mid-IR AGN colors, which are not generally associated with ULX activity \citep[e.g., Section 4.2 in ][]{Secrest2015a}. In our entire sample of mergers, there are a total of 8 dual or triple AGN candidates. We provide a detailed discussion of each individual merger in the appendix.

\begin{table*}
\caption{Nuclear Star Formation Rates, Predicted Luminosities from XRBs, and X-ray Source Luminosities}
\begin{center}
\begin{tabular}{cccccc}
\hline
\hline
\noalign{\smallskip}
Name (SDSS) & X-ray Source & Galaxy & SFR & $\rm{L}^{\rm{SF}}_{2\--10\ \rm{keV}}$ & $\rm{L}^{\rm{Abs.}}_{2\--10\ \rm{keV}}$ \vspace{-4mm} \\
\noalign{\smallskip}
 &  &  &  &  &  \vspace {-4mm}\\
\noalign{\smallskip}
 & & & $\rm{M}_\sun \ \rm{yr}^{-1}$ & $10^{40}$ erg s$^{-1}$ & $10^{40}$ erg s$^{-1} $ \\
\noalign{\smallskip}
\hline
\noalign{\smallskip}
J0122+0100 & NW & 1 & 4.74 & $0.97\pm0.22$ & $5.7\pm1.3$\\
 & SE & 2 & 1.82 & $0.38\pm0.09$ & $4.8\pm1.1$\\
J0841+0101 & E & 1 & 14.72 & $2.71\pm0.40$ & $180\pm32$\\
 & W & 2 & 2.86* & $0.78\pm0.44^{\dagger}$ & $4.5\pm4.4$\\
J0849+1114 & SE& 1 & 13.16 & $2.27\pm0.20^{\dagger}$ & $52\pm10.$\\
 & SW & 2 & 0.48 & $0.12\pm0.11$ & $5.1\pm2.9$\\
 & N & 3 & 1.79 & $0.43\pm0.20$ & $2.7\pm2.3$\\
J0859+1310 & NE & 1 & 0.18 & $0.42\pm0.40$ & $39.9\pm5.9$\\
 &\dots & 2 & $<1.3$ & $<0.3$ & \dots \\
J0905+3747 & NE& 1 & 12.32 & $2.31\pm0.38$ & $14.2\pm3.2$\\
 & \dots & 2 & 0.12* & $0.02\pm0.01$ &\dots \\
J1036+0221 & & & 11.23 & $2.09\pm0.32$ & $21.0\pm8.7$\\
J1045+3519 & W & 1 & 0.15 & $0.42\pm0.40$ & $4.7\pm1.1$\\
 & E & 2 & 0.04 & $0.34\pm0.33$ & $2.7\pm1.1$\\
J1126+1913 & & & 11.43 & $2.01\pm0.18$ & $3.9\pm3.8$\\
J1147+0945 & S & 1 & 6.20 & $1.17\pm0.30$ & $2120\pm250$\\
 & \dots & 2 & $<1.4$ & $<1.1^{\dagger}$ & \dots \\
  & \dots & 3 & 0.73* & $0.99\pm0.88$ & \dots \\
J1159+5320 & SE & 1 & 3.05 & $0.74\pm0.38$ & $4.1\pm1.7$\\
 & \dots & 2 & 4.55* & $0.99\pm0.31^{\dagger}$ &\dots \\
J1221+1137 & NE& 1 & 10.31 & $2.34\pm0.72$ & $8.5\pm2.7$\\
 & SW & 2 & 1.34* & $0.88\pm0.68^{\dagger}$ & $0.9\pm1.3$\\
J1301+2918 & NE & 1 & \dots & \dots & $0.41\pm0.50$\\
 & SW & 2 & \dots & \dots & $7.1\pm1.8$\\
J1306+0735 & NE & 1 & 1.31 & $0.37\pm0.18$ & $4.6\pm1.7$\\
 & SW & 2 & 16.48* & $2.83\pm0.18^{\dagger}$ & $15.4\pm3.6$\\
 & SE & 3 & 2.34* & $0.54\pm0.18^{\dagger}$ & $3.9\pm1.8$\\
J1356+1822 & E & 1 & $<32$* & $<6^{\dagger}$ & $65\pm12$\\
 & W & 2 & $<3.7$ & $<1.1$ & $21\pm5$\\
J2356-1016 & NW & 1 & 73.53 & $12.42\pm0.62$ & $548\pm79$\\
 & \dots & 2 & 3.82 & $0.64\pm0.04$ & \dots \\
\noalign{\smallskip}
\hline
\end{tabular}
\end{center}
\tablecomments{Column 1: Truncated merger designation. Columns 2$\--$3: X-ray source cardinal coordinates and host nucleus designation, respectively. Columns $4\--5$: Calculated SFRs and $\rm{L}^{\rm{SF}}_{2\--10\ \rm{keV}}$ using the relations from \citet{kennicutt1994} and \citet{lehmer2010}, respectively. Column 6: Observed-frame, absorbed X-ray luminosities derived from background subtracted source counts. The error bounds for $\rm{L}^{\rm{Abs.}}_{2\--10\ \rm{keV}}$ incorporate error due to the source and background counts as well as an additional 10\% error to account for systematic effects of the CCD detectors. (*): Distance assumed to be the same as the companion galaxy/source. ($\dagger$): The mass of the companion nucleus was used to compute the X-ray contribution from HMXBs.}
\label{table:sfrlums}
\end{table*}

\begin{table*}
\caption{Summary of AGN Diagnostics for Each Source}
\begin{center}
\scalebox{0.8}{
\hspace{-2cm}
\begin{tabular}{ccccccccc}
\hline
\hline
\noalign{\smallskip}
Name & X-ray Source & X-ray Detection &  & Coronal & BPT & MIR & Fe K$\alpha$ & Summary \vspace{-4mm}\\
\noalign{\smallskip}
 & & & log($\rm{L}^{\rm{Abs.}}_{2\--10\ \rm{keV}}$)$-$log($\rm{L}^{\rm{SF}}_{2\--10\ \rm{keV}}$) & & & & & \vspace{-4mm}\\
\noalign{\smallskip}
(SDSS) & & Significance &  & Lines & Class & AGN &  Line & Classification\\
\noalign{\smallskip}
\hline
\noalign{\smallskip}
J0122+0100 & & &  &  &    & Y &  & Dual AGN Candidate\\
Galaxy 1 & NW & 7.9$\sigma$ & $0.77\pm0.14$  & N & SF &  & \dots & \\
Galaxy 2 & SE & 7.3$\sigma$ & $1.11\pm0.14$  & Y & SF &  & \dots &\\ 
J0841+0101 & &  &  &  &   & N & &  Dual AGN Candidate\\
Galaxy 1 & E & 13.3$\sigma$ & $1.82\pm0.10$ & Y & AGN &  & Y & \\
Galaxy 2 & W & 1.1$\sigma$ & $0.76\pm0.49$ & N & \dots &  & \dots & \\
J0849+1114 & & &  &  &   & Y & &  Dual AGN / Triple Candidate\\
Galaxy 1 & SE & 10.2$\sigma$ & $1.36\pm0.09$ & Y & ... &  & Y & \\
Galaxy 2 & SW & 2.2$\sigma$ & $1.64\pm0.49$ & N & AGN &  & \dots &\\ 
Galaxy 3 & N & 1.4$\sigma$ & $0.80\pm0.42$ & Y & AGN &  & \dots & \\
J0859+1310 & & &  &  &   & Y &  &  Single AGN\\
Galaxy 1 & NE & 20.7$\sigma$ & $1.98\pm0.42$ &  N & AGN &  & Y & \\
Galaxy 2 & \dots & \dots & \dots  &  N & Comp. &  & \dots & \\
J0905+3747 & & &  &   &  &  Y &  & Single AGN\\
Galaxy 1 & NE & 8.0$\sigma$ & $0.79\pm0.12$  & Y & Comp. &  & \dots &\\ 
Galaxy 2 & \dots & \dots & \dots  & N & SF &  & \dots & \\
J1036+0221 & & 4.3$\sigma$ & $1.00\pm0.19$ & Y & Comp. & Y & \dots & Single AGN\\
J1045+3519 &  & &  &  &   & N &  & Dual AGN Candidate\\
Galaxy 1 & W & 4.3$\sigma$ & $1.05\pm0.43$ & N & Comp. &  & \dots &\\
Galaxy 2 & E & 2.5$\sigma$ & $0.90\pm0.46$ & N & SF &  & \dots & \\
J1126+1913 & & 2.0$\sigma$ & $0.29\pm0.43$ & Y & Comp. & Y & \dots & Single AGN\\
J1147+0945 & & &   &  &   & Y &  & Single AGN\\
Galaxy 1 & S & 56.0$\sigma$ & $3.26\pm0.12$ & N & AGN &  & Y & \\
Galaxy 2 & \dots & \dots  & \dots & N & Comp. &  & \dots & \\
Galaxy 3 & \dots & \dots  & \dots & N & \dots &  & \dots & \\
J1159+5320 & & &   &  &    & Y &  & Single AGN\\
Galaxy 1 & SE & 3.2$\sigma$ & $0.75\pm0.28$  & N & AGN &  & \dots &\\
Galaxy 2 & \dots & \dots & \dots &  N & \dots &  & \dots & \\
J1221+1137 & & &  &  &   & N &  & Dual AGN Candidate\\
Galaxy 1 & NE & 4.8$\sigma$ & $0.56\pm0.19$ & N & SF &  & \dots &\\ 
Galaxy 2 & SW & 1.1$\sigma$ & $0.01\pm0.74$ & Y & \dots &  & \dots & \\
J1301+2918 & & &   &  &   & Y &  & Dual AGN Candidate\\
Galaxy 1 & NE & 0.8$\sigma$ & \dots  & N & \dots &  & \dots & \\
Galaxy 2 & SW & 6.8$\sigma$ & \dots  & N & AGN &  & Y & \\
J1306+0735 & & &  &   &  & N &  & Dual / Triple AGN Candidate\\
Galaxy 1 & NE & 3.6$\sigma$ & $1.09\pm0.27$ & N & SF &  & \dots &\\ 
Galaxy 2 & SW & 7.4$\sigma$  &  $0.74\pm0.11$  & N & \dots &  & \dots & \\
Galaxy 3 & SE & 2.7$\sigma$ & $0.86\pm0.25$  & \dots & \dots &  & \dots &\\ 
J1356+1822 & & &  &   &   & Y &  & Dual AGN\\
Galaxy 1 & E & 12.2$\sigma$ & $>1.06$ & N & \dots &  & \dots &\\ 
Galaxy 2 & W & 6.8$\sigma$ & $>1.29$ & N & AGN &  & \dots &\\ 
J2356-1016 & & &  &   &  & Y &  & Single AGN\\
Galaxy 1 & NW & 22.8$\sigma$ & $1.64\pm0.07$ & Y & SF &  & \dots &\\ 
Galaxy 2 & \dots & \dots & \dots & N & SF &  & \dots & \\
\noalign{\smallskip}
\hline
\end{tabular}
}
\end{center}
\vspace{-4mm}
\tablecomments{Column 1: Merger and individual galaxy designations. Column 2: Statistical significance of the X-ray source detections. Column 3: Difference between the logarithm of the absorbed, observed-frame $2\--10$ keV X-ray luminosities and  predicted (absorbed) X-ray luminosities resulting from stellar processes derived using the relations in \citet{lehmer2010} and \citet{kennicutt1994}. These values represent lower limits; in reality, the unabsorbed values are likely higher. Column 4: Denotes the presence of coronal lines (or lack thereof) in each nucleus. Column 5: BPT classifications for each nucleus. Column 6: MIR classification for the combined nuclei determined via the strict three-band \WISE{} color cut given in \citet{jarrett2011}. Column 7: Indicates the presence of an Iron K$\alpha$ line in the X-ray spectrum of a source. Column 8: Our summary classification based upon the results of our analysis for each system.}
\label{table:summary}
\end{table*}

\begin{table}
\caption{Column Densities and Unabsorbed X-ray Luminosities via the $\rm{L}_{2\--10\ \rm{keV}}$ vs. $\rm{L}_{12\mu \rm{m}}$ Relationship}
\begin{center}
\begin{tabular}{ccccc}
\hline
\hline
\noalign{\smallskip}
Source & log$(\rm{L}^{\rm{Abs.}}_{2\--10\ \rm{keV}})$ & $\nh$ & log$(\rm{L}^{\rm{Unabs.}}_{2\--10\ \rm{keV}})$ \\
\noalign{\smallskip}
 & erg s$^{-1}$ & $10^{23}$ cm$^{-2}$ & erg s$^{-1}$\\
\noalign{\smallskip}
\hline
\noalign{\smallskip}
J0122+0100$^\dagger$ & 41.1 & 30.0$^{+1.6}_{-1.2}$ & 43.4 \\
J0841+0101 & 42.3 & $4.1^{+3.6 }_{-2.0}$ & 43.3 \\
J0849+1114 & 41.8 & $5.8^{+4.8 }_{-2.8}$ & 43.0 \\
J0859+1310 & 41.6 & $2.3^{+2.2 }_{-1.2}$ & 42.3 \\
J0905+3747 & 41.2 & $9.1^{+6.8}_{-4.2}$ & 42.6\\
J1036+0221$^\dagger$ & 41.4 & 9.0$^{+0.07}_{-0.04}$ & 42.8\\
J1045+3519$^\dagger$ & 41.2 & 31.0$^{+1.7}_{-1.2}$ & 43.6 \\
J1126+1913$^\dagger$ & 40.7 & 50.0$^{+2.2}_{-1.7}$ & 43.5 \\
J1147+0945 & 43.3 & \dots & \dots \\
J1159+5320 & 40.6 & $12.0^{+8.4 }_{-5.3}$ & 42.3 \\
J1221+1137$^\dagger$ & 41.5 & 27.0$^{+1.5}_{-1.1}$ & 43.7\\
J1301+2918 & 40.9 & $10.9^{+7.8 }_{-4.9}$ & 42.5  \\
J1306+0735$^\dagger$ & 41.4 & 24.0$^{+1.4}_{-1.0}$ & 43.5\\
J1356+1822 & 41.9 & $19.2^{+11.9 }_{-8.0}$ & 43.9 \\
J2356-1016 & 42.7 & $1.8^{+1.9 }_{-1.0}$ & 43.3\\
\noalign{\smallskip}
\hline
\end{tabular}
\end{center}
\tablecomments{Column 2: Total absorbed $2\--10$ keV \chandra{} X-ray luminosities, assuming a simple Galactic absorbed power-law with $\Gamma$ of 1.8. Column 3: Column densities inferred via the relationship between the X-ray $\rm{L}_{2\--10\ \rm{keV}}$ and infrared $\rm{L}_{12\mu \rm{m}}$ luminosities for each merger in this study (see Figure~\ref{fig:lxl12plot}). Column 4: Unabsorbed $2\--10$ keV X-ray luminosities obtained by taking into account the absorption in column 3. $\dagger$: Values pulled from \citet{satyapal2017}.}
\label{table:lxl12}
\end{table}

\section{Discussion}

The high incidence of AGNs in our sample demonstrates that mid-infrared color selection is a successful pre-selection strategy for finding AGNs in mergers, and is also a promising pre-selection strategy in identifying dual AGNs. There are 22 nuclei in our sample with BPT optical classifications; 14 of the 22 nuclei are optically classified as star-forming or composite galaxies. However, we can \textit{confirm} 5 out of these 14 harbor bona fide AGNs (see Table~\ref{table:summary}). Our results suggest that optical studies miss a non-negligible fraction of single and dual AGNs in advanced mergers due to large scale obscuration not associated with a torus. Note, however, the SDSS fibers are not always centered on the optical SDSS r-band nucleus, and this will impact the optical line ratios and could explain why some of the galaxy nuclei are not optically classified as AGN. Higher spatial resolution or better aligned spectroscopy centered on the nucleus would be required to obtain robust optical classifications of the nuclei.

\begin{figure*}[t]
    \centering
    \makebox[\textwidth]{\includegraphics[width=12cm]{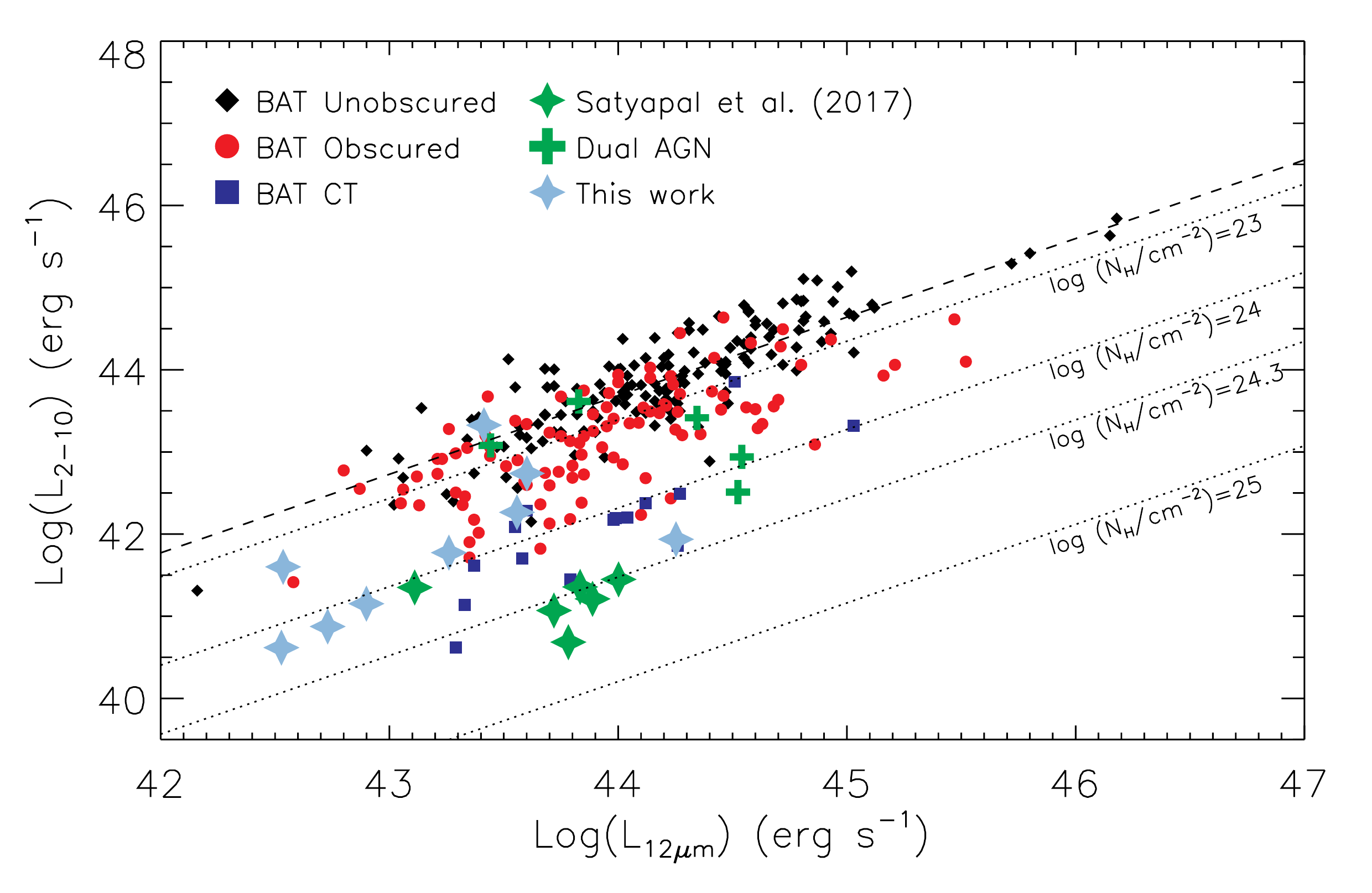}}
    \caption{The hard X-ray luminosity versus the mid-infrared luminosity for the advanced mergers from our program observed thus far, along with the sample of hard X-ray selected AGNs from the {\it Swift/BAT} survey for which spectral analysis enables a direct determination of the intrinsic absorption \citep{ricci2015,ricci2017}. The dual AGN candidates from our program are mostly located in the region of the plot occupied by the most heavily absorbed AGNs.}
    \label{fig:lxl12plot}
\end{figure*}

The large equivalent widths of the iron K$\alpha$ lines and the spectral analysis of some of our targets are consistent with large column densities of obscuration toward the X-ray sources. The mid-infrared luminosity, which is reprocessed AGN emission, and the AGN {\it unabsorbed} X-ray emission are known to follow a tight correlation over several orders of magnitude \citep{lutz2004,gandhi2009,mateos2015}. In  Figure~\ref{fig:lxl12plot}, we plot the 12~\micron\ luminosity, calculated by interpolating the W2 and W3 band luminosities, versus the hard X-ray luminosity, uncorrected for intrinsic absorption, for the advanced mergers in our sample of 15 mergers and confirmed dual AGNs in the literature \citep{bianchi2008,comerford2011,comerford2015,frey2012,fu2011,fu2015,huang2014,komossa2003,koss2011,liu2013,mazzarella2012,mcgurk2011,mullersan2015,owen1985,rodriguez2006,teng2012,woo2014,bothun1989,moran1992,secrest2017,ellison2017}, together with the sample of hard X-ray selected AGNs  from the 70 month {\it Swift/BAT} survey \citep{ricci2015,ricci2017} for which a detailed broadband spectral analysis enables a direct determination of the intrinsic absorption, showing unabsorbed ($\nh<10^{22}$~cm$^{-2}$), absorbed ($\nh=10^{22-24}$~cm$^{-2}$), and Compton-thick ($\nh>10^{24}$~cm$^{-2}$) AGNs. From Figure~\ref{fig:lxl12plot}, the X-ray luminosities (uncorrected for intrinsic absorption) are low relative to the mid-infrared luminosities for the majority of our sample, consistent with these sources being heavily absorbed or Compton-thick ($\nh>10^{24}$~cm$^{-2}$) AGNs.

The high level of obscuration suggested by our results is consistent with simulations. \cite{blecha2018} show that  the gas column densities toward the SMBHs are predicted to be high for pair separations $<$ 10~kpc, peaking just prior to coalescence, which would significantly lower the absorbed X-ray luminosity relative to the mid-infrared luminosity and generate AGN dominated mid-infrared colors, consistent with our results. The results presented in this work are consistent with other recent observations \citep{ricci2015,ricci2017,donley2018,goulding2018} suggesting that AGNs in advanced mergers are likely obscured by significant gas and dust. In our mid-infrared study of a large sample of galaxy pairs, we found  that the fraction of obscured AGNs, selected using mid-infrared color criteria, increases with merger stage relative to a rigorously matched control sample, with the most energetically dominant optically obscured AGNs becoming more prevalent in the most advanced mergers \citep{satyapal2014, ellison2015}, where star formation rates are highest \citep{ellison2016,weston2017}. 

A growing number of recent observational studies are also consistent with this scenario. For example, there is evidence from X-ray spectral analysis that there is an increase in the fraction of mergers in AGNs that are heavily absorbed or Compton-thick at moderate and high redshifts \citep{kocevski2015,lanzuisi2015, delmoro2016,koss2016}. In a recent broad-band X-ray spectral study of 52 local infrared luminous and ultraluminous galaxies, \citet{ricci2017} find that the fraction of Compton-thick AGNs in late-stage mergers is higher than in local hard X-ray selected AGNs, and the absorbing column densities are maximum when the projected separation between the two nuclei are $\approx 0.4-10.8$~kpc. Recently, \citet{donley2018} found that the majority (75\%) of the IR selected AGNs in the COSMOS/CANDELS field show disturbed morphologies compared to only 31\% of AGNs selected only via X-ray observations, strongly suggesting that major mergers play a dominant role in fueling luminous obscured AGNs. Finally, \citet{lansbury2017} find evidence of a high merger fraction in the extreme Compton-thick sources identified in the {\it NuSTAR} serendipitous survey.

\section{Conclusions}

We have presented {\it \chandra{}/ACIS} observations of thirteen advanced mergers with nuclear separations $<$ 10~kpc preselected using {\it WISE} colors, following \citet{stern2012} with a color cut of W1-W2 $>$ 0.5. Together with observations presented in Paper I, these observations represent the 15 brightest mid-infrared dual AGN candidates observed with high spatial X-ray observations.

Our main results can be summarized as follows: \\

\begin{enumerate}
\item{We detect at least one nuclear X-ray source in all 15 mergers, of which 8 exhibit at least two sources suggestive of dual AGNs. We report the detection of triple X-ray sources in two out of these 8 mergers. Note that the lack of a second detection in the seven mergers with a single X-ray source does not exclude the possibility of a fainter or Compton-thick secondary source below our detection threshold.}

\item{For 9 out of 15 of the mergers, we detect over 100 counts in the full band with either \chandra{} or \xmm{}, sufficient for direct spectral fitting. All spectra are consistent with absorbed power-law models with intrinsic absorption in the $10^{22}$ -- $10^{24}$ cm$^{-2}$ range resulting in unabsorbed X-ray luminosities in the $\rm{L}_{\mathrm{2-10~keV}} \approx 10^{42}$ --  $10^{43}\mathrm{~erg~s}^{-1}$ range. We find tentative evidence for an Fe K$\alpha$ line with equivalent width in excess of 150 eV in four targets, also suggestive of highly absorbed AGNs.}

\item{The absorbed X-ray luminosity in all but one target is significantly above that expected from star formation in the host galaxy. In a companion paper, we report the detection of near-infrared coronal line emission in 9 nuclei in our sample, providing robust evidence for an AGN in each (Constantin et al., in prep.). Based on a stringent requirement that the absorbed or unabsorbed X-ray luminosity is $\rm{L}_\mathrm{2-10~keV} > 10^{42}$~erg~s$^{-1}$, and/or the detection of a coronal line, and/or the detection of a significant Fe K$\alpha$ emission line, and/or optical spectroscopic classifications, we confirm the presence of a total of 15 AGN in our full sample, 4 of which were previously reported in the literature (Mrk 463W, Mrk 436E, J0841+0101E, NGC 4922NE), and another 4 were confirmed in Paper I. 5 of these 15 nuclei with bona fide AGN \textit{do not} exhibit AGN optical spectroscopic line ratios. Out of the 8 mergers with dual X-ray sources and/or coronal emission coincident with the galactic nuclei, and/or AGN optical classifications, we provide \textit{confirmation} for two dual AGNs using our strict definition of an AGN. The confirmed dual AGNs are J135602+1822 (Mrk 463) with separation of 4.0 kpc (a previously known dual), and J0849+1114 with separation of 5.8 kpc.}

\item{Most of the advanced mergers in our sample have absorbed $2\--10$ keV X-ray luminosities that are low relative to their mid-infrared luminosities when compared with local hard X-ray selected unabsorbed AGNs, comparable to the most obscured sources in the  {\it Swift/BAT} survey and several of the other confirmed well-known duals in the literature. This suggests heavy obscuration corresponding in some cases to intrinsic absorption $\nh$ of a few times $10^{24}$~cm$^{-2}$. }

\item{The detection of buried AGNs in advanced mergers and the demonstrated success rate of mid-infrared pre-selection in finding duals is consistent with recent observations that suggest that the most active phase in black hole growth occurs in an obscured phase. These findings are also consistent with recent hydrodynamical merger simulations which show that obscured luminous AGNs should be a natural occurrence in advanced mergers, where dual AGNs are likely to be found, and that mid-infrared color-selection is one of the best ways to select them.}

\end{enumerate}

Our results further demonstrate that mid-infrared color-selection, and in particular a color cut of W1-W2 $>$ 0.5, is a promising preselection strategy for finding single, dual, and tentatively triple AGN candidates in advanced mergers and is a complementary approach to optical and blind X-ray searches. While radio surveys do not suffer from obscuration bias, the radio emission in advanced mergers can be dominated by and indistinguishable from compact nuclear starbursts \citep{condon1991,delmoro2013}. These results imply that the merger stage characterized by the most rapid black hole growth, a key stage in the evolution of galaxies, has been missed by past studies.

\acknowledgements

This publication makes use of data products from the Wide-field
Infrared Survey Explorer, which is a joint project of the University
of California, Los Angeles, and the Jet Propulsion
Laboratory/California Institute of Technology, funded by the National
Aeronautics and Space Administration.
Funding for SDSS-III has been provided by the Alfred P. Sloan Foundation, the Participating Institutions, the National Science Foundation, and the U.S. Department of Energy Office of Science. The SDSS-III web site is \url{http://www.sdss3.org/}.

SDSS-III is managed by the Astrophysical Research Consortium for the Participating Institutions of the SDSS-III Collaboration including the University of Arizona, the Brazilian Participation Group, Brookhaven National Laboratory, Carnegie Mellon University, University of Florida, the French Participation Group, the German Participation Group, Harvard University, the Instituto de Astrofisica de Canarias, the Michigan State/Notre Dame/JINA Participation Group, Johns Hopkins University, Lawrence Berkeley National Laboratory, Max Planck Institute for Astrophysics, Max Planck Institute for Extraterrestrial Physics, New Mexico State University, New York University, Ohio State University, Pennsylvania State University, University of Portsmouth, Princeton University, the Spanish Participation Group, University of Tokyo, University of Utah, Vanderbilt University, University of Virginia, University of Washington, and Yale University.
This research has made use of the NASA/IPAC Extragalactic Database (NED) which is operated by the Jet Propulsion Laboratory, California Institute of Technology, under contract with the National Aeronautics and Space Administration. We also gratefully acknowledge the use of the software TOPCAT \citep{Taylor2005} and Astropy \citep{astropy2013}.

This publication makes use of data products from the Two Micron All Sky Survey, which is a joint project of the University of Massachusetts and the Infrared Processing and Analysis Center/California Institute of Technology, funded by the National Aeronautics and Space Administration and the National Science Foundation.

The LBT is an international collaboration among institutions in the United States, Italy and Germany. LBT Corporation partners are: The University of Arizona on behalf of the Arizona Board of Regents; Istituto Nazionale di Astrofisica, Italy; LBT Beteiligungsgesellschaft, Germany, representing the Max-Planck Society, The Leibniz Institute for Astrophysics Potsdam, and Heidelberg University; The Ohio State University, and The Research Corporation, on behalf of The University of Notre Dame, University of Minnesota and University of Virginia.

The authors thank the anonymous referee for the constructive comments and criticism provided on this manuscript.

N.J.S. held an NRC Research Associateship award at the Naval Research Laboratory for much of this work. CONICYT+PAI Convocatoria Nacional subvencion a instalacion en la academia convocatoria año 2017 PAI77170080 (C.R.). L.B. acknowledges support from the National Science Foundation (\#1715413).

S.S. and R.W.P. gratefully acknowledge support from the Chandra Guest Investigator Program under NASA grants GO6-17096X, GO7-18099X, and NNX17AD59G.

R.W.P. received a Grant-in-Aid of research (GRIAR) from Sigma Xi, the Scientific Research Society, in support of this work. R.W.P. appreciates the aid of Nathaniel Bechhofer in developing a Python code to compute small $P_{\rm{B}}$ values. R.W.P. appreciates and thanks his advisor and coauthors for their guidance and training. R.W.P. thanks Carol and Tim Pfeifle and the rest of his family, his friends, and Erin Fierro, for their love and support during this endeavor. R.W.P. dedicates this work to Colby $\--$ a loyal and loving companion who is sorrowfully missed.

\appendix

\section{Notes on Individual Systems}

The following sections detail the nature of each merger summarized in Tables 4$\--$9.

\subsection{J0122+0100: Dual AGN Candidate}
J0122+0100 was one of four mergers followed up and reexamined during Chandra Cycle 18. We report the presence of two X-ray sources, originally reported in Paper I, within the merger. The source apertures used in this study vary slightly from those obtained from the original pilot study in Paper I due to the fact that the Cycle 18 data have inherently higher signal to noise and therefore allowed for more accurate source aperture placements. The northwestern source (Galaxy 1) is detected with a significance of 7.9$\sigma$ and a hardness ratio of -0.31 while the southeastern source (Galaxy 2) is detected with a significance of 7.3$\sigma$ and hardness ratio of -0.22. SDSS classifies both galaxies as starburst galaxies, and the BPT line ratios show that both galaxies would be optically classified as starburst galaxies (see Figure~\ref{fig:sample_images}). Both sources possess absorbed X-ray luminosities above that expected for the absorbed stellar X-ray luminosity contributions. Coronal emission was detected in Galaxy 1 (see Paper I), allowing us to robustly confirm the presence of an AGN in that nucleus.

\subsubsection{J0122+0100 XMM Spectral Analysis}
\begin{figure}[h!]
    \centering
    \begin{minipage}[t]{0.45\textwidth}
        \centering
        \includegraphics[width=1.1\textwidth]{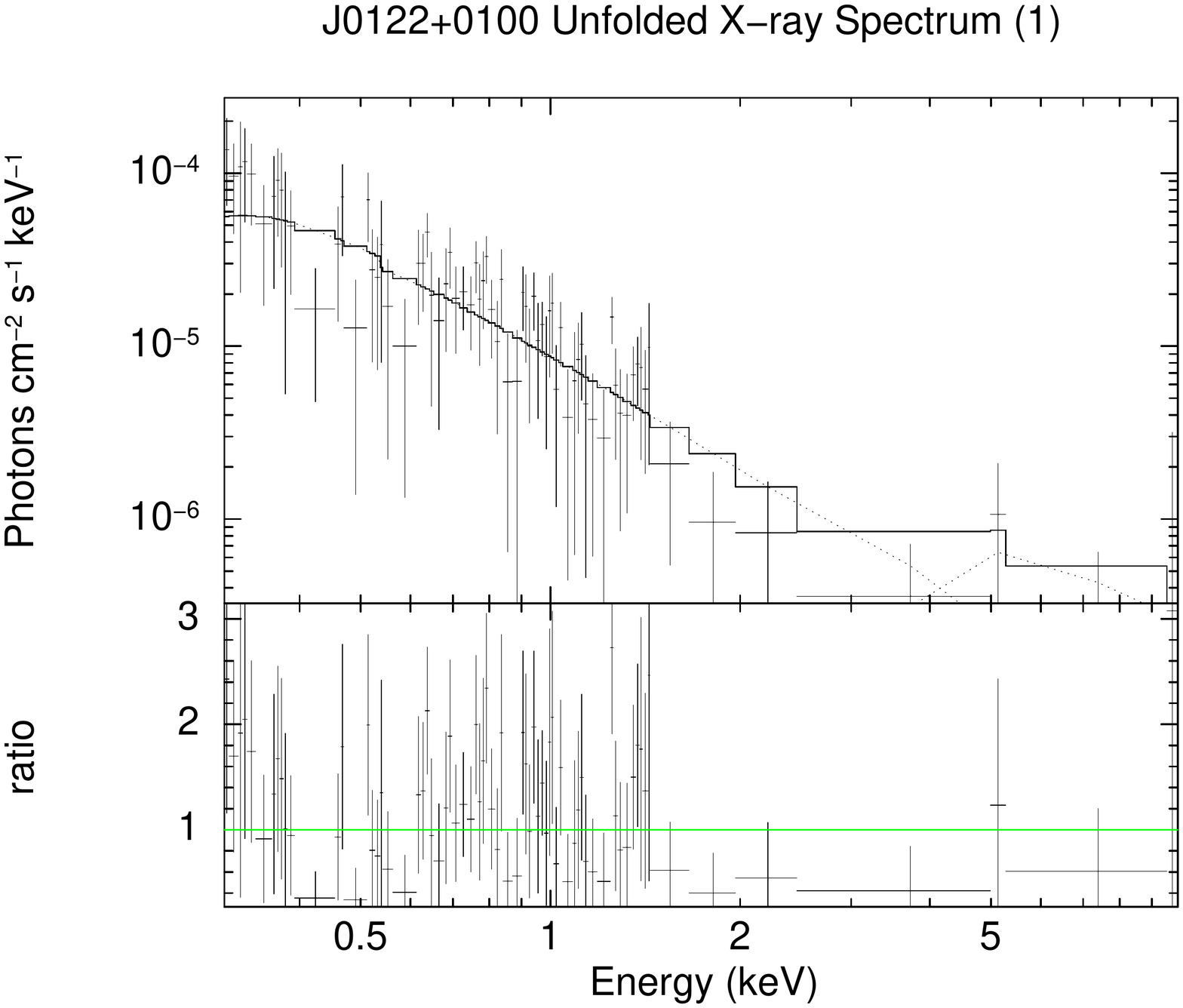} 
        \caption{The unfolded XMM X-ray spectrum for J0122+0100, modeled in \textsc{xspec} using the phenomenological approach for the full energy band $0.3\--10$ keV. The data are best fit with an absorbed power-law and a scattered power-law component.}
        \label{fig:J0122_phen}
    \end{minipage}\hspace{4mm}
    \begin{minipage}[t]{0.45\textwidth}
        \centering
        \includegraphics[width=1.1\textwidth]{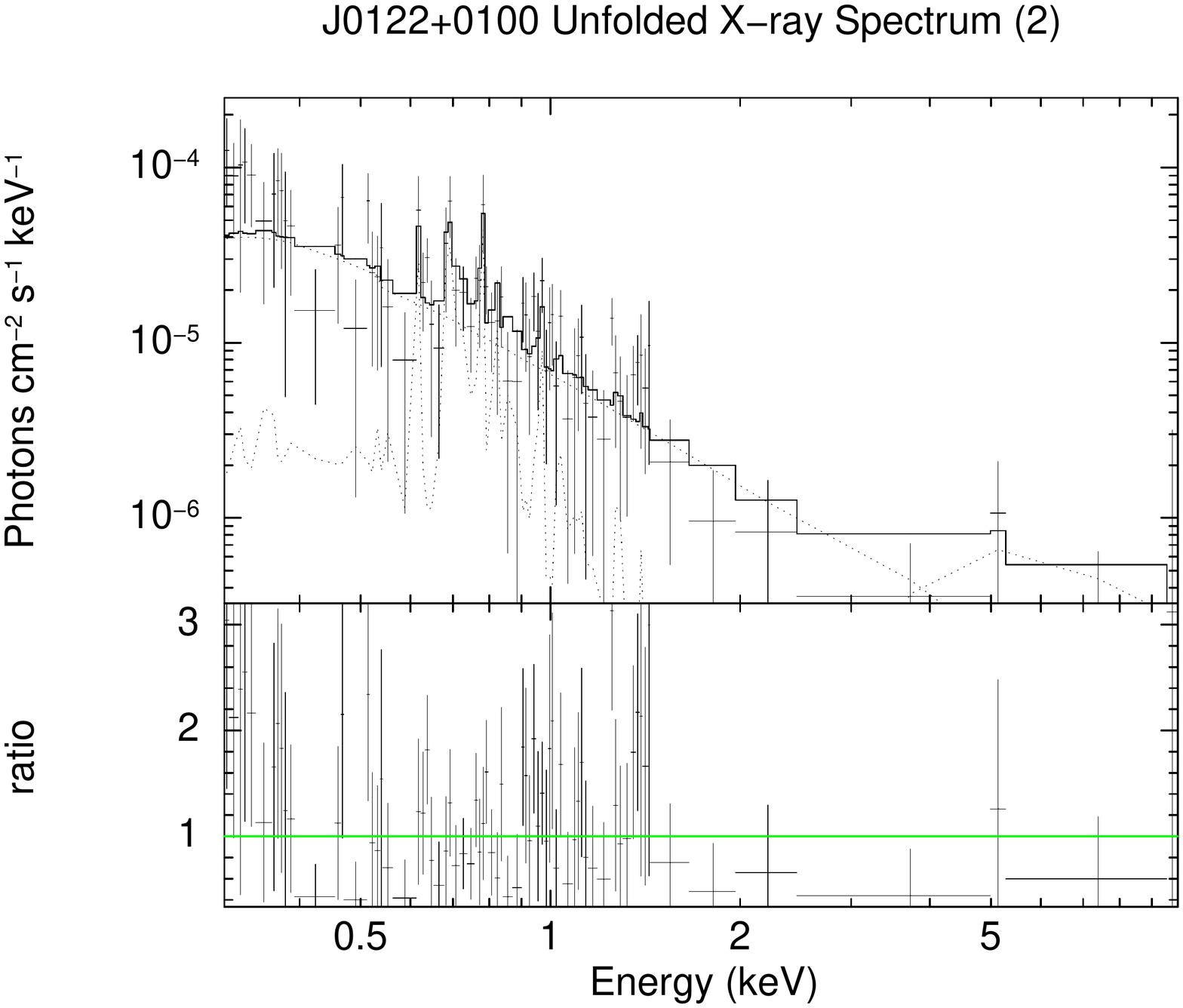} 
        \caption{The unfolded XMM X-ray spectrum for J0122+0100, modeled in \textsc{xspec} with the BNTorus approach for the full energy band $0.3\--10$ keV. The data are best fit with the base BNTorus model plus a scattered power-law and an \textsc{apec} component.}
        \label{fig:J0122_bnt}
    \end{minipage}
\end{figure}

Examining J0122+0100 with the phenomenological model, the data are best fit with an absorbed power-law with a scattered power-law component (the latter of which introduced a statistically significant change in the $\chi^2$ statistic of $4.84$ beyond the base model). This result is shown in Figure~\ref{fig:J0122_phen}. An attempt was made to incorporate \textsc{apec} into the model, however the addition of this component resulted in nonphysical values for $\Gamma$, and we therefore discarded \textsc{apec} when fitting with this approach. The model reveals high obscuration in this system, $\nh = 33^{+46}_{-16}\times10^{22}$ cm$^{-2}$, and a photon index of $\Gamma = 2.3^{+0.2}_{-0.2}$. Correcting for absorption, we find an unabsorbed luminosity of $\rm{L}_{2\--10\ \rm{keV}} = 1.3^{+1.3}_{-0.5}\times10^{42}$ erg s$^{-1}$, consistent with the presence of at least a single AGN in this system. The $\nh$ determined with this model is lower than that predicted by the relationship between the total absorbed X-ray luminosity and total 12$\mu$m luminosity for this system, which suggests a column density of $30.0^{+1.6}_{-1.2}\times10^{23}$ cm$^{-2}$. 

For the BNTorus approach we find the data are best fit with the base BNTorus model along with a scattered power-law and an \textsc{apec} component. The introduction of the scattered power-law and \textsc{apec} components both resulted in a statistically significant change to the $\chi^2$ statistic, with the combination of the two improving the $\chi^2$ statistic by $199.96$ over the base model. The best fitting model reveals a $\Gamma$ of $2.2^{+0.3}_{-0.4}$ and high obscuration, with $\nh = 26^{+51}_{-15}\times10^{22}$ cm$^{-2}$. These parameter values agree with that found by the phenomenological approach. Additionally, we find a plasma temperature of $0.50^{+0.30}_{-0.17}$ for the plasma modeled by \textsc{apec}. Correcting for absorption, the BNTorus approach yields an unabsorbed luminosity of $\rm{L}_{2\--10\ \rm{keV}} = 9.7^{+9.3}_{-4.3}\times10^{41}$ erg s$^{-1}$, slightly lower than that found with the phenomenological approach above. This result is shown in Figure~\ref{fig:J0122_bnt}. 

\subsection{J0841+0101: Dual AGN Candidate}
The eastern X-ray source (Galaxy 1) is detected with a significance of 13.3$\sigma$ and with a hardness ratio of 0.04. We report the absorbed luminosity uncorrected for intrinsic absorption, using a basic power-law model with photon index of 1.8 through PIMMS, to be $1.8\pm0.32\times10^{42}$ erg s$^{-1}$ (see Table~\ref{table:sfrlums}), which is in the range of typical AGNs. The detection of a [SiVI] coronal line in the E nucleus provided further evidence for an AGN. We report the presence of faint X-ray emission coincident with Galaxy 2 which we designate as the western source. This source possesses a formal significance of only 1.1$\sigma$, but we concluded based upon the $P_{\rm{B}}$ metric ($P_{\rm{B}} = 0.0002 < 0.002$ in the full band) that this X-ray emission does not originate from spurious background activity. Galaxy 1 has an SDSS classification of Galaxy AGN which agrees with the BPT classification (see Figure~\ref{fig:sample_images}). There is no SDSS or BPT classification for the second galaxy. 

This system was first examined by \citet{green2011} and \citet{comerford2015}, who identified it as a possible dual AGN or offset AGN system but no significant obscuration was previously reported. Using the source apertures listed for this system in Table 4, we extracted counts from the 19.8 ks 2012 archival \chandra{} data (PI: Comerford). We found no statistically significant variation in the count rates between the two data sets for either source. 

 The relationship between the total absorbed X-ray luminosity and 12$\mu$m luminosity (Figure~\ref{fig:lxl12plot}) suggests an extragalactic column density of at least $4.1^{+3.6 }_{-2.0}\times10^{23}$ cm$^{-2}$ ( Table~\ref{table:lxl12}), which agrees within the uncertainties of that found via spectral analysis (discussed below) of the eastern source.

\subsubsection{J0841+0101E Spectral Analysis Results}
\begin{figure}[h!]
    \centering
    \begin{minipage}[t]{0.45\textwidth}
        \centering
        \includegraphics[width=1.1\textwidth]{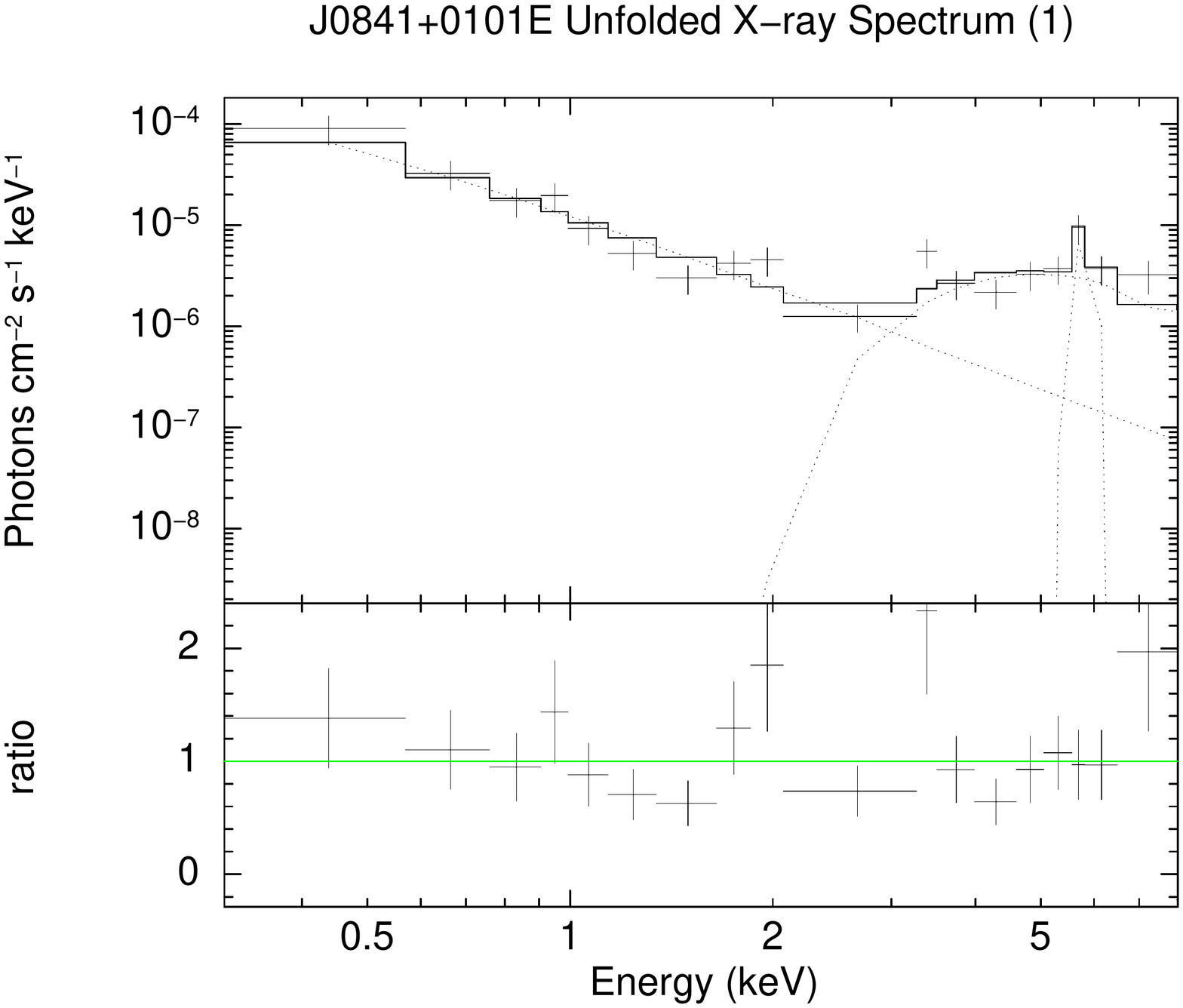} 
        \caption{The unfolded X-ray spectrum for J0841+0101E, modeled in \textsc{xspec} using the phenomenological approach for the full energy band $0.3\--8$ keV. The data are best fit with an absorbed power-law with Gaussian emission line component centered on 6.4 keV and a scattered power-law component.}
        \label{fig:J0841_phen}
    \end{minipage}\hspace{4mm}
    \begin{minipage}[t]{0.45\textwidth}
        \centering
        \includegraphics[width=1.1\textwidth]{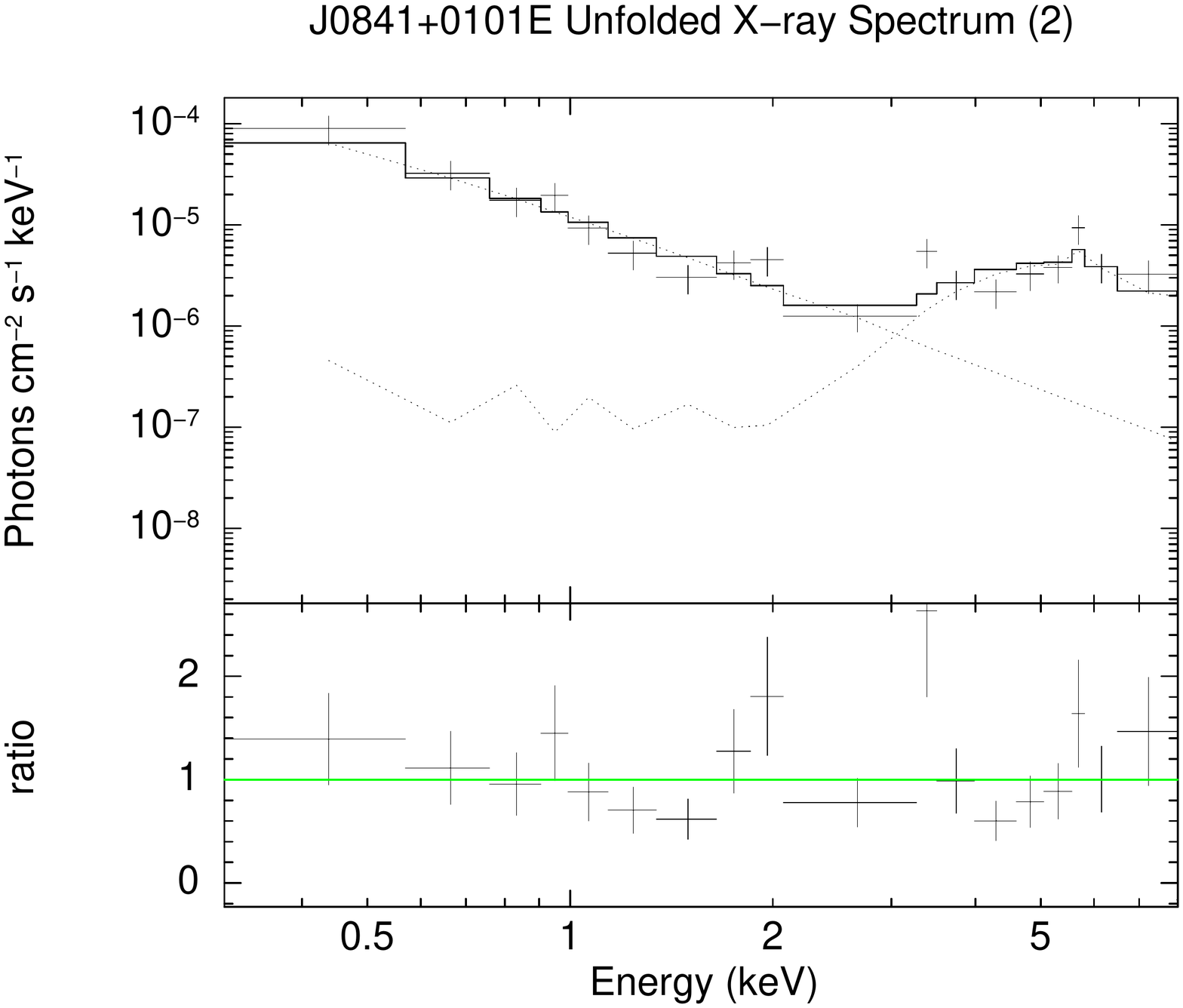} 
        \caption{The unfolded X-ray spectrum for J0841+0101E, modeled in \textsc{xspec} with the BNTorus approach for the full energy band $0.3\--8$ keV. The data are best fit with the base BNTorus model plus a scattered power-law component.}
        \label{fig:J0841_bnt}
    \end{minipage}
\end{figure}

 In analyzing this spectrum with the phenomenological model approach, introducing a scattered power-law improved the fit by $\Delta$C-Stat = $59.75 > 2.71$, and the addition of a Gaussian emission component also resulted in a statistically significant improvement to the absorbed and scattered power-laws, indicated by a $\Delta$C-Stat = $ 10.22 > 2.71$, and is suggestive of the presence of a previously unreported fluorescent iron K$\alpha$ emission line. The data for J0841+0101E are best fit using an absorbed power-law with a scattered power-law component and Gaussian emission line with line peak at 6.4 keV (see Figure~\ref{fig:J0841_phen}). The model yields a photon index of $\Gamma=2.5^{+0.4}_{-0.4}$, an obscuring column of $\nh=29.8^{+11.0}_{-9.0}\times10^{22}$ cm$^{-2}$, and we find an equivalent width of $0.75^{+2.59}_{-0.47}$ keV (see Table~\ref{table:phenspec}). Following the discussion in \citet{brightman2011} regarding the relationship between the equivalent width of the iron K$\alpha$ line and the $\nh$, this equivalent width agrees with the previously unreported high level of $\nh$ indicated by the model. Further, the high equivalent width and $\nh$ agree with the level of $\nh$ inferred through the $\rm{L}^{\rm{Abs.}}_{2\--10\ \rm{keV}}$ vs. 12$\mu$m luminosity (see  Table~\ref{table:lxl12}). Correcting for absorption, this model indicates an unabsorbed X-ray luminosity of $L_{2\--10\ \rm{keV}} = 2.3^{+0.7}_{-1.5}\times10^{43}$ erg s$^{-1}$ which is consistent with the idea that the source is an AGN.

When fitting with the BNTorus approach (Figure~\ref{fig:J0841_bnt}), we found that introducing a scattered power-law (scattering fraction $1.4^{2.0}_{-0.8}\%$) to the base model resulted in a statistically significant improvement to the fit ($\Delta$C-Stat$ = 47.56 > 2.71$). All fits attempted with the \textsc{apec} component yielded less statistically significant results and thus were discarded. We therefore find for this method the data are best fit using the BNTorus model with a scattered power-law component. The model indicates a photon index of $\Gamma=2.5^{+\dots}_{-0.4}$, for which an upper bound could not be computed, and an obscuring column of $\nh=35.2^{+15}_{-9.5}$ cm$^{-2}$. Correcting for intrinsic absorption, this model indicates an unabsorbed X-ray luminosity of $\rm{L}_{2\--10\ \rm{keV}} = 3.1^{+0.9}_{-2.6}\times10^{43}$ erg s$^{-1}$. These results agree with the level of obscuration predicted from the relationship between the infrared 12$\mu$m and $2\--10$ keV absorbed X-ray luminosity for this merger (see Figure~\ref{fig:lxl12plot}) which is $\nh \gtrsim 10^{23}$ cm$^{-2}$.

When fitting with the MYTorus model, we found a best fit using the MYTorus zeroth-order continuum paired with the MYTorus fluorescent emission line table, a scattered power-law, as well as an \textsc{apec} component. The results of the MYTorus model agree with that found by BNTorus and the phenomenological model with the exception of $\Gamma$, which is pegged at 1.4 by MYTorus. 

\subsection{J0849+1114: Dual AGN / Triple AGN Candidate}
We report the presence of three X-ray sources in the merger J0849+1114. We identify the southeastern source with the nucleus of Galaxy 1, the southwestern source with the nucleus of Galaxy 2, and the northern X-ray source appears coincident with a third galaxy (Galaxy 3) or a spiral arm of Galaxy 1 (see SDSS \textit{rgi} image in Figure~\ref{fig:sample_images}.) The northern source is well separated from the other sources, with an angular separation of 5.8$\arcsec$ from the southeastern source, while the southeastern and southwestern sources are textbf{in closer proximity}, with the southeastern source being significantly brighter than the southwestern counterpart. We find the angular separation of the extraction apertures for these two sources to be roughly 3.3$\arcsec$. Archival data were available for this system (PI: Liu, 2013, exposure time of 19.8ks), for which we extracted counts using the same apertures used for our data set. We discuss the archival data alongside our results (exposure time of 21.9 ks) in this section. This system was included in a sample of optically selected multi-AGN mergers in \citet{liu2011}.

The SE X-ray source is detected robustly with a significance of $10.2\sigma$ and with a hardness ratio of -0.06. We report in Table~\ref{table:sfrlums} the absorbed X-ray luminosity of $5.2\pm1.0\times10^{41}$ erg s$^{-1}$. A [SiVI] coronal emission line was detected in this galaxy nucleus and therefore we robustly confirm the presence of an AGN in the nucleus of Galaxy 1. As discussed below in subsection 7.3.1, our models indicate an AGN with an unabsorbed luminosity in excess of $10^{42}$ erg s$^{-1}$. The SW X-ray source is detected with a significance of $2.2\sigma$ and hardness ratio of -0.85. The N X-ray source is detected with a significance of only $1.4\sigma$. Despite the low-count nature of the north source, we note that the $P_{\rm{B}}$ value for this source ($P_{\rm{B}} = 0.000003 < 0.002$ in the full band) rules out the possibility that this emission arises from spurious background activity. All three X-ray sources exhibited absorbed luminosities roughly an order of magnitude higher than that expected from star formation, suggesting stellar processes alone cannot account for the absorbed X-ray emission. Further, a [SiVI] coronal line was detected in the N nucleus, robustly confirming the presence of an AGN. For all three sources, we see no statistically significant variability between the 2013 data and the 2016 data. 

With the coronal line detections, optical diagnostics, and the results of the X-ray modeling, we conclude that two AGN are robustly detected in this merger. As a result of the presence of an additional candidate AGN in this system, we designate this merger a \textit{triple AGN candidate}. BPT optical line ratios were available for the N and SW sources, both of which are classified as AGNs. However, due to the positioning of the SDSS fiber, as shown in Figure~\ref{fig:sample_images}, is it unclear if the AGN in the SE nucleus is contributing to the line fluxes observed near the SW nucleus. Therefore, we cannot unambigously claim the presence of an AGN in the SW nucleus. The true nature of this system will be the focus of a forthcoming publication, Pfeifle et al. (2019b, in preparation). An SDSS classification was available only for the N source, which classified the region as a galaxy starburst. No BPT or SDSS classification was available for the region occupied by the SE X-ray source. We infer from the relationship between the total absorbed X-ray luminosity and the 12$\mu$m luminosity of this merger a column density of $5.8^{+4.8}_{-2.8}\times10^{23}$ cm$^{-2}$ along the line of sight, which is in agreement with theoretical predictions for AGNs in advanced mergers such as this system. 

\subsubsection{J0849+1114SE Spectral Analysis Results}
\begin{figure}[h!]
    \centering
    \begin{minipage}[t]{0.45\textwidth}
        \centering
        \includegraphics[width=1.1\textwidth]{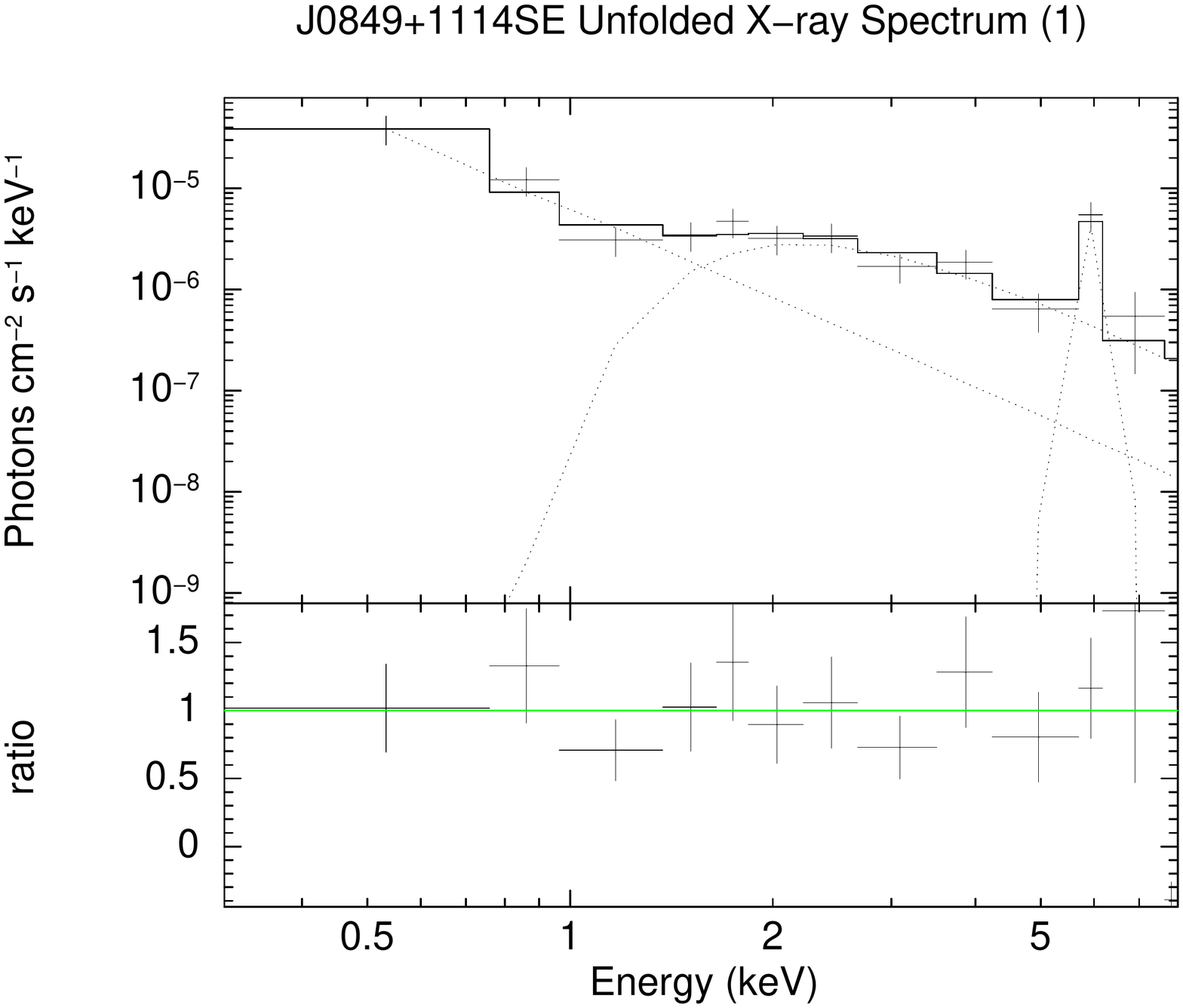} 
        \caption{The unfolded X-ray spectrum for J0849+1114SE, modeled in \textsc{xspec} with the phenomenological approach for the full $0.3\--8$ keV energy band. The data are best fit with an absorbed power-law with Gaussian emission line component centered on 6.4 keV and a scattered power-law component.}
        \label{fig:J0849_phen}
    \end{minipage}\hspace{4mm}
    \begin{minipage}[t]{0.45\textwidth}
        \centering 
        \includegraphics[width=1.1\textwidth]{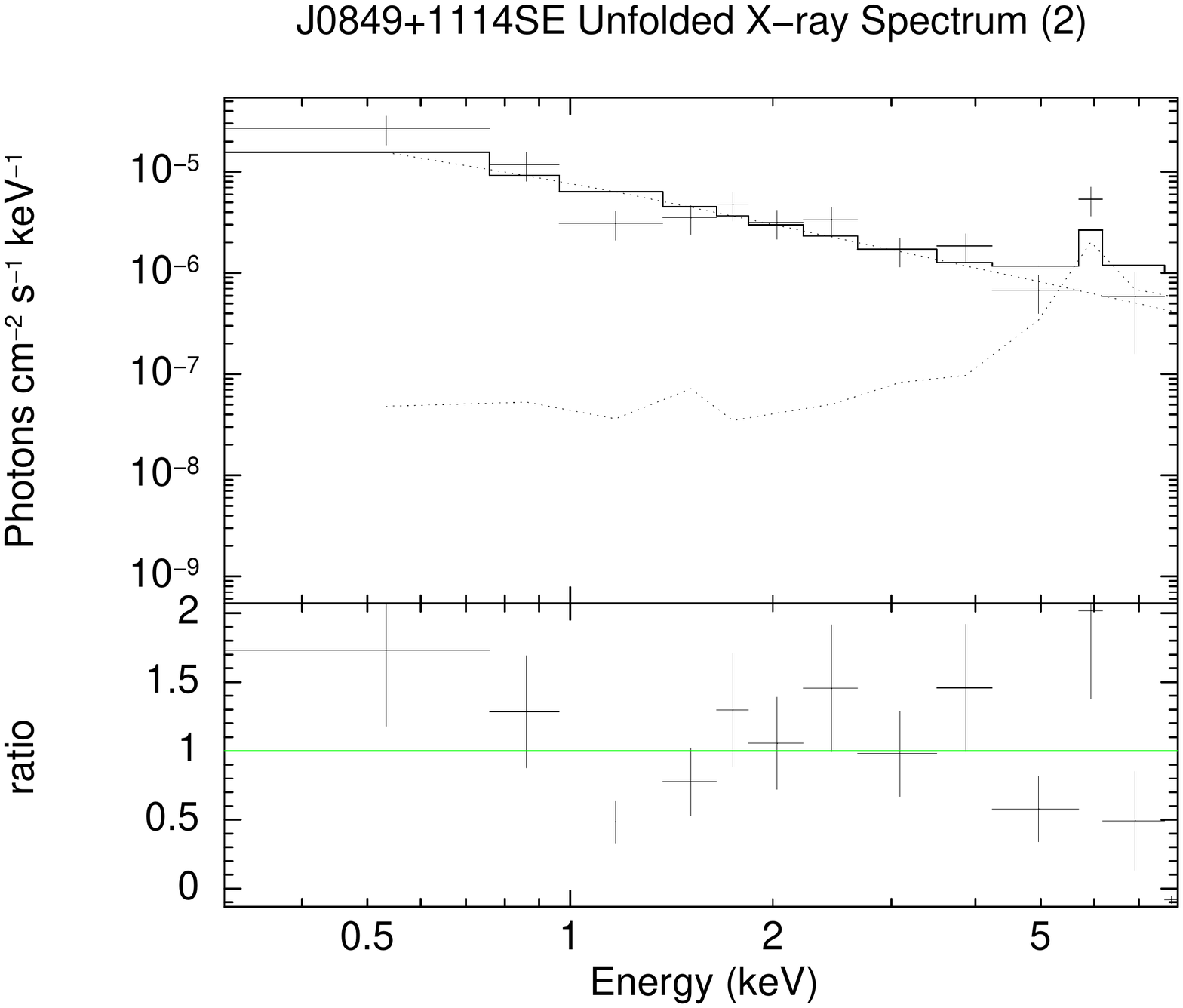} 
        \caption{The unfolded X-ray spectrum for J0849+1114SE, modeled in \textsc{xspec} with the BNTorus approach for the full $0.3\--8$ keV energy band. The data are best fit with the base BNTorus model plus a scattered power-law component.}
        \label{fig:J0849_bnt}
    \end{minipage}
\end{figure}

Using the phenomenological approach, we found that introducing a scattered power-law  to the base model resulted in a statistically significant improvement to the fit ($\Delta$C-Stat$ = 6.81 > 2.71$). The addition of a Gaussian emission component also resulted in a statistically significant improvement over the absorbed and scattered power-laws, indicated by a $\Delta$C-Stat = $12.3 > 2.71$. We therefore find for this method the data are best fit using an absorbed power-law with a scattered power-law component and Gaussian emission line with line peak at 6.4 keV, suggestive of an Fe K$\alpha$ emission line (see Figure~\ref{fig:J0849_phen}). The model indicates a photon index of $\Gamma=3.0^{+0.8}_{-0.9}$, an obscuring column of $\nh=4.0^{+1.8}_{-1.5}\times10^{22}$ cm$^{-2}$, and we find an equivalent width of $4.37^{+11.78}_{-2.82}$ keV (see Table~\ref{table:phenspec}). Following the discussion in \citet{brightman2011} regarding the relationship between the equivalent width of the iron K$\alpha$ line and the $\nh$, we report the equivalent width conflicts with the result for $\nh$ and actually suggests the column density could be higher, on the order of $10^{23}\--10^{24}$ cm$^{-2}$; this is in fact the case when using the BNTorus or MYTorus models. Further, we found that initial fits identified high levels of obscuration, on the order of $\sim 10^{23}$ cm$^{-2}$, and a lower $\Gamma$, but running the \textsc{xspec} error commands finds a best fit with low $\nh$ and high $\Gamma$ - it is likely that the S/N of the data is to blame for this apparent degeneracy.  Correcting for intrinsic absorption, this model indicates an unabsorbed X-ray luminosity of $\rm{L}_{2\--10\ \rm{keV}} = 1.4^{+0.5}_{-0.9}\times10^{42}$ erg s$^{-1}$. The equivalent width of the emission line component agrees with the level of obscuration predicted from the relationship between the infrared and absorbed X-ray $2\--10$ keV luminosity for this merger (see Figure~\ref{fig:lxl12plot}), which is $\nh \sim10^{23}$ cm$^{-2}$.
 
Using the BNTorus model approach, we found that introducing a scattered power-law (scattering fraction $4.7^{+51.4}_{-4.4}\%$) to the base model resulted in a statistically significant improvement to the fit ($\Delta$C-Stat$ = 36 > 2.71$). All fits attempted with the \textsc{apec} component yielded either nonphysical values for parameters or were less statistically significant and thus were discarded. We therefore find for this method the data are best fit using the BNTorus model with a scattered power-law component (Figure~\ref{fig:J0849_bnt}). The model indicates a photon index of $\Gamma=1.5^{+0.6}_{-\dots}$, for which a lower bound could not be computed, and an obscuring column of $\nh=114^{+\dots}_{-93}\times10^{22}$ cm$^{-2}$, for which the upper bound was pegged at the maximum value and thus invalid. Correcting for intrinsic absorption, this model indicates an unabsorbed X-ray luminosity $\rm{L}_{2\--10\ \rm{keV}} = 1.4^{+0.7}_{-0.1}\times10^{43}$ erg s$^{-1}$. These results are in agreement with the level of obscuration predicted from the relationship between the infrared 12$\mu$m and observerd X-ray $2\--10$ keV luminosity for this merger (see Figure~\ref{fig:lxl12plot}), which is $\nh \sim10^{23}$ cm$^{-2}$.
 
When fitting with the MYTorus model, we found a best fit using the MYTorus zeroth-order continuum paired with the MYTorus fluorescent emission line table and a scattered power-law component. The results of the MYTorus model largely agree with that found by BNTorus with the exception of $\Gamma$, which is pegged at 1.4 by MYTorus. We note specifically that MYTorus finds $\nh = 77^{+57}_{-45}\times10^{22}$ cm$^{-2}$, which is slightly lower than that found by BNTorus, but still agrees with the results of BNTorus within the uncertainties. As a result of this, however, BNTorus finds an unabsorbed luminosity an order of magnitude higher ($\sim 10^{43}$ erg s$^{-1}$) than that determined via MYTorus ($\sim 10^{42}$ erg s$^{-1}$).

\subsection{J0859+1310: Single AGN}
The northeastern \chandra{} source (Galaxy 1) represents a firm X-ray point source detection with a significance of 20.7$\sigma$ and with a hardness ratio of 0.97. We find no evidence for an X-ray point source above the background for Galaxy 2. We infer from the relationship between the absorbed X-ray luminosity and 12$\mu$m luminosity an obscuring column density of $2.3^{+2.2 }_{-1.2}\times10^{23}$ cm$^{-2}$. Based upon the two available SDSS spectra for this merger, which coincide with the galaxy nuclei, both Galaxy 1 and Galaxy 2 are classified as galaxies. The optical line ratios depicted on the BPT diagram (see Figure~\ref{fig:sample_images}) show Galaxy 1 classified as an AGN while Galaxy 2 falls within the composite region.  The X-ray luminosity of the source in Galaxy 1 is two orders of magnitude higher than that expected from star formation (Table~\ref{table:sfrlums}). While no X-ray source was detected in Galaxy 2, the optical line ratios shown in the BPT diagram for this galaxy in Figure~\ref{fig:sample_images} place it quite close to the \citet{kewley2001} demarcation, suggesting there could be an AGN in this nucleus. Additional follow-up optical  spectroscopy centered more accurately on Galaxy 2's nucleus could shed light on this issue.  

\subsubsection{J0859+1310NE Spectral Analysis Results}
\begin{figure}[h!]
    \centering
    \begin{minipage}[t]{0.45\textwidth}
        \centering
        \includegraphics[width=1.1\textwidth]{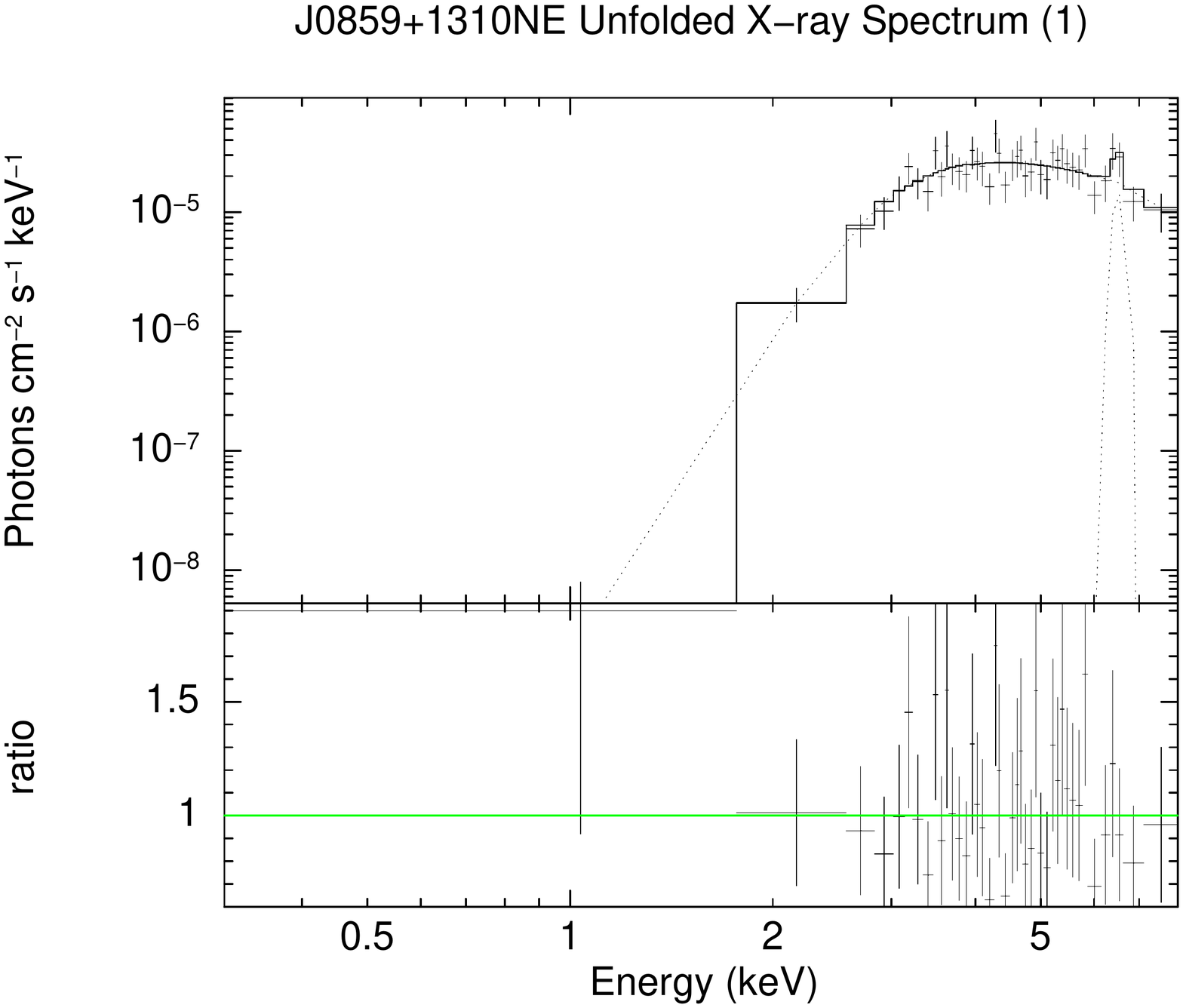} 
        \caption{The unfolded X-ray spectrum for J0859+1310NE modeled in \textsc{xspec} with the phenomenological approach for the full $0.3\--8$ keV energy band. The data are best fit with an absorbed power-law along with a Gaussian emission line component with a line energy of 6.7 keV.}
        \label{fig:J0859_phen}
    \end{minipage}\hspace{4mm}
    \begin{minipage}[t]{0.45\textwidth}
        \centering
        \includegraphics[width=1.1\textwidth]{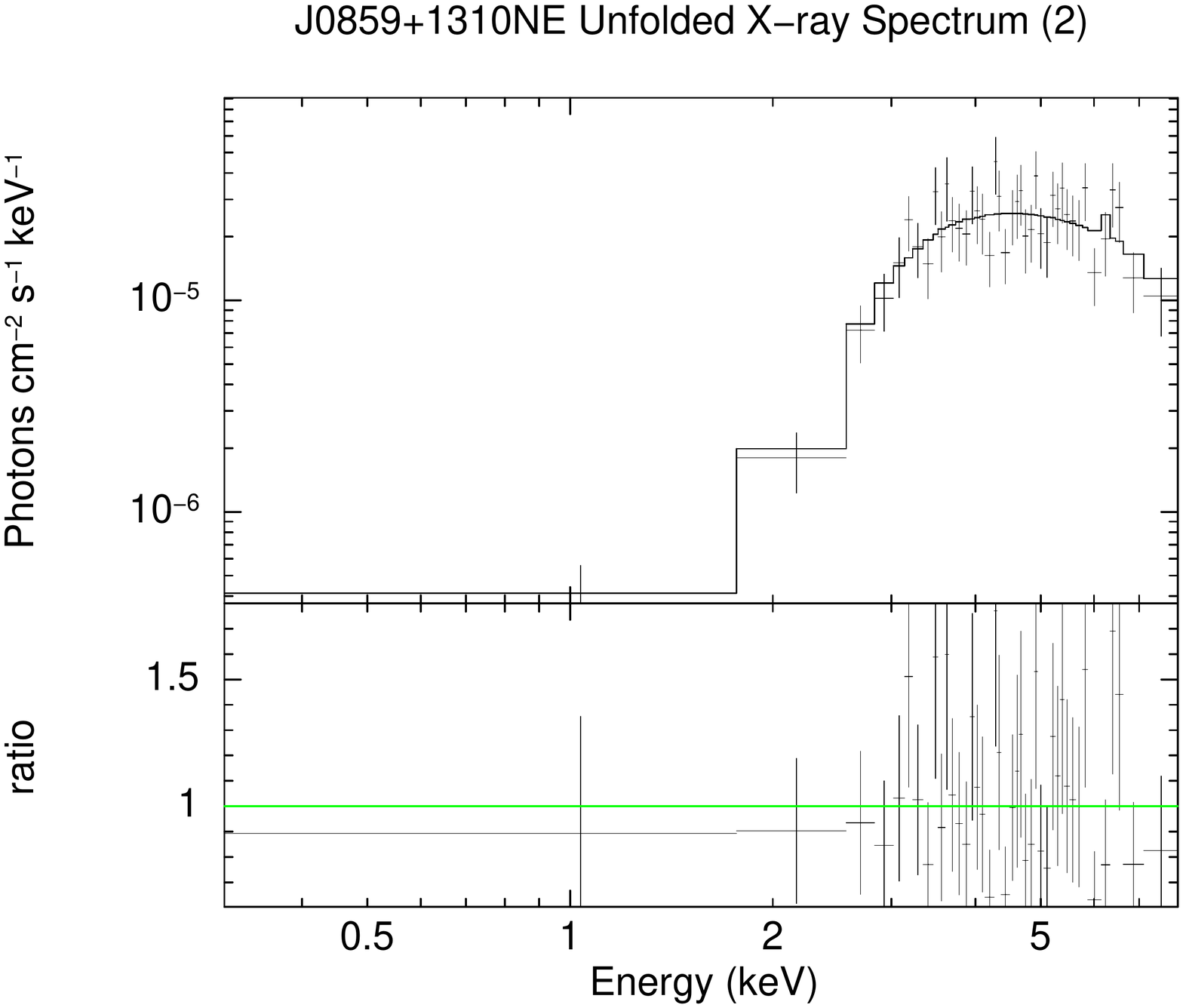} 
        \caption{The unfolded X-ray spectrum for J0859+1310NE modeled in \textsc{xspec} with the BNTorus approach for the full $0.3\--8$ keV energy band. The data are best fit with the base BNTorus model.}
        \label{fig:J0859_bnt}
    \end{minipage}
\end{figure}

The spectrum for J0859+1310NE is highly depleted in the soft X-ray energies (see Figures ~\ref{fig:J0859_phen} and ~\ref{fig:J0859_bnt}). In analyzing this spectrum through the phenomenological model, we report that a scattered power-law component did not introduce a statistically significant improvement beyond the absorbed power-law (indicated by $\Delta$C-Stat = $1.63 < 2.71$), while the Gaussian emission component did introduce a statistically significant improvement beyond the absorbed power-law, with $\Delta$C-Stat = $3.18 > 2.71$. The data for J0859+1310NE are best fit using an absorbed power-law with a Gaussian emission line component with line peak energy at 6.7 keV (Figure~\ref{fig:J0859_phen}), suggestive of an ionized iron K$\alpha$ emission line. The model indicates a photon index of $\Gamma=2.4^{+0.9}_{-0.8}$ and an obscuring column of $\nh=17.4^{+5.0}_{-4.3}\times10^{22}$ cm$^{-2}$, and we find an equivalent width of $0.23^{+\dots}_{-\dots}$ keV; we could not, however, constrain the error on the equivalent width using the \textsc{xspec} \textsc{err} command. This equivalent width agrees with the $\nh$ determined through this model. Correcting for intrinsic absorption, this model indicates an unabsorbed X-ray luminosity $\rm{L}_{2\--10\ \rm{keV}} = 7.9^{+1.1}_{-8.5}\times10^{42}$ erg s$^{-1}$, which indicates a robust detection in the X-rays of an AGN in this nucleus. 

Examining the system with the BNTorus model approach, we found that the introduction of a scattered power-law to the base model did not result in a statistically significant improvement to the fit  ($\Delta$C-Stat$ = 0.35 < 2.71$). All fits attempted with the \textsc{apec} component yielded either nonphysical values for parameters or were less statistically significant and thus were discarded. We therefore find for this method the data are best fit using the base BNTorus model (Figure~\ref{fig:J0859_bnt}). The model indicates a photon index of $\Gamma=2.2^{+0.6}_{-0.6}$ and an obscuring column of $\nh=14.9^{+3.9}_{-2.7}\times10^{22}$ cm$^{-2}$. Correcting for intrinsic absorption, this model indicates an unabsorbed X-ray luminosity of $\rm{L}_{2\--10\ \rm{keV}} = 6.7^{+0.9}_{-0.7}\times10^{42}$ erg s$^{-1}$. These results agree with the results found using the phenomenological approach above and with the level of obscuration predicted from the relationship between the infrared 12$\mu$m and absorbed X-ray $2\--10$ keV luminosity for this merger (see Figure~\ref{fig:lxl12plot}), which is $\nh \sim10^{23}$ cm$^{-2}$.

Attempts to fit this model using MYTorus yielded lower photon indexes and slightly lower $\nh$ values. Further, with MYTorus we could not identify the presence of a statistically significant iron line. A similar unabsorbed luminosity, $\rm{L}_{2\--10\ \rm{keV}} \sim 10^{42}$ is found using the MYTorus zeroth-order continuum.

\subsection{J0905+3747: Single AGN}
 The northeastern \chandra{} source (Galaxy 1) was detected with a significance of 8.0$\sigma$ and a hardness ratio of 0.45. No source was detected in Galaxy 2. The relationship between the X-ray luminosity and the infrared $\rm{L}_{12\mu \rm{m}}$ luminosity (see Figure~\ref{fig:lxl12plot}) suggests a column density of $9.1^{+6.8}_{-4.2}\times10^{23}$ cm$^{-2}$. Two SDSS spectra were available for the merger, coinciding with the galaxy nuclei, which classify Galaxy 1 as a starburst galaxy and Galaxy 2 as a star-forming galaxy. Examination of the optical line ratios depicted in the BPT diagram (see Figure~\ref{fig:sample_images}) show that Galaxy 1 falls within the composite region of the diagram while Galaxy 2 is classified as a star-forming galaxy.  The absorbed X-ray luminosity in Galaxy 1 is an order of magnitude greater than that expected from stellar processes. A [SiVI] coronal line was detected in Galaxy 1, allowing us to confirm the presence of at least one AGN in this merger.

\subsection{J1045+3519: Dual AGN Candidate}
J1045+3519 was another of four systems followed up during \chandra{} Cycle 18 and was previously identified in Paper I as a dual AGN candidate system. The merger was observed across two time periods and the data were merged together using the \textsc{ciao} \textsc{reproject\_obs} command. The apertures used for extraction of counts in the Cycle 17 data vary slightly from those used to extract counts from the Cycle 15 data. The western X-ray source (Galaxy 1) was detected in Cycle 18 with a significance of 4.3$\sigma$, a hardness ratio of -0.50, and a $P_{\rm{B}} < 0.002$, ruling out the possibility that the emission arose from spurious background activity. The eastern X-ray source (Galaxy 2) was detected with a significance of 2.5$\sigma$, a hardness ratio of -0.14, and a $P_{\rm{B}} < 0.002$ which signifies this source was also not the result of spurious background activity. Coronal line emission was not detected in either galaxy nucleus, but the absorbed X-ray luminosity (see Table~\ref{table:sfrlums}) for both sources is an order of magnitude higher than expected from stellar processes in each nucleus. SDSS classifies Galaxy 1 as a starburst galaxy and Galaxy 2 as a star-forming galaxy. The BPT diagram for this system shows that Galaxy 1 is optically classified as a star-forming galaxy while Galaxy 2 is classified as a composite galaxy (Figure~\ref{fig:sample_images}). Based upon the relationship between the total X-ray luminosity and 12 micron infrared luminosity of this merger (see Figure~\ref{fig:lxl12plot} and  Table~\ref{table:lxl12}, with values adopted from Paper I), we infer an obscuring column density of approximately $31.0^{+1.7}_{-1.2}\times10^{23}$ cm$^{-2}$, which results in a total unabsorbed luminosity in excess of $10^{43}$ erg s$^{-1}$. 
 
\subsection{J1147+0945: Single AGN}
The J1147+0945 system hosts three galaxy nuclei. The \chandra{} data revealed one bright X-ray point source in the southern galaxy (Galaxy 1) whose emission covered all three galaxy nuclei. This system was more closely examined to determine if the other nuclei exhibited X-ray emission in excess of that seen extending from the first nucleus. Drawing on the technique of \citet{ellison2017}, we placed apertures around the southern nucleus at roughly the same radii as the northeastern and northwestern nuclei. We then sampled the counts in these regions and compared the regions coincident with the NE and NW nuclei with regions which did not overlap with a nucleus. We found no statistically significant difference between the regions coincident with the NW and NE nuclei and regions placed at other positions within the emission area of the southern nucleus. We therefore cannot conclude that any other X-ray sources are present in this system at this time.

The S X-ray source (Galaxy 1) is detected with a significance of 56.0$\sigma$, providing a firm detection of an X-ray source, and a hardness ratio of 0.58. Based upon the absorbed luminosity alone, determined to be $2.12\pm0.25\times 10^{43}$ erg s$^{-1}$ via PIMMS with a basic absorbed power-law with photon index of 1.8, we can robustly confirm this X-ray source as an AGN as it has an absorbed X-ray luminosity in excess of $10^{43}$ erg s$^{-1}$. There are SDSS classifications for the S galaxy nucleus (Galaxy 1) and the NE galaxy (Galaxy 2), which identify Galaxy 1 as a QSO AGN broadline while Galaxy 2 is identified as a galaxy. The BPT plot identifies Galaxy 1 as an AGN and Galaxy 3 as a composite galaxy. No BPT or SDSS classifications were available for the NW nucleus (Galaxy 3). There is an additional emission region resolved by the \textsc{LBT} data which is not seen in the SDSS images. It is unclear if this additional object is simply a resolved component within Galaxy 1 or if it is its own separate entity positioned between Galaxy 1 and Galaxy 3. We do not include this object in our analysis. This merger was also included in a sample of optically selected multi-AGN mergers in \citet{liu2011}.

\subsubsection{J1147+0945S Spectral Analysis Results}
\begin{figure}[h!]
    \centering
    \begin{minipage}[t]{0.45\textwidth}
        \centering
        \includegraphics[width=1.1\textwidth]{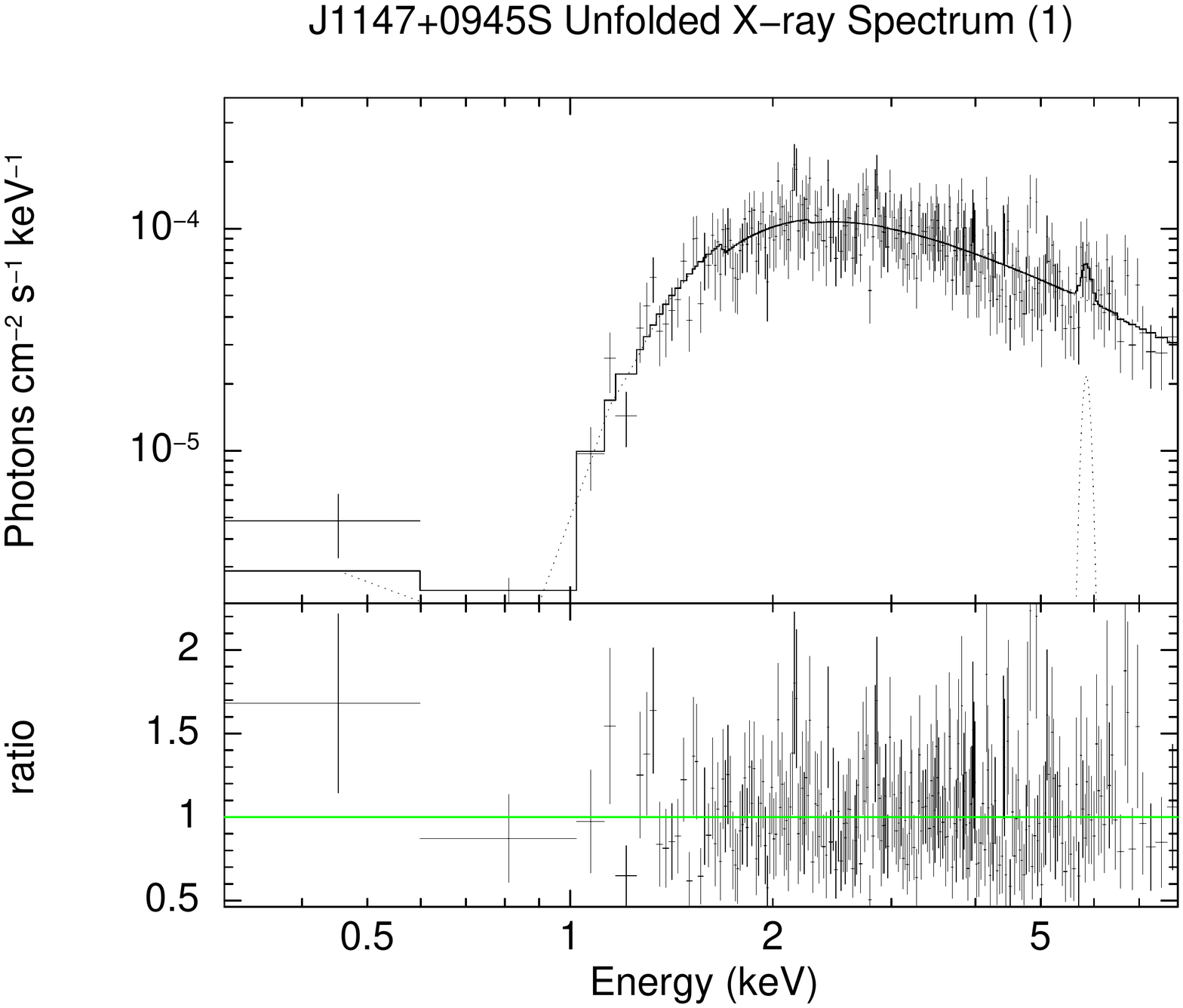} 
        \caption{The unfolded X-ray spectrum for J1147+0945S fit in \textsc{xspec} using the phenomenological approach for the full $0.3\--8$ keV energy band. The data are best fit with an absorbed power-law and Gaussian emission line component with line energy of 6.4 keV.}
        \label{fig:J114_phen}
    \end{minipage}\hspace{4mm}
    \begin{minipage}[t]{0.45\textwidth}
        \centering
        \includegraphics[width=1.1\textwidth]{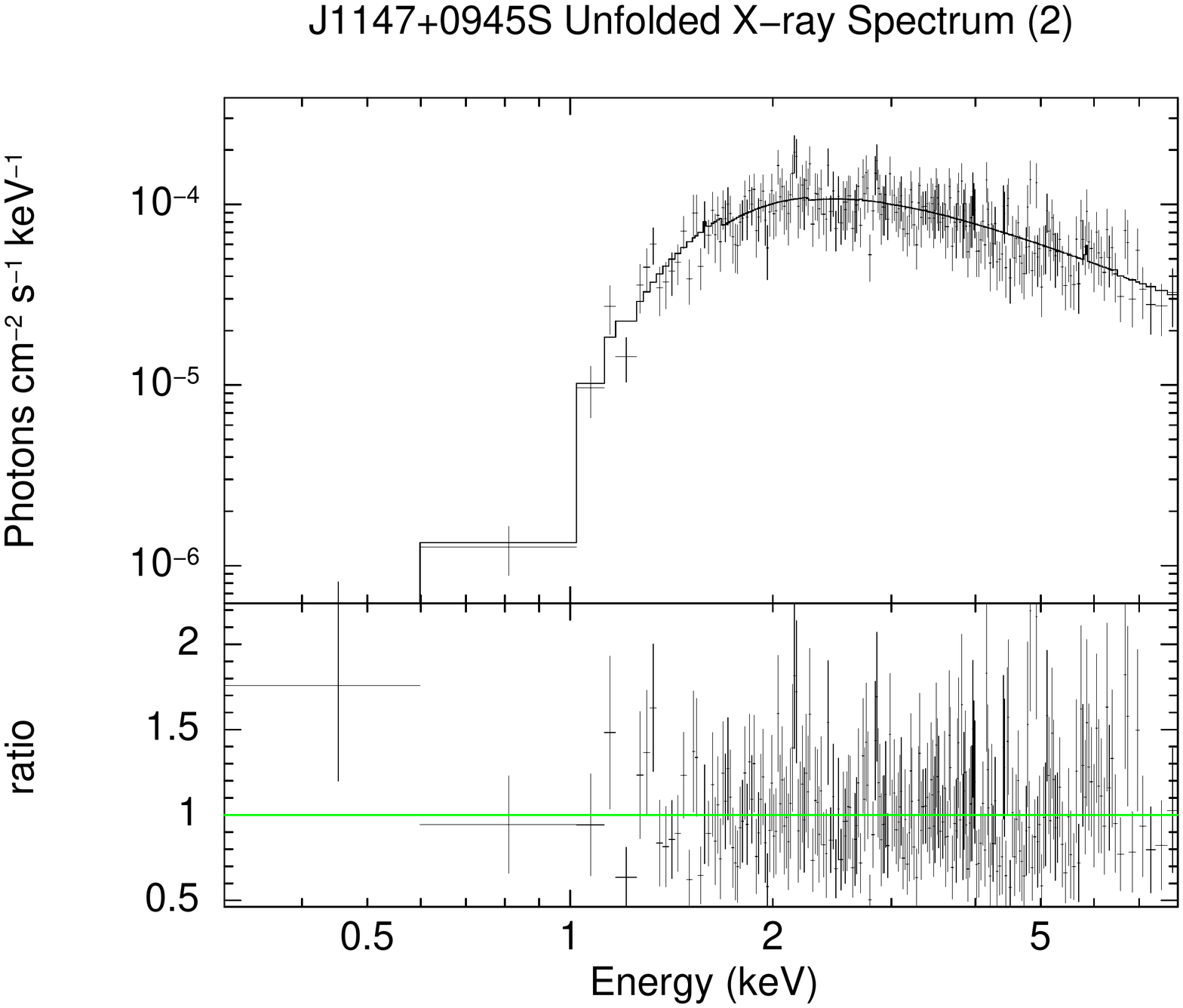} 
        \caption{The unfolded X-ray spectrum for J1147+0945S fit in \textsc{xspec} with the BNTorus approach for the full $0.3\--8$ keV energy band. The data are best fit with the base BNTorus model.}
        \label{fig:J114_bnt}
    \end{minipage}
\end{figure}

The spectrum for J1147+0945S (Figure~\ref{fig:J114_phen} and~\ref{fig:J114_bnt}) exhibits heavy depletion of the soft X-ray energies $\--$ which could suggest a high level of obscuration $\--$ and an excess above the absorbed power-law component is present around $6\--7$ keV. In analyzing this spectrum with the phenomenological approach, the Gaussian emission component introduced a statistically significant improvement beyond the absorbed power-law, with $\Delta$C-Stat = $4.91 > 2.71$. Adding a scattered power-law component does not introduce a statistically significant improvement beyond the absorbed power-law and Gaussian component (indicated by $\Delta$C-Stat = $ 0.29 < 2.71$), and was rejected. The spectrum is therefore best fit using an absorbed power-law with a Gaussian emission line component with line peak energy at 6.4 keV (Figure~\ref{fig:J114_phen}). The model yields a photon index of $\Gamma=1.5^{+0.2}_{-0.1}$, an obscuring column of $\nh=2.6^{+0.3}_{-0.3}\times10^{22}$ cm$^{-2}$, and an iron line equivalent width of $0.11^{+0.09}_{-0.09}$ keV (see Table~\ref{table:phenspec}). This equivalent width agrees with the $\nh$ determined with this model. From this level of $\nh$, we conclude this is a Compton-thin obscured AGN. These results also agree with that suggested by plotting the infrared 12$\mu$m luminosity and absorbed X-ray $2\--10$ keV luminosity (shown in Figure~\ref{fig:lxl12plot}), which shows a level of obscuration of only $\nh\sim 10^{22}$ cm$^{-2}$. Correcting for intrinsic absorption, this model indicates an unabsorbed X-ray luminosity of $\rm{L}_{2\--10\ \rm{keV}} = 8.9^{+1.3}_{-1.5}\times10^{43}$ erg s$^{-1}$. 

Using the BNTorus model approach, we found that introducing a scattered power-law or \textsc{apec} component resulted in fits no more statistically significant than the base BNTorus model. We therefore find for this method the data are best fit using the base BNTorus model (Figure~\ref{fig:J114_bnt}). The model indicates a photon index of $\Gamma=1.5^{+0.1}_{-0.1}$ and an obscuring column of $\nh=2.3^{+0.2}_{-0.2}\times10^{22}$ cm$^{-2}$. Correcting for intrinsic absorption, this model indicates an unabsorbed X-ray luminosity of $\rm{L}_{2\--10\ \rm{keV}} = 8.9^{+1.2}_{-1.7}\times10^{43}$ erg s$^{-1}$. These results agree with that found using the phenomenological model, as well as the level of obscuration predicted from the relationship between the infrared 12$\mu$m and absorbed X-ray $2\--10$ keV luminosity for this merger (see Figure~\ref{fig:lxl12plot}), which is $\nh \sim10^{22}$ cm$^{-2}$. The results found for BNTorus and the phenomenological approach agree with that obtained using a MYTorus zeroth-order continuum paired with the MYTorus fluorescent emission line table to account for the Iron K$\alpha$ line.

\subsection{J1159+5320: Single AGN}
The southeastern \chandra{} source (Galaxy 1) is detected with a significance of 3.2$\sigma$, with a majority of counts originating from the hard band and leading to a hardness ratio of 0.80. We note the absorbed X-ray luminosity (Table~\ref{table:sfrlums}), $\rm{L}_{2\--10\ \rm{keV}} = $ $4.1\pm1.7\times10^{40}$ erg s$^{-1}$, while not high enough to provide unambiguous proof of the existence of an AGN, is nearly an order of magnitude higher than that expected from XRBs. No source was detected in the northwestern nucleus (Galaxy 2). Comparing the absorbed X-ray luminosity with the 12$\mu$m luminosity (see  Table~\ref{table:lxl12} and Figure~\ref{fig:lxl12plot}), we infer an obscuring column density for this system of at least $12.0^{+8.4 }_{-5.3}\times10^{23}$ cm$^{-2}$. Correcting for absorption inferred by the $\rm{L}^{\rm{Abs.}}_{2\--10\ \rm{keV}}$ vs. $\rm{L}_{12\mu \rm{m}}$ relationship, we report an unabsorbed X-ray luminosity for this system in excess of $10^{42}$ erg s$^{-1}$. One SDSS spectrum was available which coincided with the nucleus of Galaxy 1 and classifies it as an AGN. The BPT diagram (see Figure~\ref{fig:sample_images}) also classifies Galaxy 1 as an AGN. 

\subsection{J1221+1137: Dual AGN Candidate}
J1221+1137 was another system followed up during \chandra{} Cycle 18 and previously reported as a candidate dual AGN system in Paper I. In this analysis we adopted identical source regions to those used in the pilot study. The northeastern source is detected with a significance of 4.8$\sigma$ and with a hardness ratio -0.45. We report that the absorbed X-ray luminosity of the NE source is over three times that expected from stellar processes and we therefore rule out stellar processes as the sole origin of this emission. The southwestern source is detected with a significance of only 1.1$\sigma$, but we note that the $P_{\rm{B}}$ value obtained for this source ($P_{\rm{B}} = 0.00001 < 0.002$) does indicate the X-ray emission is not the result of spurious background activity. The absorbed X-ray luminosity exhibited by the SW source is very similar to that expected from stellar X-ray contributions if (1) one assumes the mass of Galaxy 2 is the same as Galaxy 1 (no mass for Galaxy 2 was available) and (2) that all of the Pa$\alpha$ flux arises from gas ionized by stellar processes alone, although in reality some of this flux would arise from gas ionized by any potential AGN. However, as reported in Paper I, coronal line emission was detected in Galaxy 2. We therefore conclude that this system hosts at least one robustly confirmed AGN and is also a dual AGN candidate with a secondary X-ray source in the northeastern nucleus. SDSS classifies Galaxy 1 as a starburst galaxy, which agrees with the classification based upon the optical line ratios in the BPT plot shown in Figure~\ref{fig:sample_images}. Neither BPT line ratios or an SDSS classification was available for Galaxy 2. From the relationship between the total absorbed X-ray luminosity and 12$\mu$m luminosity of the system, we infer an obscuring column density of $\nh = 27.0^{+1.5}_{-1.1}\times10^{23}$ cm$^{-2}$, which was previously reported in Paper I. We report this value in  Table~\ref{table:lxl12}.

\subsubsection{J1221+1137 XMM Spectral Analysis}

\begin{figure}[h!]
    \centering
    \begin{minipage}[t]{0.45\textwidth}
        \centering
        \includegraphics[width=1.1\textwidth]{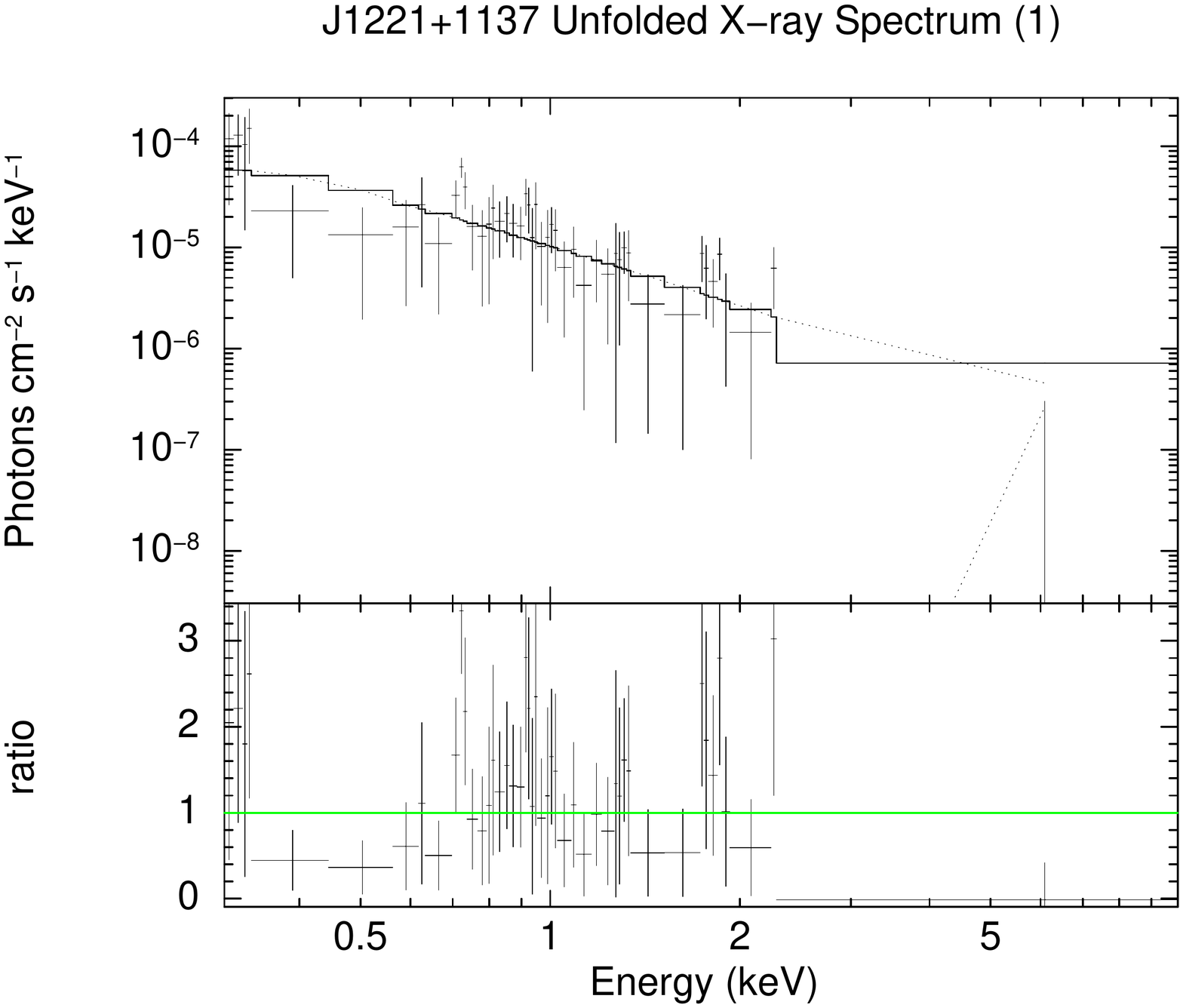} 
        \caption{The unfolded XMM X-ray spectrum for J1221+1137, modeled in \textsc{xspec} using the phenomenological approach for the full energy band $0.3\--10$ keV. The data are best fit with an absorbed power-law and scattered power-law component.}
        \label{fig:J1221_phen}
    \end{minipage}\hspace{4mm}
    \begin{minipage}[t]{0.45\textwidth}
        \centering
        \includegraphics[width=1.1\textwidth]{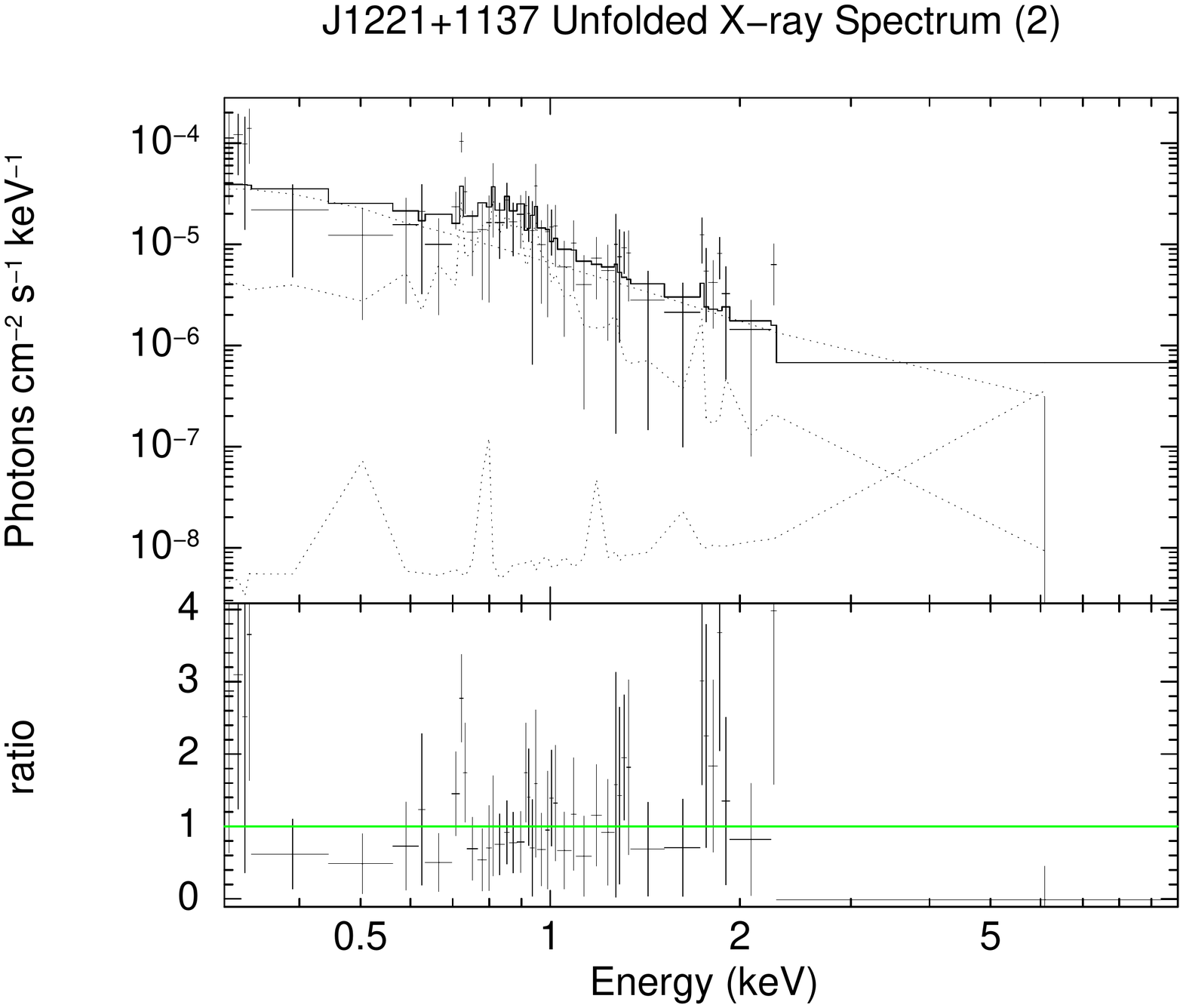} 
        \caption{The unfolded XMM X-ray spectrum for J1221+1137, modeled in \textsc{xspec} with the BNTorus approach for the full energy band $0.3\--10$ keV. The data are best fit with the base BNTorus model plus a scattered power-law and an \textsc{apec} component.}
        \label{fig:J1221_bnt}
    \end{minipage}
\end{figure}

Examining J1221+1137 with the phenomenonological model, we found that adding a scattered power-law provided no statistically significant change in the $\chi^2$ statistic, in fact the statistic remained nearly unchanged by the addition of the parameter. However, we did find that without the scattered power-law, the model missed the hard X-rays detected by XMM and fit only the soft X-rays. Therefore, based on the physical assumption that the soft X-rays are the result of scattered photons, we included a scattered power-law in addition to the base absorbed power-law. With this noted, the data are best fit with an absorbed power-law and a scattered power-law component.This result is shown in Figure~\ref{fig:J1221_phen}. An attempt was made to incorporate \textsc{apec} into the model, however the addition of this component resulted in nonphysical values for $\nh$, and we therefore discarded \textsc{apec} when fitting with this approach. The best fitting model reveals high obscuration in this system, $\nh = 65^{+\dots}_{-38}\times10^{22}$ cm$^{-2}$ (for which an upper bound could not be computed), and a photon index of $\Gamma = 2.0^{+0.3}_{-0.3}$. Correcting for absorption, we find an unabsorbed luminosity of $\rm{L}_{2\--10\ \rm{keV}} = 3.3^{+9.1}_{-1.7}\times10^{42}$ erg s$^{-1}$, consistent with the presence of at least a single AGN in this system. The $\nh$ determined with this model is lower than that predicted by the relationship between the total absorbed X-ray luminosity and total 12$\mu$m luminosity for this system, which suggests a column density of $\sim 2.7\times10^{24}$ cm$^{-2}$. 

For the BNTorus approach we find the data are best fit with the base BNTorus model along with a scattered power-law and an \textsc{apec} component (Figure~\ref{fig:J1221_bnt}). The introduction of the scattered power-law ($\Delta \chi^2 = 103.2$) and \textsc{apec} ($\Delta \chi^2 = 10.16$ beyond the base model plus scattered power-law) both proved a statistically significant change over the base model and were included in the final fitting. The best fitting model reveals a $\Gamma$ of $2.0^{+0.5}_{-\dots}$ (a lower bound could not be computed) and high obscuration, with $\nh = 43^{+\dots}_{-\dots}\times10^{23}$ cm$^{-2}$ (error bounds were pegged at the maximum and minimum values and thus could not be computed). This result is shown in Figure 16. Additionally, we find a plasma temperature of $0.92^{+0.25}_{-0.33}$ keV for the optically-thin plasma modeled by \textsc{apec}. Correcting for absorption, the BNTorus approach yields an unabsorbed luminosity of $\rm{L}_{2\--10\ \rm{keV}} = 2.0^{+3.5}_{-1.0}\times10^{42}$ erg s$^{-1}$. These parameter values agree with that found by the phenomenological approach. However, the obscuration and unabsorbed luminosity found with the BNTorus model are still lower than that expected from the total absorbed X-ray luminosity and total 12$\mu$m luminosity relationship for this system. 

\subsection{J1301+2918: Dual AGN Candidate}
The northeastern \chandra{} source (Galaxy 2) is detected with a significance of 6.8$\sigma$ with a hardness ratio of -0.16. For Galaxy 2 (the southwestern \chandra{} source), we identified a few counts which coincided with the galaxy nucleus. This X-ray emission holds a significance of only 0.8$\sigma$ above the background. We note, however, that the $P_{\rm{B}}$ value found for this X-ray emission ($P_{\rm{B}} = 0.0002 < 0.002$ in the full band) indicates this emission does not originate from spurious background activity.  Examining the total X-ray luminosity vs. 12$\mu$m luminosity relationship (see Table~\ref{table:lxl12} and Figure~\ref{fig:lxl12plot}), we infer a high column density for this system of $10.9^{+7.8 }_{-4.9}\times10^{23}$ cm$^{-2}$. The absorbed luminosity of the NE source is $\rm{L}_{2\--10\ \rm{keV}} = 7.1\pm1.8 \times10^{40}$ erg s$^{-1}$ while the SW source absorbed luminosity is only $\rm{L}_{2\--10\ \rm{keV}} = 0.41\pm0.50 \times10^{40}$ erg s$^{-1}$, both of which are lower than expected for AGNs. SDSS spectra were available for Galaxy 2, which classifies it as an AGN. The optical line ratios depicted in the BPT diagram (see Figure~\ref{fig:sample_images}) also show that Galaxy 2 would be optically classified as an AGN. Pa$\alpha$ line fluxes were lost due to atmospheric absorption, and thus calculation of the SFR and potential contribution by X-ray binaries was not possible. J1301+1822, also known as NGC 4922, was examined in thorough detail previously in \citet{alonso1999} and \citet{ricci2017mnras}, the latter of which found previous evidence via NuSTAR for a buried AGN in the NE nucleus.

Archival data were available in addition to our Cycle 17 observation to further the analysis of this system. J1301+2918 (NGC 4922) was observed in \chandra{} Cycle 5 (2005, PI: Salzer) and Cycle 14 (2014, PI: Sanders) with exposure times of 3.8 ks and 14.9 ks, respectively. Using the apertures for source extraction from the Cycle 17 data (photometric results listed in Table 4), we extracted source counts for the two other data sets. We found no statistically significant change in the count rates for the two sources across the three data sets. 

\subsubsection{J1301+2918NW Spectral Analysis Results}
\begin{figure}[h!]
    \centering
    \begin{minipage}[t]{0.45\textwidth}
        \centering
        \includegraphics[width=1.1\textwidth]{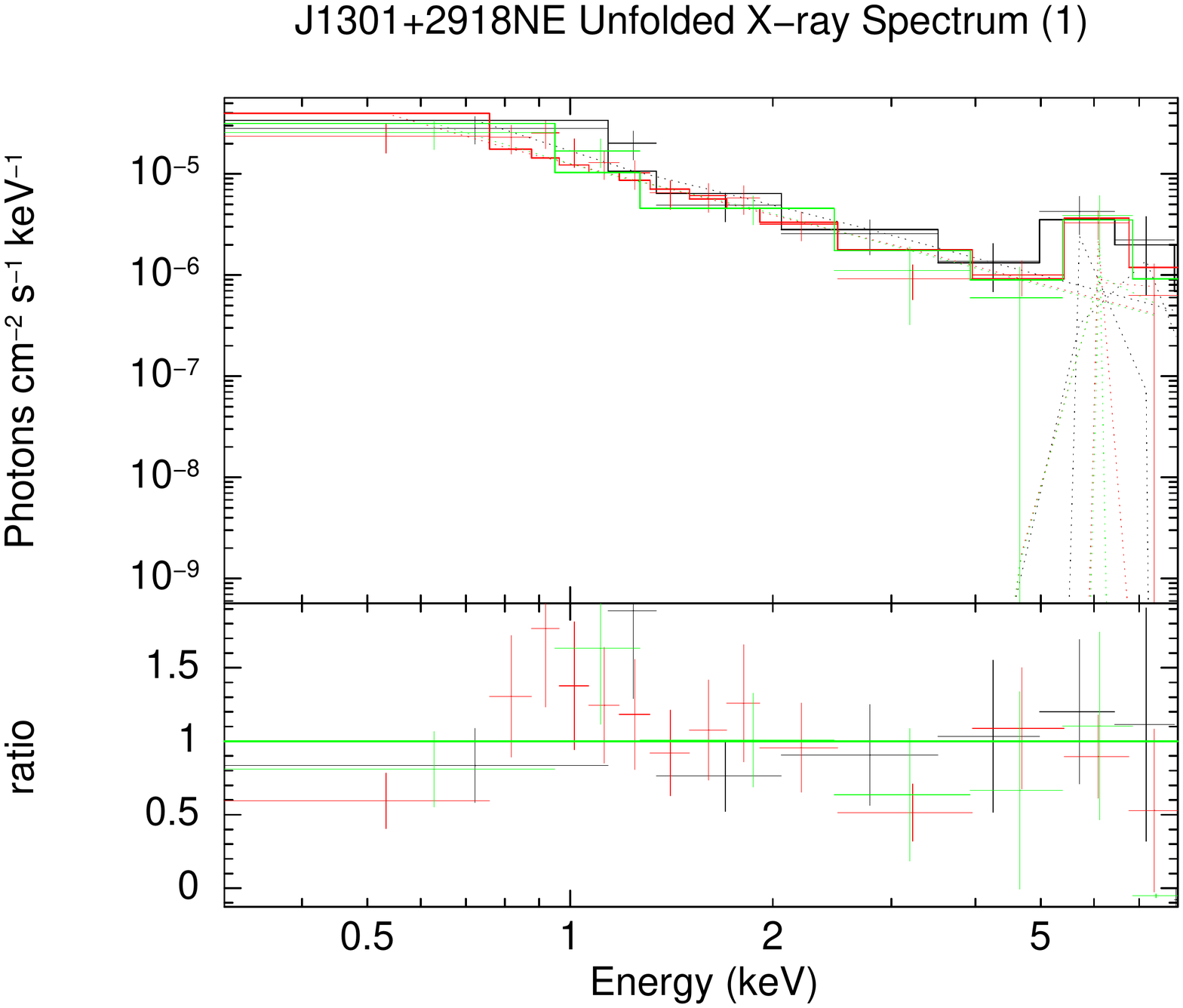} 
        \caption{The unfolded X-ray spectrum for J1301+2918NE fit in \textsc{xspec} with the phenomenological model. The data are best fit using an absorbed power-law with a scattered power-law and a Gaussian emission line at 6.4 keV. Black data points correspond to \chandra{} data taken in 2016 (PI: Satyapal), red points correspond to data taken in 2012 (PI: Sanders), and green points correspond to data from 2014 (PI: Salzer.)}
        \label{fig:J1301_phen}
    \end{minipage}\hspace{4mm}
    \begin{minipage}[t]{0.45\textwidth}
        \centering
        \includegraphics[width=1.1\textwidth]{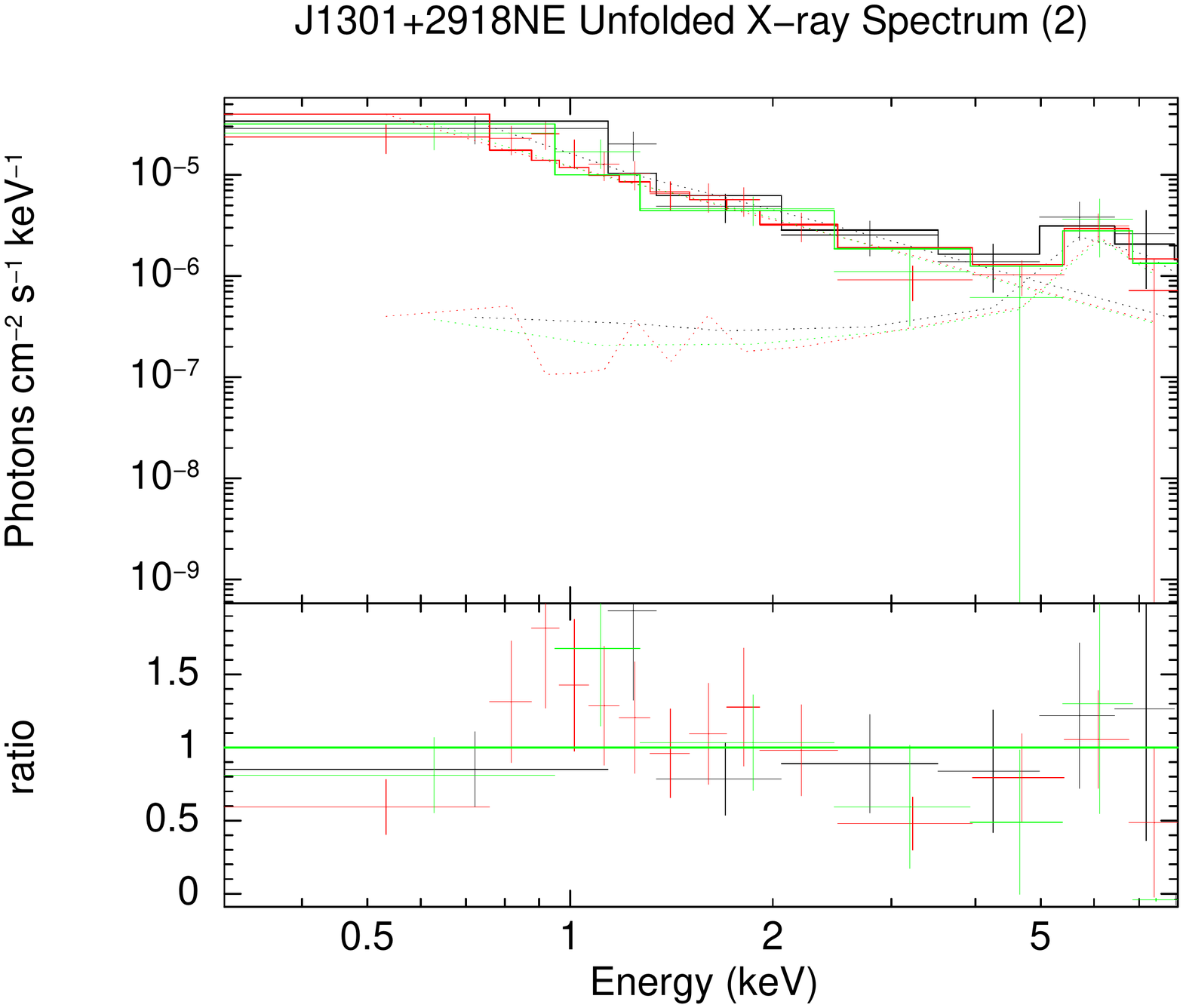} 
        \caption{The unfolded X-ray spectrum for J1301+2918NE fit in \textsc{xspec} with the BNTorus model. The data are best fit using the BNTorus base model plus a scattered power-law. Black data points correspond to \chandra{} data taken in 2016 (PI: Satyapal), red points correspond to data taken in 2012 (PI: Sanders), and green points correspond to data from 2014 (PI: Salzer.)}
        \label{fig:J1301_bnt}
    \end{minipage}
\end{figure}

To analyze the spectrum for J1301+2918NW, we obtained spectral files for three separate observations: \chandra{} cycle 5 (2005, PI: Salzer), cycle 14 (2014, PI: Sanders), and cycle 17 (2016, PI: Satyapal). We fit these data simultaneously in \textsc{xspec} (see Figure~\ref{fig:J1301_phen} and \ref{fig:J1301_bnt} for spectrum.) Using the phenomenological model approach, we found that a scattered power-law component does introduce a statistically significant improvement beyond the absorbed power-law (indicated by $\Delta$C-Stat = $36.85 > 2.71$), and was therefore included for further fitting. A Gaussian emission component introduced a statistically significant improvement beyond the absorbed and scattered power-laws, with $\Delta$C-Stat = $14.32 > 2.71$ and was therefore included in the final fitting. The data are therefore best fit with an absorbed power-law, a scattered power-law, and a Gaussian emission line component with line peak at 6.4 keV (Figure~\ref{fig:J1301_phen}). The model yields a photon index of $\Gamma=1.7^{+0.2}_{-0.2}$, an obscuring column of $\nh=438.3^{+945.7}_{-294.6}\times10^{22}$ cm$^{-2}$, and an iron line equivalent width of $2.06^{+7.58}_{-0.67}$ keV (see Table~\ref{table:phenspec}). Despite the large error bounds in this work, these results do agree with the measured $\nh$ and Iron K$\alpha$ equivalent width reported in \citet{ricci2017mnras} for the NE source. A comparison between the infrared luminosity to the X-ray $2\--10$ keV luminosity (see Figure~\ref{fig:lxl12plot}) indicates a column density of at least $10^{24}$ cm$^{-2}$, which agrees with that inferred via the equivalent width of the Fe K$\alpha$ line and the $\nh$ determined with the phenomenological model. Finally, this model yields an unabsorbed luminosity of $6.9^{+0.7}_{-5.4}\times10^{45}$ erg s$^{-1}$.

Examining the system with the BNTorus model (Figure~\ref{fig:J1301_bnt}, we found that the introduction of a scattered power-law to the base model resulted in a statistically significant improvement to the fit  ($\Delta$C-Stat $ = 130.21 > 2.71$). We note, too, that the addition of an \textsc{apec} component introduced a statistically significant improvement to the base model ($\Delta$C-Stat$ = 104.71$) , as well as to the base model plus the scattered power-law ($\Delta$C-Stat$ = 18.05$), but it also also introduced non-monotonicity issues when determining parameter values and error bounds for the latter model. Further, we could no longer determine both upper and lower error bounds for $\Gamma$ when \textsc{apec} was included. As a result, we elected to reject \textsc{apec} from our final best fitting model. We therefore find for this method the data are best fit using the base BNTorus model and a scattered power-law component only. The model indicates a photon index of $\Gamma=1.8^{+0.3}_{-0.2}$, a scattering fraction of $1.2^{+2.3}_{-0.7}\%$, and an obscuring column of $\nh=148^{+75}_{-53}\times 10^{22}$ cm$^{-2}$. Correcting for intrinsic absorption, this model indicates an unabsorbed X-ray luminosity of $\rm{L}_{2\--10\ \rm{keV}} = 5.2^{+2.0}_{-0.5}\times10^{42}$ erg s$^{-1}$. These results are similar to the results found using the phenomenological approach above and agree with the level of obscuration predicted from the relationship between the infrared 12$\mu$m and absorbed X-ray $2\--10$ keV luminosity for this merger (see Figure~\ref{fig:lxl12plot}), which is $\nh \sim10^{24}$ cm$^{-2}$. This model approach, though, finds a slightly lower value for $\nh$ and a lower unabsorbed luminosity than the phenomenological approach.

Attempts to fit this model using a MYTorus zeroth-order continuum, fluorescent emission line table, and scattered power-law yielded lower photon indexes and $\nh$ values an order of magnitude lower than the approaches above. With MYTorus we could identify a statistically significant Fe K$\alpha$ emission line with equivalent width $\sim 1.8^{+0.9}_{-0.9}$ keV, which agrees with that found by the phenomenological model. The unabsorbed luminosity determined by MYTorus is an order of magnitude lower, however, on the order od $\rm{L}_{2\--10\ \rm{keV}} \sim 10^{41}$ erg s$^{-1}$ - this is likely due to the lower value of $\nh$ determined by MYTorus.

\subsection{J1306+0735: Dual AGN Candidate}
Previously reported in Paper I as a dual AGN candidate, J1306+0735 was followed up during the \chandra{} observation Cycle 18. We report the presence of three regions of X-ray emission in the merger, two of which were previously reported and the third of which has now appeared in the higher-exposure observations. The merger was observed across three different time periods and the data were then merged together using the \textsc{ciao} \textsc{reproject\_obs} command. For the northeastern and southwestern sources, we adopted in this study identical source regions for count extractions as the pilot study (Paper I), but we used the Gaussian smoothing technique in ds9 for aperture positioning of the new southeastern source aperture. 

The NE source (previously reported) was detected with a significance of 3.6$\sigma$ and hardness ratio of -0.35. The SW source (previously reported) was detected with a significance of 7.4$\sigma$ and hardness ratio of -0.12. The third, previously unreported SE source, was detected with a significance of 2.7$\sigma$ and with a hardness ratio of -0.74. We note that for all three sources the calculated $P_{\rm{B}}$ value indicates the reported sources are not the result of spurious background activity.

No coronal emission was detected in the NE, SW, or SE regions. The absorbed X-ray emission originating from each of the three X-ray sources is roughly an order of magnitude higher than that expected from star formation, suggestive of the presence of AGNs. Comparing the X-ray and 12$\mu$m luminosity for this system, we infer a column density of $24.0^{+1.4}_{-1.0}\times10^{23}$ cm$^{-2}$ along the line of sight.

SDSS classifies Galaxy 1 as a starburst galaxy which agrees with the optical classification of star-forming galaxy from the BPT diagram (see Figure~\ref{fig:sample_images}). No SDSS or BPT classifications were available for the other extraction regions. 

\subsection{J1356+1822: Dual AGN}
The eastern X-ray source (Galaxy 1) is detected with a significance of 12.2$\sigma$ and a hardness ratio of -0.09. The western source (Galaxy 2) is detected with a significance of 6.8$\sigma$ and a hardness ratio of 0.32. The absorbed X-ray luminosities for both sources are roughly an order of magnitude greater than those expected from stellar processes. There were no SDSS classifications for the two galaxies in this merger, but the BPT optical line ratios (see Figure~\ref{fig:sample_images}) show the western galaxy (Galaxy 2) is classified as an AGN. No BPT line ratios were available for Galaxy 1. Based on the presented diagnostics and the spectral analysis below, we confirm the presence of a dual AGN in this merger.

J1356+1822 was, in fact, previously analyzed in \citet{bianchi2008} and is also known as Mrk 463. \citet{bianchi2008} identified two X-ray sources in the 50 ks exposure \chandra{} archival data (PI: Predahl, 2004) and confirmed the presence of two AGNs in this merger. Our results agree with that found by \citet{bianchi2008}, as discussed in the following Spectral Analysis subsection. We extracted counts from the archival 50 ks exposure from 2004 and found a statistically significant ($>3\sigma$) decrease in the count rates for the E source between 2004 and 2016. The variability of the source is limited to the soft band (0.3$\--$2 keV). We see no evidence for statistically significant variability in the hard band (2$\--$10  keV). 

Examining the combined X-ray luminosity and the 12$\mu$m luminosity for the system, however, we can infer a total column density (see  Table~\ref{table:lxl12} and Figure~\ref{fig:lxl12plot}) of approximately $19.2^{+11.9 }_{-8.0}\times10^{23}$ cm$^{-2}$. This estimated $\nh$ agrees with theoretical predictions for obscuration within advanced mergers, but we note this value is higher than that found through X-ray modeling in previous works as well as in the following spectral analysis.

\subsubsection{J1356+1822E Spectral Analysis Results}
\begin{figure}[h!]
    \centering
    \begin{minipage}[t]{0.45\textwidth}
        \centering
        \includegraphics[width=1.1\textwidth]{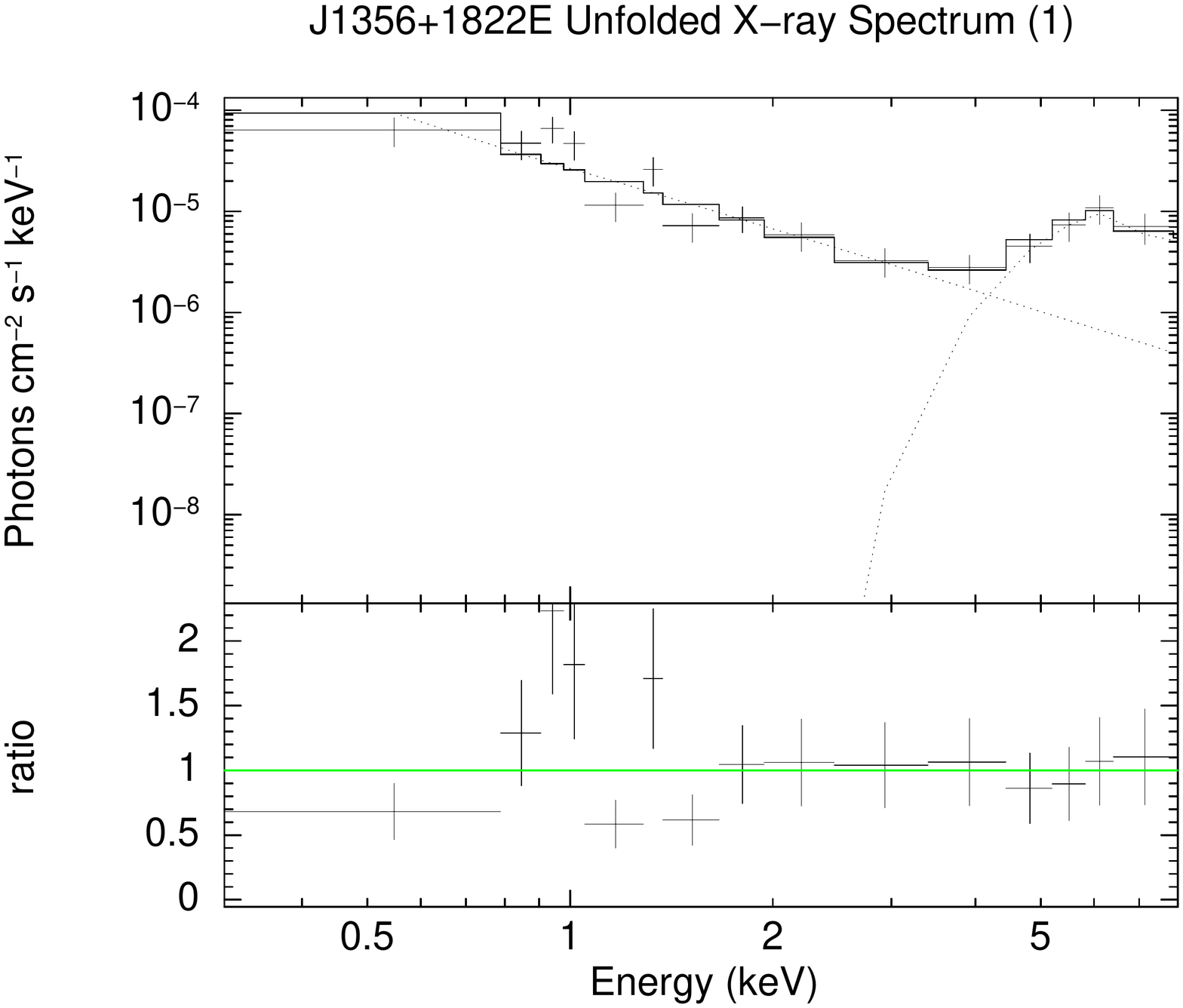} 
        \caption{The unfolded X-ray spectrum for J1356+1822E modeled in \textsc{xspec} with the phenomenlogical approach for the full $0.3\--8$ keV band. The data are best fit using an absorbed power-law with scattered power-law component.}
        \label{fig:J135_phen}
\end{minipage}\hspace{4mm}
    \begin{minipage}[t]{0.45\textwidth}
        \centering
        \includegraphics[width=1.1\textwidth]{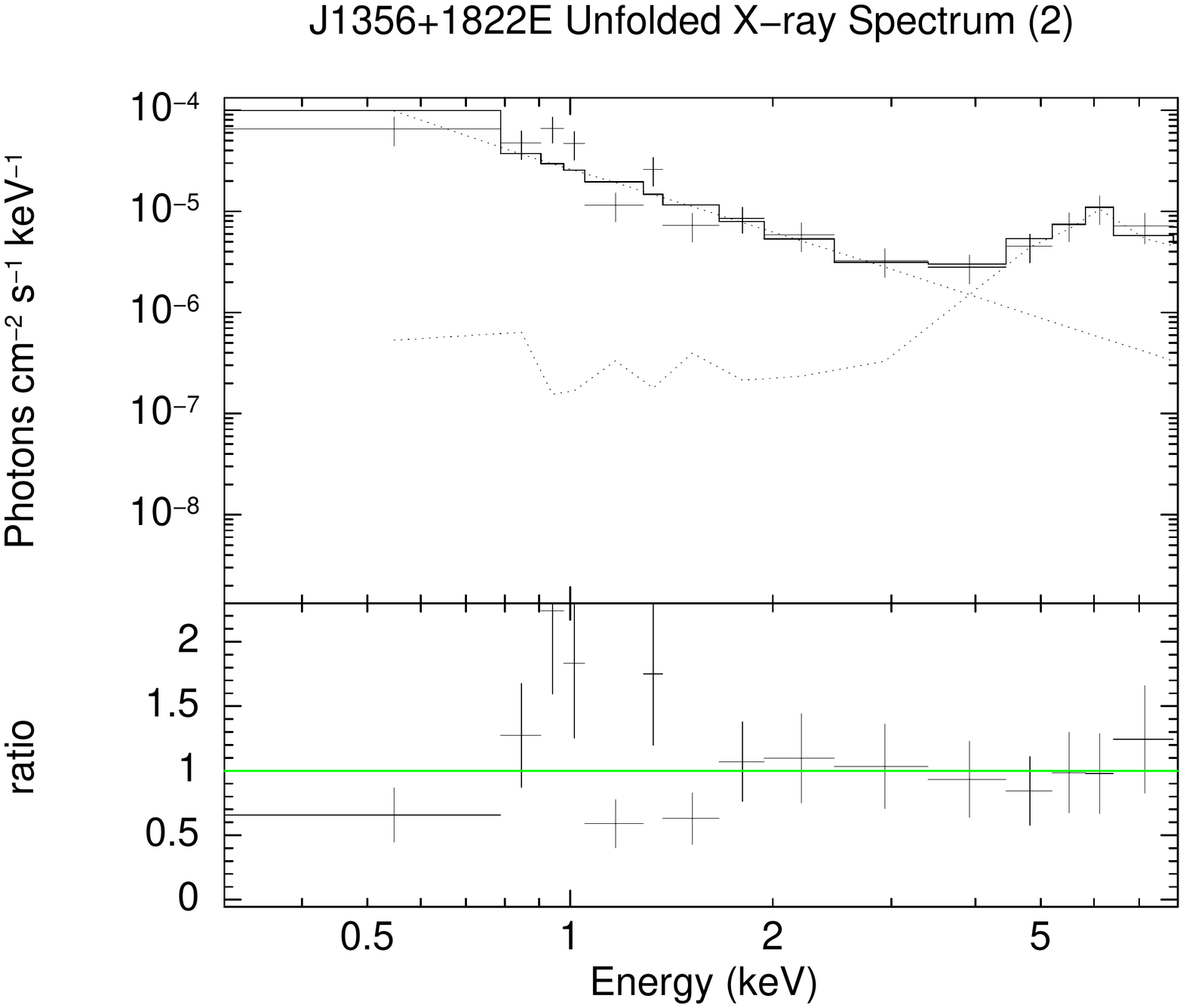} 
        \caption{The unfolded X-ray spectrum for J1356+1822E modeled in \textsc{xspec} with the BNTorus approach for the full $0.3\--8$ band. The data are best fit using the base BNTorus model plus a scattered power-law component.}
        \label{fig:J135_bnt}
    \end{minipage}
\end{figure}

We present the spectrum for J1356+1822E in Figure~\ref{fig:J135_phen}. During the analysis with the phenomenological model, a scattered power-law did provide a statistically significant improvement to the base model ($\Delta$C-Stat = $49.36 > 2.71$), but we found no excess above the absorbed power-law component in the range of $6\--7$ keV, and we report (fitting with and without Gaussian emission line component) a $\Delta$C-Stat = $0.02 < 2.71$, indicating a Gaussian emission component is statistically insignificant to the absorbed and scattered power-law model. The data are therefore best fit using an absorbed power-law along with a scattered power-law component. The model yields a photon index of $\Gamma=2.1^{+0.4}_{-0.3}$ and an obscuring column of $\nh=75.2^{+42.7}_{-28.3}\times10^{22}$ cm$^{-2}$ (see Table~\ref{table:phenspec}).  Correcting for intrinsic absorption, this model indicates an unabsorbed X-ray luminosity of $\rm{L}_{2\--10\ \rm{keV}} = 3.9^{+1.4}_{-3.6}\times10^{43}$ erg s$^{-1}$. Though our exposure time was a factor of 5 lower than the data reported in \citet{bianchi2008}, the results for $\Gamma$ and $\nh$ do agree within the uncertainties, with the exception that an iron line at 6.4 keV was previously reported. We see no evidence of an iron line based upon a $\sim$10 ks exposure with this specific model. We note also that the obscuration determined through this model agrees with that predicted ($\nh \sim 10^{23}$ cm$^{-2}$) by the relationship between the infrared luminosity and X-ray $2\--10$ keV luminosity (see Figure~\ref{fig:lxl12plot}).

Examining the system with the BNTorus approach (Figure~\ref{fig:J135_bnt}), we found that the introduction of a scattered power-law to the base model resulted in a statistically significant improvement to the fit  ($\Delta$C-Stat$ = 86.5 > 2.71$). We also note that the addition of an \textsc{apec} component to the base model was a statistically significant improvement of $\Delta$C-Stat$ = 94.77$, which is a more statistically significant improvement than the scattered power-law. However, when employing the \textsc{apec} component, we found the model pegged $\Gamma$ at the maximum value allowed by BNTorus, 2.8, and we could no longer constrain error bounds for this parameter - as a result we have elected to reject the model containing \textsc{apec} and accept the former model. Combining BNTorus with a scattered power-law \textit{and} \textsc{apec} did not yield a more statistically significant model. We therefore find for this method the data are best fit using the BNTorus model with a scattered power-law. The model indicates a photon index of $\Gamma=2.1^{+0.4}_{-0.4}$, scattering fraction of $1.7^{+3.0}_{-1.2}\%$, and an obscuring column of $\nh=56^{+33}_{-20}\times 10^{22}$ cm$^{-2}$. Correcting for intrinsic absorption, this model indicates an unabsorbed X-ray luminosity of $\rm{L}_{2\--10\ \rm{keV}} = 1.9^{+0.6}_{-0.2}\times10^{43}$ erg s$^{-1}$. These results agree with the results found using the phenomenological approach above as well as the results of \citet{bianchi2008}, within the uncertainties. These results are also agree with the level of obscuration predicted from the relationship between the infrared 12$\mu$m and absorbed X-ray $2\--10$ keV luminosity for this merger system (see Figure~\ref{fig:lxl12plot}), which is $\nh \sim10^{23}$ cm$^{-2}$.

Attempts to fit this model using the MYTorus zeroth-order continuum, fluorescent emission line table, and scattered power-law yielded a lower photon index and slightly lower $\nh$ values ($\sim 3\times10^{23}$ cm$^{-2}$). With MYTorus we could identify the presence of a statistically significant iron line, with an equivalent width of roughly 200 eV, which is agrees with the level of $\nh$ determined here \textit{and} the equivalent width found by \citet{bianchi2008}. A similar unabsorbed luminosity, $\rm{L}_{2\--10\ \rm{keV}} \sim 10^{42}$ erg s$^{-1}$, is found using MYTorus.

\subsection{J2356-1016: Single AGN}
The northwestern \chandra{} source (Galaxy 1), detected with a significance of 22.8$\sigma$ with hardness ratio of 0.79, represents a firm detection of an X-ray point source in this merger with an absorbed X-ray luminosity of $\rm{L}_{2\--10\ \rm{keV}} = 5.48\pm0.79 \times10^{42}$ erg s$^{-1}$, which is over an order of magnitude higher than that expected from contributions by XRBs. Examining the relationship between the absorbed X-ray luminosity and the 12 micron luminosity (see  Table~\ref{table:lxl12} and Figure~\ref{fig:lxl12plot}), we infer an obscuring column of $1.8^{+1.9 }_{-1.0}\times10^{23}$ cm$^{-2}$, slightly higher than that inferred via spectral analysis (as discussed below). We also report the detection of one coronal line in this galaxy nucleus, [SiVI], providing robust confirmation of an AGN in the nucleus of Galaxy 1. No X-ray point source was detected for Galaxy 2. The optical line ratios of the BPT diagram (shown in Figure~\ref{fig:sample_images}) indicate that both Galaxy 1 and 2 are star-forming galaxies rather than AGN hosts, which is at odds with our identification of an AGN in Galaxy 1. There are two SDSS spectra available for this merger coinciding with the two galaxy nuclei which classify Galaxy 1 as a QSO starburst broadline while Galaxy 2 is classified as a starburst galaxy. 

\begin{figure}[h!]
    \centering
    \begin{minipage}[t]{0.45\textwidth}
        \centering
        \includegraphics[width=1.1\textwidth]{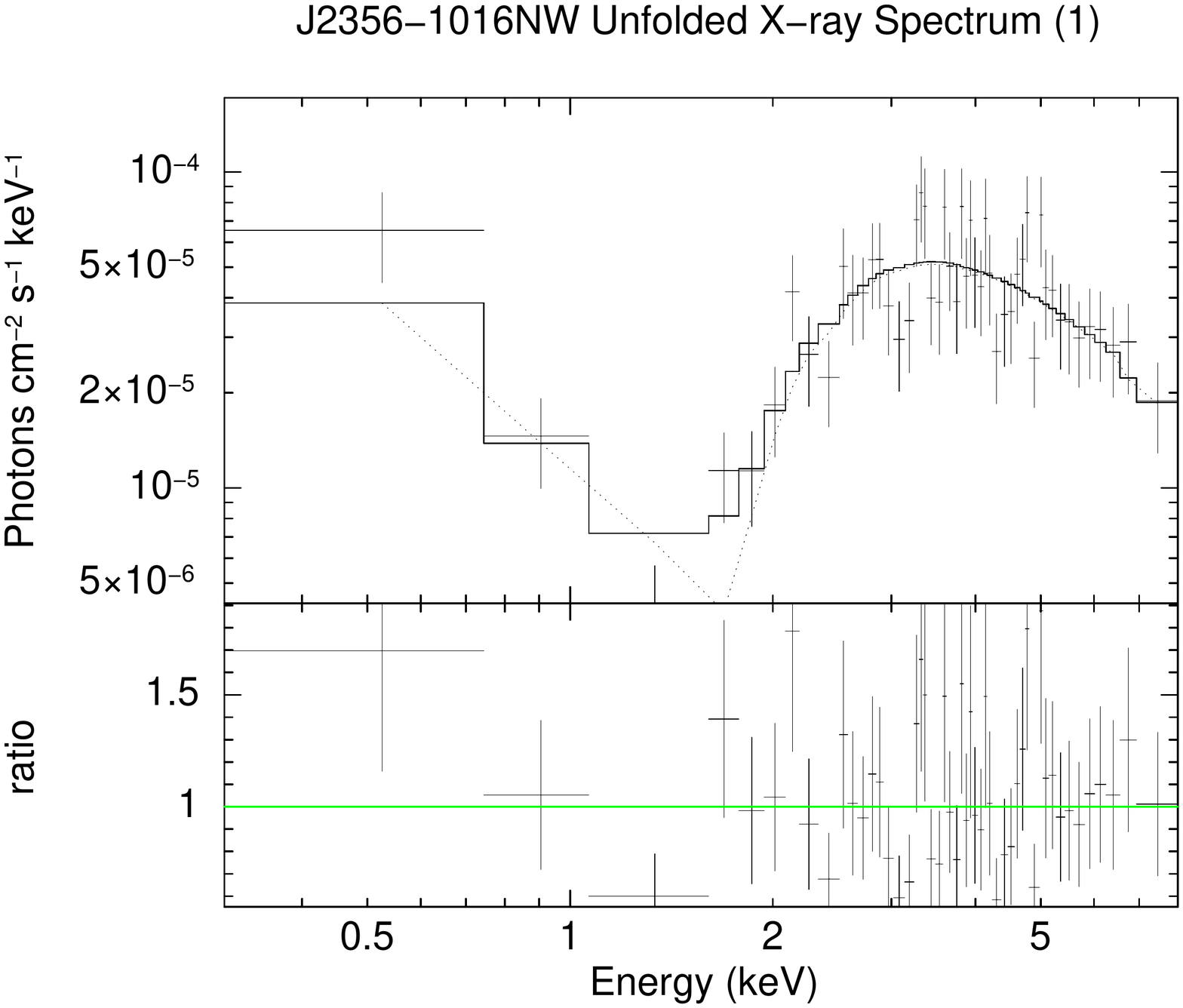} 
        \caption{The unfolded $0.3\--8$ keV X-ray spectrum of J2356-1016NW fit in \textsc{xspec} using the phenomenological approach. The data are best fit with an absorbed power-law with a scattered power-law component.}
        \label{fig:J235_phen}
    \end{minipage}\hspace{4mm}
    \begin{minipage}[t]{0.45\textwidth}
        \centering
        \includegraphics[width=1.1\textwidth]{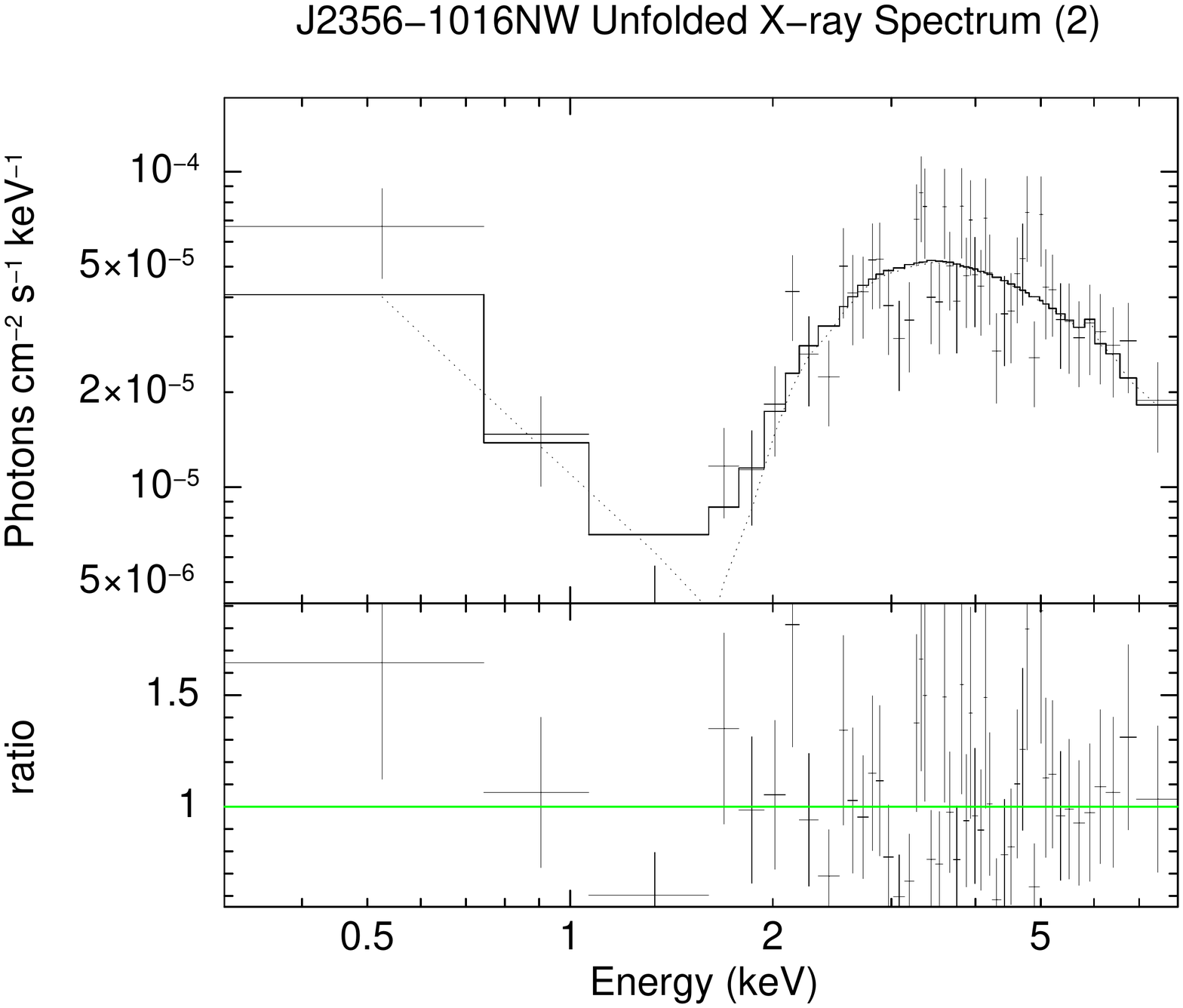} 
        \caption{The unfolded $0.3\--8$ keV X-ray spectrum of J2356-1016NW fit in \textsc{xspec} using the BNTorus approach. The data are best fit with the base BNTorus model with a scattered power-law component.}
        \label{fig:J235_bnt}
    \end{minipage}
\end{figure}

\subsubsection{J2356-1016NW Spectral Analysis Results}
We present the spectrum for J2356-1016NW in Figure~\ref{fig:J235_phen}, which shows heavy depletion of the soft X-ray energies ($0.3\--2$ keV), which is a typical sign of heavy obscuration. Using the phenomenological model, we found that adding a scattered power-law component to the basic absorbed power-law improved the fit by $\Delta$C-Stat = 63.75, demonstrating that a scattered power-law was statistically significant to the fit. We found no excess above the absorbed power-law component in the range of $6\--7$ keV, and we report (by fitting with and without a Gaussian emission line component) a $\Delta$C-Stat = $0 < 2.71$, indicating a Gaussian emission component was statistically insignificant for this fit. The data are therefore best fit using an absorbed power-law with a scattered power-law component. The model exhibits a photon index of $\Gamma=2.0^{+0.4}_{-0.4}$ and an obscuring column of $\nh=8.6^{+1.7}_{-1.6}\times10^{22}$ cm$^{-2}$ (see Table~\ref{table:phenspec}).  While the column density is slightly lower than that predicted by theoretical models \citep{blecha2018}, we note that the $\nh$, within the uncertainties, does enter into the $\sim10^{23}$ cm$^{-2}$ regime. Correcting for intrinsic absorption, this model indicates an unabsorbed X-ray luminosity of $\rm{L}_{2\--10\ \rm{keV}} = 4.7^{+0.8}_{-2.5}\times10^{43}$ erg s$^{-1}$, with which we robustly confirm this source as an AGN. Comparing the absorbed X-ray $2\--10$ keV luminosity to the infrared $\rm{L}_{12\mu \rm{m}}$ luminosity (see Figure~\ref{fig:lxl12plot}) we expect the $\nh$ to be at least $10^{23}$ cm$^{-2}$.

Examining the system with the BNTorus model (Figure~\ref{fig:J235_bnt}), we found that the introduction of a scattered power-law to the base model yielded a statistically significant improvement to the fit  ($\Delta$C-Stat$ = 60.93 > 2.71$). All fits attempted with the \textsc{apec} component yielded either nonphysical values for parameters or were less statistically significant and thus were discarded. We therefore find for this method the data are best fit using the base BNTorus model plus a scattered power-law component. The model indicates a photon index of $\Gamma=2.2^{+0.5}_{-0.4}$, a scattering fraction of $0.6^{+0.8}_{-0.4}\%$, and an obscuring column of $\nh=8.2^{+1.4}_{-1.6}\times 10^{22}$ cm$^{-2}$. Correcting for intrinsic absorption, this model indicates an unabsorbed X-ray luminosity of $\rm{L}_{2\--10\ \rm{keV}} = 4.7^{+0.7}_{-2.4}\times10^{43}$ erg s$^{-1}$. These results agree with the results found using the phenomenological approach above. We do find $\nh$ to be slightly lower here than that expected from the relationship between the infrared 12$\mu$m and absorbed X-ray $2\--10$ keV luminosity for this merger (see Figure~\ref{fig:lxl12plot}), which is $\nh \sim10^{23}$ cm$^{-2}$.

Attempts to fit this model using a MYTorus zeroth-order continuum with a scattered power-law component yielded a lower photon index, a higher scattering fraction, and a slightly lower $\nh$ value. A similar unabsorbed luminosity, $\rm{L}_{2\--10\ \rm{keV}} \sim 10^{42}$ erg s$^{-1}$, is found using MYTorus.

\bibliographystyle{yahapj}

\end{document}